\documentclass[preprint,authoryear,12pt]{elsarticle}

\usepackage[utf8]{inputenc}

\usepackage{subcaption}
\usepackage{graphicx}
\usepackage{amsmath}
\usepackage{bbm}
\usepackage{amsfonts}
\usepackage{array}
\usepackage{multirow}
\usepackage[table]{xcolor}
\usepackage{dsfont}

\usepackage{scalerel}
\usepackage{hyperref}

\usepackage{setspace}
\usepackage{caption}
\usepackage{etoolbox}
\AtBeginEnvironment{table}{\doublespacing}
\AtBeginEnvironment{figure}{\doublespacing}

\biboptions{authoryear}

\DeclareMathAlphabet{\mathpzc}{OT1}{pzc}{m}{it}

 \usepackage{pdfrender}

\newtheorem{theorem}{Theorem}
\newtheorem{corollary}{Corollary}
\newtheorem{definition}{Definition}
\newtheorem{lemma}{Lemma}

\allowdisplaybreaks

\begin{document}

\begin{frontmatter}

\title{Modeling Higher-Order Interactions in Sparse and Heavy-Tailed Neural Population Activity}

\author[1]{Ulises Rodr\'iguez-Dom\'inguez\corref{cor1} \href{https://orcid.org/0000-0001-9192-584X}{\includegraphics[scale=0.5]{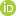}}}
\ead{ulises.rodriguez.dominguez@ciencias.unam.mx}
\author[2,3]{Hideaki Shimazaki \corref{cor2}
\href{https://orcid.org/0000-0001-7794-3064}{\includegraphics[scale=0.5]{ORCID.png}}}
\ead{h.shimazaki@i.kyoto-u.ac.jp}

\address[1]{Faculty of Science, National Autonomous University of Mexico (UNAM), Mexico}
\address[2]{Graduate School of Informatics, Kyoto University, Japan}
\address[3]{Center for Human Nature, Artificial Intelligence, and Neuroscience (CHAIN), Hokkaido University, Japan}

\begin{abstract}
Neurons process sensory stimuli efficiently, showing sparse yet highly variable ensemble spiking activity involving structured higher-order interactions. Notably, while neural populations are mostly silent, they occasionally exhibit highly synchronous activity, resulting in sparse and heavy-tailed spike-count distributions. However, its mechanistic origin  — specifically, what types of nonlinear properties in individual neurons induce such population-level patterns — remains unclear. In this study, we derive sufficient conditions under which the joint activity of homogeneous binary neurons generates sparse and widespread population firing rate distributions in infinitely large networks. We then propose a subclass of exponential family distributions that satisfy this condition. This class incorporates structured higher-order interactions with alternating signs and shrinking magnitudes, along with a base-measure function that offsets distributional concentration, giving rise to parameter-dependent sparsity and heavy-tailed population firing rate distributions. Analysis of recurrent neural networks that recapitulate these distributions reveals that individual neurons possess threshold-like nonlinearity followed by supralinear activation that jointly facilitates sparse and synchronous population activity. These nonlinear features resemble those in modern Hopfield networks, suggesting a connection between widespread population activity and the network's memory capacity. The theory establishes sparse and heavy-tailed distributions for binary patterns, forming a foundation for developing energy-efficient spike-based learning machines. 
\end{abstract}

\begin{keyword}
Sparse distribution\sep widespread distribution\sep binary patterns\sep higher-order interactions\sep heavy-tailed distribution \sep exponential family distribution \sep neural population activity\sep nonlinear activation function
\end{keyword}

\end{frontmatter}

\section{Introduction}

One of the fundamental constraints placed on neural systems operating in natural environments is efficiency \citep{attneave1954some, barlow1961possible}, 
which often imposes sparseness on their activities. 
Neurons therefore exhibit sparsity in various aspects of their activity patterns \citep{willmore2001characterizing,WillmoreMazerGallant_2011} such as in the distribution of individual neuron responses to multiple stimuli (lifetime sparseness) \citep{VinjeGallant_2000} and the response distribution of the activity in a population of neurons (population sparseness) \citep{VinjeGallant_2000,YenBakerGray_2007,Froudakaris_etal_2014}. 
For continuous distributions, sparsity is characterized using the statistical moments such as kurtosis or the coefficient of variation \citep{field1994goal, rolls1995sparseness, willmore2001characterizing}. Many parametric sparse distributions for continuous values have been proposed, often within the context of the Bayesian prior for sparse coding \citep{OLSHAUSEN_FIELD_1997}. However, sparse distributions for spiking activities of neurons have not been established. Further, how the sparse profiles in the population distributions arise from the collective activity of interacting neurons remains elusive.

One approach to understanding cooperative spiking activities of neurons involves analyzing near-simultaneous activities 
by binarizing the spiking activity within short time windows. When expressed by the exponential family distributions with interactions of multiple orders, this analysis can reveal interactions among subsets of neurons in the population. Interactions among more than two neurons are often termed higher-order interactions (HOIs). A model that lacks non-zero HOIs is obtained by constructing a distribution that maximizes Shannon entropy while constraining activity rates of individual neurons and joint activity rates of neuron pairs \citep{jaynes1957information, Schneidmain_etal_2006}. This model, in which all HOIs are fixed at zero, is called a pairwise maximum entropy (MaxEnt) model (a.k.a., the spin-glass or Ising model in statistical physics and the Boltzmann machine in machine learning). The pairwise MaxEnt model highlights the role of HOIs. The joint activity of more than two neurons produced by this model appears as chance coincidences expected from the activity rates of individual neurons and neuron pairs. Consequently, if nonzero HOIs exist, they indicate deviations in the joint activities of more than two neurons from these chance coincidences.

There is considerable evidence suggesting that HOIs are necessary for characterizing the activity of neural populations. Early in vitro studies reported that the pairwise MaxEnt model accounted for approximately 90\% of activity patterns of small populations \citep{Schneidmain_etal_2006, Shlens_etal_2006}, implying that HOIs made only marginal contributions. However, HOIs may become more prominent as the population size increases \citep{roudi2009pairwise, Ganmor_etal_2011, Tkacik_etal_2014, Barreiro_etal_2014}. Accordingly, significant HOIs were later found ubiquitously in both in vitro and in vivo neurons \citep{Yu_etal_2011, Ohiorhenuan_etal_2010, Tkacik_etal_2014, Montani_etal_2009, Shimazaki_etal_2015}. Analyzing HOIs enabled researchers to uncover the underlying circuitry \citep{shomali_etal_2023_uncovering} and provided insights into their stimulus coding \citep{CaycoGajic_etal_2015, ZylberbergSheaBrown_2015, Ohiorhenuan_etal_2010, Montani_etal_2009, Shimazaki_etal_2012}. 

One of the most striking features of neural activities involving HOIs is the widespread and heavy-tailed nature of their distribution. Here, by the widespread or non-concentrated distribution, we mean one whose density takes on a non-zero and finite value at every point within its support. Spike-count histograms for the number of simultaneously active neurons often exhibit widespread distributions, with notably distinct tails reflecting probabilities of highly synchronous states compared to independent or pairwise MaxEnt models \citep{Tkacik_etal_2014, Montani_etal_2009, Ganmor_etal_2011}.  
Another important feature involving HOIs is sparseness. In this study, we consider nonnegative, right-tailed distributions sparse if their dominant peak is concentrated at zero. 
Evidence of the sparse activity can be found in the spike-count histogram of individual neurons, such as retinal ganglion cells \citep{Berry_etal_1997_retinal_ganglion}, V1 neurons \citep{Schwartz_etal_2004}, and primary auditory cortex neurons \citep{MOSHITCH_NELKEN_2014}. Population-level histograms of neural activity also display sparse profiles. Neurons are only sparsely active over time, with the duration of a state in which all neurons are silent being significantly longer than the prediction made by the pairwise MaxEnt model in both in vitro \citep{Ganmor_etal_2011, Tkacik_etal_2014, Shimazaki_etal_2015} and in vivo \citep{Montani_etal_2009, Ohiorhenuan_etal_2010, Chettih_and_Harvey_2019, shomali_etal_2023_uncovering} studies. 
Figure \ref{fig:MODELS_MARRE_MAX_ENT} illustrates a population spike-count histogram of neural population activity, along with predictions from the independent and pairwise homogeneous MaxEnt models. Notably, the MaxEnt models struggle to capture the entire distribution, including the frequency of the total silence and the tail, emphasizing the necessity of modeling structured HOIs.

\begin{center}
  \includegraphics[width=0.8 \textwidth]{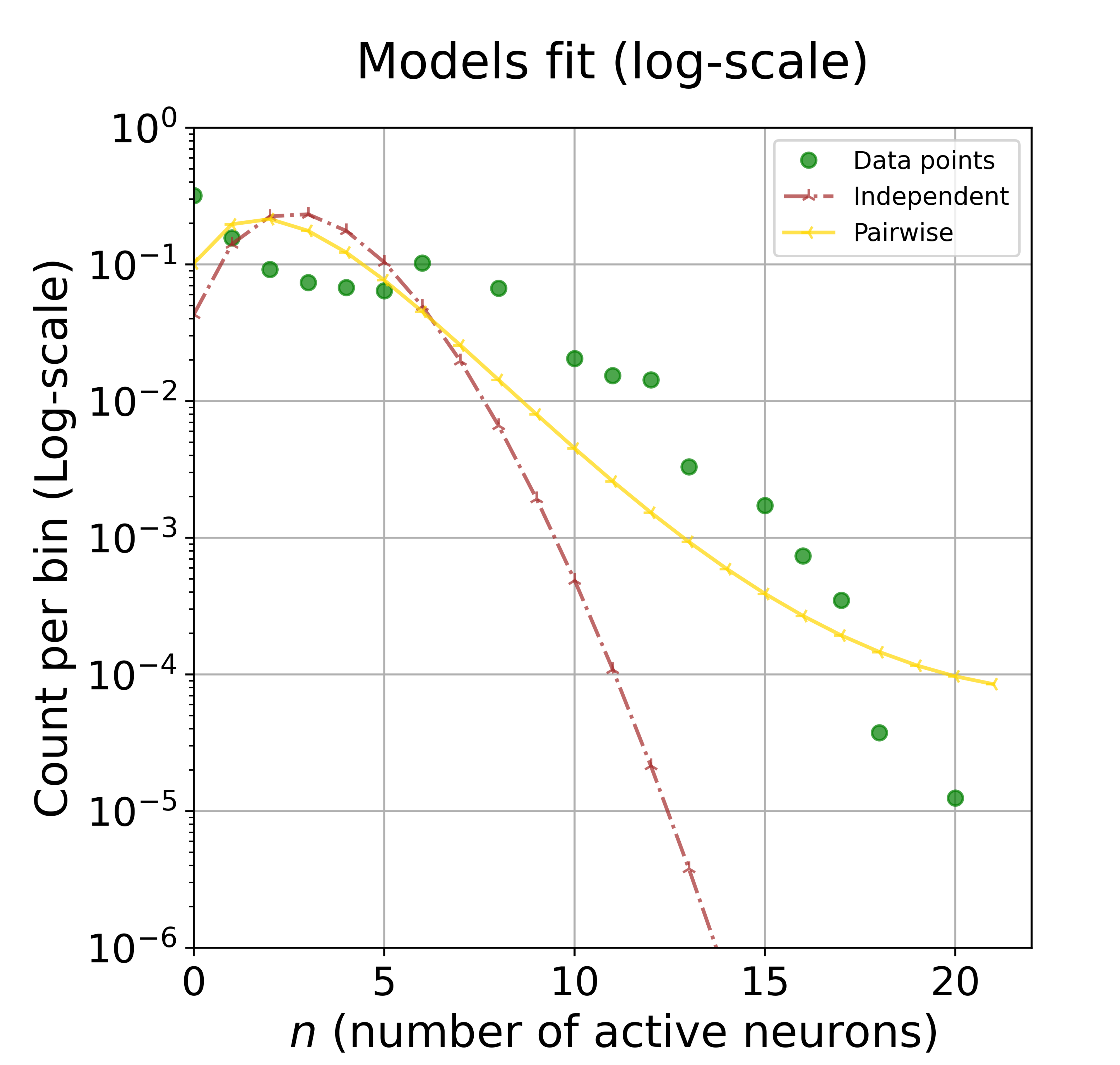}
\end{center}
\captionof{figure}{The spike-count population histogram of salamander retinal ganglion cells (N=40) responding to visual stimuli from a natural movie \citep{marre2017multi}. The histogram (green) in a logarithmic scale shows the frequency of observing patterns having $n$ active neurons in the binary spike data (time window $20$ ms). The independent (brown) and pairwise (yellow) homogeneous MaxEnt models are fitted to the data under the maximum likelihood principle.}
\label{fig:MODELS_MARRE_MAX_ENT}

It is, therefore, evident that HOIs play a critical role in shaping sparse, widespread, and heavy-tailed spike-count distributions of neuronal population activity. However, sufficient conditions and parametric models for such distributions remain unestablished. The necessity of non-zero HOIs in constructing widespread distributions was first pointed out by \cite{Amari_etal_2003}. 
In the current study, we present specific constraints on the spiking activities of a homogeneous neural population that result in distributions with all the properties above. Specifically, to ensure the sparse and widespread properties hold regardless of the population size (scale invariance), we propose to include the entropy-canceling base measure function as part of our modeling choice for neural activity distributions. We show that the standard MaxEnt model, which lacks the entropy-cancellation term, leads to a concentrated distribution in the limit of large neuronal populations due to the combinatorial explosion of patterns dominating the probabilities. Moreover, we introduce a new theoretical definition for heavy-tailed distributions within a compact support, highlighting the crucial role of the HOIs in shaping highly synchronous states in the tail of the population activity distribution. 
Building on these theoretical foundations, we introduce new parametric models belonging to the exponential family distribution, characterized by structured alternating and shrinking higher-order interactions. These models yield sparse and widespread population rate distributions in the limit of a large number of neurons while also exhibiting heavy tails. 

Interpreting the dynamics of a recurrent neural network that behaves as a sampler of the sparse distributions described above reveals that neuronal activation functions exhibit threshold-like nonlinearity, followed by a supralinear increase. This nonlinearity, promoting the sparse and heavy-tailed population activity, may arise from computations performed at the dendrites on top of the spike-generation nonlinearity at the axon hillock of the soma. In the context of machine learning, the proposed models with HOIs are variants of the modern Hopfield networks or dense associative memories \citep{krotov2016dense,demircigil2017model,krotov2023new}, which enhance their capacity by capturing the higher-order dependency of inputs through specific nonlinearities that are also related to Transformers \citep{ramsauer2020hopfield} and diffusion models \citep{ambrogioni2023search}. We show that the nonlinearity of our proposed models closely resembles that of the original dense associative memories. Thus, the present theory, which establishes sparse distributions for binary patterns, provides insights into the nonlinearity for enhancing modern learning machines and will serve as a building block for developing prior distributions for energy-efficient spike-based sparse coding in machine learning.

The paper is organized as follows. In Section \ref{SECTION_HOMOGENEOUS_POPULATION}, we describe a probability function of $(0,1)$ binary patterns using the exponential family distribution, assuming homogeneity over the neurons, and construct a population-count histogram, a distribution of the total activity in the population. We provide sufficient conditions that make a distribution widespread with its peak at a population spike count of zero in the limit of a large number of neurons. In this section, we additionally formalize the heavy-tailedness of a sparse and compactly supported distribution. In Section \ref{SECTION_ENTROPY_DOMINATED_CASE}, we show how a common scenario for homogeneous MaxEnt models leads to the widespread property being hindered due to entropy domination in the PMF for a large number of neurons. Then, in Section \ref{SECTION_ALTERNATING_SHRINKING_MODEL}, we introduce our alternating shrinking higher-order interaction models, whose probability densities become widespread and remain sparse in the limit of a large number of neurons. In this section, we fit the proposed models to the spiking activity of in vitro neurons and evaluate the predictive performance of the proposed models against alternatives. We also provide a perspective on these models from the viewpoint of recurrent neural networks with extended nonlinearity and summarize their key statistical properties. Finally, we conclude with a discussion in Section \ref{SECTION_DISCUSSION}.

\section{Homogeneous sparse population of neurons} \label{SECTION_HOMOGENEOUS_POPULATION}

The activity of $N$ neurons is represented by a set of binary random variables, using a column vector, $\mathbf{X}=[X_1,X_2,\ldots,X_N]^{\textsf{T}}$ where $X_i \in \{0,1\}$ and for which we assume stationarity. The $i$th neuron activity $X_i$ is $1$ if the neuron is active 
and $0$ otherwise. The probability of generating binary activity patterns $\mathbf{x}=[x_1,x_2,\ldots,x_N]^{\textsf{T}}$, where $x_i \in \{0,1\}$, is given as $\mathcal{P}\left( \mathbf{X} = \mathbf{x} \right)$. This probability mass function (PMF) can be written in the form of an exponential family distribution:

\begin{align}
    \mathcal{P}\left(\mathbf{X} = \mathbf{x} \right) &=
     \frac{h\left(\mathbf{x}\right)}{Z}\exp\left[ \sum\limits_{i=1}^{N} \theta_i x_i + \sum\limits_{i_1<i_2}  \theta_{i_1 i_2} x_{i_1} x_{i_2}
     + \sum\limits_{i_1<i_2<i_3}  \theta_{i_1 i_2 i_3} x_{i_1} x_{i_2} x_{i_3}
     \right.
     \nonumber\\
     &\phantom{======================}   + \hdots + \theta_{12 \hdots N} x_{1} x_{2} \cdots x_{N}  \Bigg ],
     \label{eq_binary_interactions_pmf}
\end{align}
\noindent where $Z$ is a normalization term, and the parameters $\{ \theta_i \}_{i=1}^{N}$, $\{ \theta_{i_1 i_2} \}_{i_1<i_2}$, $\ldots$, $\theta_{1 \hdots N}$ are called natural parameters. They characterize interactions among subset neurons indicated by the subscript indices \citep{Martignol_etal_2000, NakaharaAmari_2002}. The exponential family distribution allows the base measure function $h\left( \mathbf{x} \right)$ to be a general nonnegative function of the vector pattern $\mathbf{x}$. Here, we will assume that $h\left(\mathbf{x}\right)$ is a function of the total activity, $\sum_{i=1}^{N} x_i$. Although Eq.~\eqref{eq_binary_interactions_pmf}  can realize arbitrary probabilities for all possible patterns even if $h\left(\mathbf{x}\right)=1$, as we will show, the introduction of an appropriate base measure function simplifies the conditions for the sparse widespread distributions and for modeling the neural interactions.  For simplicity, we use $\mathcal{P}\left(\mathbf{x} \right)$ to represent $\mathcal{P}\left(\mathbf{X} = \mathbf{x} \right)$.

We study the activity of a population of homogeneous neurons. Homogeneity is an important assumption that ignores specific preference over some neural activity patterns. Nonetheless, homogeneity allows us to change the analysis focus from a local to a global view of the sparse neural population activity in a region, facilitating the identification of theoretical properties. The binary activity of the homogeneous population is described by using single parameters $\theta_k$ ($k=1,2,\ldots, N$) for all the combinatorial $k$-th order interactions in Eq.~\eqref{eq_binary_interactions_pmf}
\begin{align}
    \mathcal{P}\left( \mathbf{x} | \boldsymbol{\theta}_N \right) = &
  \frac{h\left( \sum_{i=1}^{N}  x_i \right)}{Z}\exp\left[ \theta_1 \sum_{i=1}^{N} x_i + \theta_2 \sum_{i_1<i_2} x_{i_1} x_{i_2} + \hdots + \theta_N x_{i_1} x_{i_2} \hdots  x_{i_N} \right],
  \label{eq_binary_homogeneous_interactions_pmf}
\end{align}
where $\boldsymbol{\theta}_N = (\theta_1, \theta_2, \ldots, \theta_N)$. This model extends the theoretical work by \cite{Amari_etal_2003}
, where $h\left( \sum_{i=1}^{N} x_i \right) = 1$. 
The population activity of the homogeneous neurons is characterized by the distribution of the number of active neurons in the population. For homogeneous neurons, any individual binary pattern where $n$ neurons are active has the same probability. Therefore, the probability of having $n$ active neurons in the population is given by
\begin{align}
  \mathcal{P}\left( \left . \sum_{i=1}^{N} X_i = n \right|  \boldsymbol{\theta}_N \right) 
  = & \left( \begin{array}{c}
       N\\
       n 
  \end{array} \right) \mathcal{P}\left( x_{1}=1, \hdots, x_{n}=1, x_{n+1}=0, \hdots, x_{N}=0 \;| \; \boldsymbol{\theta}_N \right) \nonumber \\
  =& \left( \begin{array}{c}
       N\\
       n 
  \end{array} \right) \frac{h\left(n\right) }{Z} \exp\left[ \left( \begin{array}{c}
       n\\
       1 
  \end{array} \right) \theta_1 + \left( \begin{array}{c}
       n\\
       2 
  \end{array} \right) \theta_2 + \hdots + \left( \begin{array}{c}
       n\\
       n 
  \end{array} \right) \theta_n \right] \nonumber \\
  =& 
\left( \begin{array}{c}
       N\\
       n 
  \end{array} \right) \frac{h\left( n \right) }{Z}
  \exp\left[ \sum_{k=1}^{n} \left( \begin{array}{c}
       n\\
       k 
  \end{array} \right) \theta_k \right].
  \label{eq_binary_homogeneous_interactions_pmf_02}
\end{align}

Let the fraction of active neurons (or population rate) be $R_N = \frac{1}{N} \sum_{i=1}^{N} X_i$. Using Eq.~\eqref{eq_binary_homogeneous_interactions_pmf_02}, the PMF of the random variable $R_N$, $\mathcal{P}\left( R_N = r_N | \boldsymbol{\theta}_N \right)$, where $r_N \in S_r$ with $S_r \equiv \left\{ 0, \frac{1}{N}, \frac{2}{N}, \hdots, 1 \right\}$, is 
\begin{align}
  \mathcal{P}\left( R_N = r_N | \boldsymbol{\theta}_N \right)
  = & \mathcal{P}\left(\left. \sum_{i} X_i = N r_N \right| \boldsymbol{\theta}_N \right) \nonumber \\
  = & \left( \begin{array}{c}
   N\\
   N r_N 
\end{array} \right) \frac{ h\left( N r_N \right) }{Z} \exp\left[ \sum_{k=1}^{N r_N} \left( \begin{array}{c}
   N r_N \\
   k 
\end{array} \right) \theta_k \right].
\label{eq_homogeneous_r_pmf}
\end{align}
We call this a PMF of the discrete population rate, and we rewrite it as 
\begin{align}
  \mathcal{P}\left( R_N = r_N | \boldsymbol{\theta}_N \right)
= & \frac{1}{Z} \exp\left[ N G_N(r_N; \boldsymbol{\theta}_N) \right],
\label{eq_homogeneous_r_pmf_g_function}
\end{align}
where
\begin{align}
  G_{N}\left(r_N ; \boldsymbol{\theta}_N \right) = \frac{1}{N} \log
  \left( \begin{array}{c}
   N\\
   N r_N 
\end{array} \right)
  + \frac{1}{N} \log h\left( N r_N \right) + \frac{1}{N} Q_N\left(r_N ; \boldsymbol{\theta}_N \right),
\label{eq_g_r_function}
\end{align}
using a polynomial term defined as 
\begin{equation}
  Q_{N}\left(r_N ; \boldsymbol{\theta}_N \right) = \sum_{k=1}^{N r_N} \left( \begin{array}{c}
   N r_N\\
   k 
\end{array} \right) \theta_k.
\label{eq_polynomial_r}
\end{equation}
We note that the new underlying base measure function for such population rate distribution (Eq.~\eqref{eq_homogeneous_r_pmf}) consists of the binomial term multiplied by the $h\left( \cdot \right)$ function, i.e., $\left( \begin{array}{c}
   N\\
   N r_N 
\end{array} \right) h\left( N r_N \right)$. As we stated before, such a base measure function could alternatively be represented in a different way inside the (possibly non-polynomial) function $Q_{N}\left(\cdot \right)$ as a function of the active neurons given the canonical parameters. Nonetheless, the representation we chose facilitates analysis in the limit of a large population of neurons, as we will see. In the following, we use $\mathcal{P}\left(r_N | \boldsymbol{\theta}_N \right)$ to represent the PMF above.

We are interested in the behavior of the PMF (Eq.~\eqref{eq_homogeneous_r_pmf_g_function}) in the limit of a large number of neurons ($N \rightarrow \infty$): Namely, the probability density function (PDF) given through the relation $p\left(r | \boldsymbol{\lambda} \right) dr = \lim_{N \to \infty} \mathcal{P}\left( r_N | \boldsymbol{\theta}_N \right)$, where $r$ is the continuous population rate defined in the support $[0,1]$ and $\boldsymbol{\lambda}$ is a set of parameters for the PDF. 
We wish to know the conditions under which this PDF is sufficiently concentrated, near its peak of $0$. Such a PDF would be relevant to model experimentally observed sparse population activity across different cortical populations, where arbitrarily low firing rates were exhibited by most neurons \citep{Wohrer_etal_2013_PopWideDistr}. 
Following Amari et al.'s framework to construct widespread distributions \citep{Amari_etal_2003}, we provide a new theorem below giving sufficient conditions for sparse and widespread distributions.

\begin{theorem}
\label{theorem_widespread_zero_peaked}
Let $G_N \left(r_N ; \boldsymbol{\theta}_N \right)$ be a non-positive strictly decreasing function with finite values for $r_N \in S_r$. If $N G_N \left(r_N ; \boldsymbol{\theta}\right)$ is asymptotically bounded with respect to $N$ as follows
\begin{equation}
    \label{eq_theorem_widespread_zero_peaked_order}
    \mathcal{O}\left(N G_N \left(r_N ; \boldsymbol{\theta}_N \right) \right) = \mathcal{O}\left(1\right),
\end{equation}
then the corresponding probability density function given through $\; p\left(r | \boldsymbol{\lambda} \right) dr = \lim_{N \to \infty}\ \mathcal{P}\left( r_N | \boldsymbol{\theta}_N \right) \;$ is widespread in $(0,1]$ with a single non-concentrated maximum at $0$.
\end{theorem}
For a proof, the reader can refer to \ref{APPENDIX_THEOREM_PROOF}. 

Specifically, in this study, we propose to use the following form of the entropy-cancelling base measure function $h(\mathbf{x})$: 
\begin{align}
    h\left( \sum\displaystyle_{i=1}^{N} x_i \right)  
    &= 1 \left/ \left( \begin{array}{c}
   N\\
   \sum_{i=1}^{N} x_i 
\end{array} \right) \right.
\label{eq_pmf_h_function}
\end{align}
With this form, the first two terms in Eq.~\eqref{eq_g_r_function} cancel out, resulting in 
\begin{equation}
\label{eq_G_r_function_to_Q_r_function}
G_N\left(r_N ; \boldsymbol{\theta}_N \right) = \frac{1}{N} Q_N \left(r_N ; \boldsymbol{\theta}_N \right).
\end{equation}

Thus, we obtain the following corollary:  
\begin{corollary}
\label{corollary_h_function_g_polynomial}
Let $h(\mathbf{x})$ be given by Eq.~\eqref{eq_pmf_h_function}. If the polynomial term $Q_N \left(r_N ;\boldsymbol{\theta}_N\right)$ satisfies
\begin{equation}
    \label{eq_corollary_Q_order}
    \mathcal{O}\left(Q_N\left(r_N ; \boldsymbol{\theta}_N \right) \right) = \mathcal{O}\left(1\right),
\end{equation}
and if $q\left(r; \boldsymbol{\lambda} \right) = \lim_{N \to \infty} Q_N \left(r_N ; \boldsymbol{\theta}_N \right)$ is a non-positive strictly decreasing function, then the probability density function $p\left(r | \boldsymbol{\lambda} \right)$ is widespread in $(0,1]$ with a single non-concentrated maximum at $0$.
\end{corollary}

We emphasize that canceling the entropy term using the base measure function is a key aspect of our modeling choice. This approach simplifies the conditions required to derive sparse, widespread distributions, as demonstrated in Corollary 1. Without this cancellation, the adjustment would need to be incorporated into the polynomial term $Q_N\left(r_N ; \boldsymbol{\theta}_N\right)$, significantly increasing its complexity. We also note that the non-positivity of $G_N \left(r_N ; \boldsymbol{\theta}_N \right)$ is not required for the proof of Theorem 1, provided that the integrability of the limiting density is preserved. Similarly, non-positivity of $q\left(r; \boldsymbol{\lambda} \right)$ is not required in Corollary 1. However, for analytical simplicity and modeling convenience, we impose the non-positivity conditions when constructing sparse and widespread distributions.

We introduce the two simplest homogeneous sparse models in the following subsection (Subsection \ref{SUBSECTION_INDEPENDENT_SECOND_ORDER}).
Then, in Section \ref{SECTION_ENTROPY_DOMINATED_CASE}, we present a common scenario to which Theorem \ref{theorem_widespread_zero_peaked} does not apply, resulting in a concentrated distribution. 
In Section \ref{SECTION_ALTERNATING_SHRINKING_MODEL}, we present our proposed model whose population rate PMF is a particular case of Eq.~\eqref{eq_homogeneous_r_pmf_g_function}. The distribution satisfies the conditions in Corollary \ref{corollary_h_function_g_polynomial} and converges to a widespread continuous distribution with parameter-dependent sparsity.

\subsection{First and second-order homogeneous sparse models} \label{SUBSECTION_INDEPENDENT_SECOND_ORDER}

Here, we introduce the simplest homogeneous model that produces sparse population activity, i.e., a homogeneous population of binary neurons with only the first-order parameters ($\theta_2=\theta_3=\ldots=\theta_N=0$). Using $\theta_1=-\mathpzc{f}/N$, the PMF of the binary population is given as 
\begin{equation}
\label{eq_homogeneous_first_order_negative_interactions_binary_pmf}
    \mathcal{P}\left( \mathbf{x} | \mathpzc{f} \right) = \frac{ h\left( \sum\displaystyle_{i=1}^{N} x_i \right) }{ Z } \exp\left[ -\mathpzc{f} \frac{\sum\displaystyle_{i=1}^{N} x_i}{N}  \right],
\end{equation}
\noindent where we assume that $\mathpzc{f}>0$ and the function $h\left(\cdot\right)$ is given by Eq.~\eqref{eq_pmf_h_function}. With this $h\left(\cdot\right)$ function that cancels out with the binomial term in Eq.~\eqref{eq_g_r_function}, the corresponding population rate PMF (Eq.~\eqref{eq_homogeneous_r_pmf_g_function}) is given as 
\begin{align}
   \mathcal{P}\left(r_N | \theta_1 \right) 
   &= \frac{1}{Z} e^{-\mathpzc{f} r_N}.
\label{eq_homogeneous_first_order_negative_interactions_r_pmf}
\end{align}

The corresponding continuous PDF is obtained through
\begin{align}
p\left(r | \mathpzc{f} \right) dr = \lim_{N \to \infty} \mathcal{P}\left( r_N | \theta_1 \right) = \frac{1}{Z} e^{-\mathpzc{f} r} dr,  
\label{eq_homogeneous_first_order_negative_interactions_r_pdf}
\end{align}
where the normalization constant is obtained as
\begin{align}
Z = \int_{0}^{1} e^{-\mathpzc{f} r }  dr 
 = \frac{1 - e^{-\mathpzc{f}}}{\mathpzc{f}}.
\label{eq_homogeneous_first_order_negative_interactions_r_pdf_Z}
\end{align}
Since $\mathcal{O}\left(-\mathpzc{f} r_N\right) = \mathcal{O}\left(1\right)$ and $q\left(r ; \mathpzc{f} \right) = - \mathpzc{f} r$ is a strictly decreasing function, the PDF in Eq.~\eqref{eq_homogeneous_first_order_negative_interactions_r_pdf} is widespread in $(0,1]$ with a single non-concentrated maximum at $0$ (Corollary \ref{corollary_h_function_g_polynomial}). The sparsity in such PDF is controlled by the $\mathpzc{f}$ parameter. See 
\ref{APPENDIX_SUBSECTION_INDEPENDENT_HOM_MODEL} for the distribution's mean and variance.

This density corresponds to the PDF of an exponential distribution with parameter $\mathpzc{f} > 0$ but with a compact support in $[0,1]$, instead of the support in $[0,\infty)$, and also serves as a baseline for investigating the effect of the pairwise and the HOIs in shaping the sparse population activity distribution. 
Additionally, the first-order model serves as a reference baseline to determine if other distributions are heavy-tailed or not in this compact support $[0,1]$, similarly to how the exponential distribution is used to evaluate the heavy-tailedness of other distributions with support in $\mathbb{R}$ (see Subsection \ref{SUBSECTION_HEAVY_TAILEDNESS}).

As the second simplest homogeneous model, we obtain the model having up to the second-order parameters. The canonical parameters given by $\theta_1 = \frac{\mathpzc{f}_1}{N} +  \frac{\mathpzc{f}_2}{N^2}$ and $\theta_2 = 2  \frac{\mathpzc{f}_2}{N^2}$ lead to a polynomial $Q_{N}\left( r_N ; \boldsymbol{\theta}_N \right) = \mathpzc{f}_1 r_N + \mathpzc{f}_2 r_N ^2 $. Using again Eq.~\eqref{eq_pmf_h_function} for $h\left(\cdot\right)$ we obtain a continuous PDF as
\begin{align}
p\left(r | \mathpzc{f}_1, \mathpzc{f}_2 \right) dr = \lim_{N \to \infty} \mathcal{P}\left( r_N | \theta_1, \theta_2 \right) = \frac{1}{Z} \exp\left[ {\mathpzc{f}_1 r + \mathpzc{f}_2 r^2 } \right] dr,  
\label{eq_homogeneous_second_order_sparse_r_pdf}
\end{align}
See \ref{APPENDIX_SUBSECTION_SECOND_HOM_MODEL} for the normalization constant.

\subsection{Heavy-tailedness of compactly supported sparse homogeneous distributions} \label{SUBSECTION_HEAVY_TAILEDNESS}

Traditionally, a distribution is known to be heavy-tailed if its tail is heavier than the tail of any exponential distribution \citep{Nair_Wierman_Zwart_2022}. For right-tailed distributions, the 
precise definition involves taking the ratio between the right limits of the test and the exponential distributions. When a tail of a test distribution decays slower than that of an exponential distribution, the ratio becomes $+\infty$ in the limit. If this ratio is less than $+\infty$, the test distribution is considered not heavy-tailed. 

The exponential distribution arises as the simplest sparse distribution in our compactly supported framework when only the first-order interactions are present. Following the tailedness definition in the unbounded support, we consider the exponential distribution as the baseline for the tail comparison. More precisely, we consider that if the tail of a sparse distribution, i.e., its complementary cumulative distribution function $\bar{F}_{\boldsymbol{\lambda}}$, has values larger than those of the tail of an exponential distribution $\bar{F}_{\mathpzc{f}} ^{exp}$ under a sparsity parameter value $\mathpzc{f}$ along its support (not including the extremes at 0 and 1), then such distribution is heavy-tailed.

In the homogeneous framework defined over the compact support in $[0,1]$, multiple parameter-dependent sparse distributions belonging to the exponential family can be constructed using our base measure function. For an $\mathpzc{f}$-sparse distribution function $F_{\boldsymbol{\lambda}}$ whose density $p(r | \boldsymbol{\lambda})$ satisfies Corollary \ref{corollary_h_function_g_polynomial}, we define the heavy-tailedness as follows.

\begin{definition}
\label{definition_heavy_tail_distr_in_r}
An $\mathpzc{f}$-sparse distribution function $F_{\boldsymbol{\lambda}}$ with support in $[0,1]$ is heavy-tailed if and only if for $\mathpzc{f} \in \left(0,\infty\right)$
\begin{equation}
    \label{eq_definition_heavy_tail_distr_in_r}
    \bar{F}_{\boldsymbol{\lambda}}\left(r\right) > \bar{F}^{exp} _{\mathpzc{f}}\left(r\right) \;\;\forall r \in \left(0,1\right),
\end{equation}
where $\boldsymbol{\lambda} \supseteq \mathpzc{f}$ and $\bar{F}$ indicates the complementary cumulative distribution function. 
\end{definition}

Since $\bar{F}\left(r \right) = 1 - F\left( r \right)$, Eq.~\eqref{eq_definition_heavy_tail_distr_in_r} can be alternatively stated for $\mathpzc{f} \in \left(0,\infty\right)$ as
\begin{equation}
    \label{eq_definition_heavy_tail_distr_in_r_02}
    F_{\boldsymbol{\lambda}}\left(r\right) < F^{exp} _{\mathpzc{f}}\left(r\right) \;\;\forall r \in \left(0,1\right).
\end{equation}
The above inequality is an instance of the first-order stochastic dominance of $F_{\boldsymbol{\lambda}}\left(r\right)$ over $F^{exp} _{\mathpzc{f}}\left(r\right)$ \citep{Mulero_etal_2017_stochastic_dom,Yildiz_2015_mit}. Using any non-decreasing function $u(\cdot)$, it is equivalent to 
\begin{align}
    \label{eq_expectations_comparison_case_01_main}
    \mathbb{E}_{R | \boldsymbol{\lambda} } \left[ u\left(r\right) \right] > \mathbb{E}^{exp} _{R | \mathpzc{f}}\left[ u\left( r \right) \right] \;\;\forall r \in \left(0,1\right),
\end{align}
See \ref{APPENDIX_HEAVY_TAILEDNESS} for more discussion. 

We note that the alternating shrinking models, to be introduced in Section \ref{SECTION_ALTERNATING_SHRINKING_MODEL}, differ from $F^{exp} _{\mathpzc{f}}\left(r\right)$ only in the forms of the shrinking coefficients, while sharing both the sparsity-inducing parameter $\mathpzc{f}$ and the base measure function, in accordance with Corollary \ref{corollary_h_function_g_polynomial}. In Section \ref{SECTION_EXP_NEURAL_STAT}, we will employ these definitions to prove that the proposed distributions are heavy-tailed.

\section{Entropy-dominated homogeneous population} \label{SECTION_ENTROPY_DOMINATED_CASE}

This section demonstrates that the condition $h\left(\mathbf{x}\right) = 1$, commonly used in the standard pairwise MaxEnt model, fails to achieve entropy cancellation, leading to a concentrated distribution. 

We now analyze the behavior of the homogeneous PMF (Eq.~\eqref{eq_homogeneous_r_pmf_g_function}) with $h\left( N r_N \right) = 1$ as the number of neurons $N$ grows to infinity while keeping the order of the polynomial part $Q_N\left(r_N ; \boldsymbol{\theta}_N\right)$ constant in $N$, i.e.,
\begin{equation}
    \label{eq_polynomial_r_order}
    \mathcal{O}\left(Q_N\left(r_N ; \boldsymbol{\theta}_N \right)\right) = \mathcal{O}\left( 1 \right).
\end{equation}

Coupled with the Stirling formula for factorials using the order notation, i.e.,
\begin{equation}
    \label{eq_Stirling_identity_factorials}
    N ! = \sqrt{2 \pi N} \left(\frac{N}{e}\right)^{N}  \left( 1 + \mathcal{O}\left(\frac{1}{N}\right)\right),
\end{equation}
the function $G_N\left( \cdot ; \boldsymbol{\theta}_N \right)$ from the PMF (Eq.~\eqref{eq_homogeneous_r_pmf_g_function}) for $r_N \neq 0$ and $r_N \neq 1$ becomes (see 
\ref{APPENDIX_ENTROPY_DOMINATED_CASE} for the details)
\begin{align}
    \label{eq_G_r_function_02}
    G_N\left( r_N ; \boldsymbol{\theta}_N \right) &= -\frac{1}{N}\log\sqrt{2 \pi N r_N \left(1-r_N\right)} + H\left(r_N\right) + \frac{1}{N} Q_N\left(r_N ; \boldsymbol{\theta}_N \right) +\frac{1}{N} \mathcal{O}\left( \frac{1}{N} \right), 
\end{align}
where $H\left( r_N \right)$ is the entropy term, defined as
\begin{equation}
    \label{eq_entropy_r}
    H\left(r_N \right) = -r_N \log\left(r_N\right) - \left( 1 - r_N\right) \log\left(1 - r_N\right).
\end{equation}

Because the entropy order $\mathcal{O}\left( H\left(r_N\right)\right) = \mathcal{O}\left( 1\right)$ is constant and considering Eq.~\eqref{eq_polynomial_r_order}, then the order of the function $N G_N \left( r_N ; \boldsymbol{\theta}_N \right)$ is
\begin{align}
 \mathcal{O} &\left(  -\log \sqrt{2 \pi N r_N \left(1 - r_N\right)} +  N H\left(r_N\right)  + Q_N\left(r_N ; \boldsymbol{\theta}_N \right) +  \mathcal{O}\left( \frac{1}{N} \right)   \right) \nonumber \\
&= \mathcal{O}\left( - \sqrt{N} + N  + 1 + \frac{1}{N} \right) \nonumber \\
&= \mathcal{O}\left(N \right).
    \label{eq_entropy_dominating_case_NG_order}
\end{align}
Eq.~\eqref{eq_entropy_dominating_case_NG_order}
is the linear order of $N$ because the entropy term dominates over the other terms for large $N$.

The dominance of the entropy as $N \to \infty$ leads to the following delta PDF
\begin{align}
    \lim_{N \rightarrow \infty}  \mathcal{P}\left( r_N | \boldsymbol{\theta}_N \right)   &=
    p\left( r | r^{*} \right) dr\nonumber \\
    &= \delta\left( r - r^{*} \right)dr,
    \label{eq_P_r_limit_delta_entropy_peak} 
\end{align}
\noindent whose peak is concentrated at its maximum $r^{*}$ in the region dominated by the entropy. The corresponding distribution function is
\begin{align}
    \lim_{N \rightarrow \infty}  F\left( r_m | r^{*}\right) = & 
    \int_{0}^{r_m} \delta \left( r - r^{*} \right) dr \nonumber \\
    = & u\left( r_m - r^{*} \right),
    \label{eq_F_r_limit_heavyside_entropy_peak}
\end{align}
\noindent where $u\left( \cdot  \right)$ denotes the Heaviside step function. For a proof of Eqs.~\eqref{eq_P_r_limit_delta_entropy_peak} and \eqref{eq_F_r_limit_heavyside_entropy_peak} see 
\ref{APPENDIX_ENTROPY_DOMINATED_CASE}.

The result above proves that, when $h\left( \mathbf{x} \right)=1$, the distribution concentrates as it does not cancel the entropy, unlike when using the base measure function in Eq.~\eqref{eq_binary_homogeneous_interactions_pmf_02}. 
Different base measure functions in the exponential family for the binary patterns correspond to different base measure functions of the homogeneous population models (discrete or continuous). We summarize these correspondences in Table \ref{Table01}. Note that the base measure function of the homogeneous continuous population rate model approaches the delta function if we use $h\left(\mathbf{x}\right)=1$. In this case, the limiting PDF is written by this base measure function alone, Eq.~\eqref{eq_P_r_limit_delta_entropy_peak}.

We also note that most existing models, e.g., the K-pairwise maximum entropy model by Tkacik et al. \citep{Tkacik_etal_2014, Tkacik_etal_2015_thermodynamics} and the dichotomized Gaussian (DG) model \citep{Amari_etal_2003} that result in the widespread distributions, do not explicitly define a base measure function for the binary patterns. According to our theory, when the discrete homogeneous distributions exhibit the widespread property (in the continuous limit), their corresponding function must contain an equivalent component that cancels with the entropy.

\noindent
\begin{center}
\bgroup
\def\arraystretch{2.0}
\begin{tabular}{|c|c|c|}
\hline
\multirow{2}{*}{ $\begin{array}{c}
     h\left( \cdot \right) \text{ function} \\
      \text{in } \mathcal{P}\left( \mathbf{x} | \boldsymbol{\theta}_N \right)
\end{array}$ } & \multicolumn{2}{c|}{Base measure function} \\
\cline{2-3}
& $\mathcal{P}\left( r_N | \boldsymbol{\theta}_N \right)$ & $p\left( r | \boldsymbol{\lambda}\right)$ \\
\hline
$1 \left/ \binom{N}{\sum_{i=1}^{N}x_i}\right.$
& $1$ & $1$ \\
\hline
$1$ & $\frac{1}{\sqrt{2 \pi r_N \left( 1 - r_N \right)}} \exp\left[ N H\left( r_N \right) \right]$ & $\delta \left( r - r^{*} \right)$ \\
\hline
\end{tabular}
\egroup
\end{center}
\vspace{1ex}
\noindent\captionof{table}{The base measure functions of the population rate models for two choices of the $h\left( \cdot \right)$ function.}
\label{Table01}

Alternatively, we may introduce a parameter that relatively weights the entropy term and the pattern probabilities and consider that the models with widespread distributions are realized at a precisely tuned parameter. The authors in \citep{Macke_etal_2011} analyzed the homogeneous DG model by introducing a non-canonical parameter $\beta$ that scales the pattern distribution as $\mathcal{P}_\beta \left( \mathbf{x} | \boldsymbol{\theta}_N \right) = \mathcal{P}\left( \mathbf{x} | \boldsymbol{\theta}_N \right)^\beta / Z_\beta$, which sets an imbalance between the entropy term $H\left( r_N \right)$ and the one that comes from $h\left( N r_N \right)^\beta$. The widespread distribution is only possible at $\beta=1$ in the limit of $N$, and they reported it as a phase transition along the parameter.

\section{The model with alternating and shrinking higher-order interactions} \label{SECTION_ALTERNATING_SHRINKING_MODEL}

Neuronal populations exhibit a significant excess rate of simultaneous silence \citep{Ohiorhenuan_etal_2010, Ganmor_etal_2011,Tkacik_etal_2014, Shimazaki_etal_2015}, where all neurons become inactive, compared to the chance level predicted by the pairwise MaxEnt models. When expressed in $(0,1)$ patterns, the probability of silence of all neurons is captured by the feature given by $\prod_{i=1}^{N} (1-x_i)$ of the exponential family distribution. The model of the excessive simultaneous silence was proposed by adding this term to the pairwise maximum entropy model, $p(\mathbf{x}|\boldsymbol{\theta}) \propto \exp [\sum_i \theta_i x_i + \sum_{i,j} \theta_i x_i x_j + \theta_0 \prod_{i=1}^{N} (1-x_i)]$ \citep{Shimazaki_etal_2015}, where the expansion of the silence feature leads to the HOIs with alternating signs for each order,
\begin{align}
    \theta_0 \prod_{i=1}^{N} (1-x_i) = \theta_0 - \theta_0 \sum_i x_i + \theta_0 \sum_{i,j} x_i x_j - \theta_0 \sum_{i,j,k} x_i x_j x_k + \theta_0 \sum_{i,j,k,l} x_i x_j x_k x_l - \cdots .
\end{align}
However, the simultaneous silence model is limited in that it captures only a state of total silence, whose measure becomes negligibly small for a large number of neurons and does not yield a widespread population rate distribution in the limit of a large number of neurons. Instead, we show that the following model with the alternating higher-order interactions, whose strength shrinks as the order increases with an appropriate base measure function, can result in a sparse, widespread population rate distribution.

To construct a sparse model with the widespread property in the large $N$ limit, we consider the following distribution for the activity patterns of the homogeneous population of binary neurons: 
\begin{equation}
    \label{eq_alternating_shrinking_binary_pmf}
    \mathcal{P}( \mathbf{x} | \boldsymbol{\omega} ) = \frac{ h\left( \sum_{i=1}^{N} x_i \right) }{Z}
    \exp\left[ -\mathpzc{f} \sum_{j=1}^{N} \left(-1\right)^{j+1} C_j \left(\frac{\sum_{i=1}^{N} x_i}{N}\right)^{j}   \right],
\end{equation}
where $\boldsymbol{\omega} = \left\{ \mathpzc{f}, C_1, C_2, \ldots, C_N \right\}$ is the set of parameters and $Z$ is its partition function. We assume that $\mathpzc{f} > 0$ and $C_j$ are positive ($C_j>0$) and decreasing with respect to $j$, $C_j < C_{j-1} \;\;\forall  j=2,...,N$. Combined with $\left(-1\right)^{j+1}$, such coefficients impose an alternating structure whose magnitude shrinks as the order of interaction increases. We will provide specific choices of $C_j$ that make the alternating terms in the exponent converge to a decreasing function with respect to $r_N$. In this model, we use Eq.~\eqref{eq_pmf_h_function} for the base measure function $h(\mathbf{x})$. Using such a function is one of the sufficient conditions required for the distribution to become widespread in the limit of a large number of neurons (Theorem \ref{theorem_widespread_zero_peaked}). Therefore, the population rate PMF becomes (see Eq.~\eqref{eq_G_r_function_to_Q_r_function})
\begin{align}
  \mathcal{P}\left(  r_N | \boldsymbol{\theta}_N \right)
= & \frac{1}{Z} \exp\left[ Q_N\left(r_N; \boldsymbol{\theta}_N\right) \right],
\label{eq_homogeneous_r_pmf_Q_function}
\end{align}
where $Q_N\left(r_N; \boldsymbol{\theta}_N\right)$ is a polynomial given by Eq.~\eqref{eq_polynomial_r}, which will be calculated as follows.

The canonical form of the homogeneous population activity is given by Eq.~\eqref{eq_binary_homogeneous_interactions_pmf}. From Eq.~\eqref{eq_alternating_shrinking_binary_pmf}, the canonical parameters ($\theta_k$, with the interaction of the order $k$) of the alternating and shrinking interaction model are computed as 
\begin{align}
  \theta_1 =& \sum_{l=1}^{N} \left( -1 \right)^{l} \frac{\mathpzc{f} C_l }{ N^{l} } , \nonumber \\
  \theta_2 =& \sum_{l=2}^{N} \left(-1\right)^{l} \frac{ \mathpzc{f} C_l }{ N^{l} } \sum\limits_{\substack{
                  k_{1} + k_{2} = l\\
                  k_{1} > 0, k_{2} > 0
                  }}  \left( \begin{array}{c}
       l\\
       k_1 , k_2 
  \end{array} \right), \nonumber \\
  \theta_3 =& \sum_{l=3}^{N} \left(-1\right)^{l} \frac{ \mathpzc{f} C_l }{ N^{l} } \sum\limits_{\substack{
                  k_{1} + k_{2} + k_{3} = l\\
                  k_{1} > 0, k_{2} > 0, k_{3} > 0
                  }}  \left( \begin{array}{c}
       l\\
       k_1 , k_2, k_3
  \end{array} \right), \nonumber \\
  & \vdots \nonumber \\
  \theta_N = &
  \left(-1\right)^{N} \frac{\mathpzc{f} C_N }{ N^{N} } N!. \label{eq_alternating_shrinking_binary_pmf_canonical_coordinates}
\end{align}
See \ref{APPENDIX_CANONICAL_COORDINATES} for the detailed derivation. 
We note that in addition to the alternating shrinking behavior for the coefficients in the PMF from Eq.~\eqref{eq_alternating_shrinking_binary_pmf}, the canonical parameters for the homogeneous PMF (Eq.~\eqref{eq_binary_homogeneous_interactions_pmf}) also show the alternating shrinking property. This can be seen from Eq.~\eqref{eq_alternating_shrinking_binary_pmf_canonical_coordinates}, where for each $\theta_k$, its coefficient with the largest magnitude (or dominant) is at $l=k$ for large $N$. Then, the dominant term for $\theta_1$ is negative, for $\theta_2$ is positive and with smaller magnitude than that of $\theta_1$, and so on for $\theta_3$ until $\theta_N$.

The PMF of the discrete population rate (Eq.~\eqref{eq_homogeneous_r_pmf_Q_function}) will be specified by using these canonical parameters, $\boldsymbol{\theta}_N=(\theta_1,\ldots,\theta_N)$, where these parameters appear in the polynomial term, $Q_N\left(r_N ; \boldsymbol{\theta}_N\right)$ (Eq.~\eqref{eq_polynomial_r}). Consequently, Eq.~\eqref{eq_alternating_shrinking_binary_pmf_canonical_coordinates} is required to obtain (non-trivially) the following form of the polynomial term, computed as
\begin{align}
Q_N \left( r_N ; \boldsymbol{\theta}_N \right) &= 
- \mathpzc{f} \sum\limits_{j=1}^{N r_N} \left( -1 \right)^{j+1}  C_j \left( r_N \right)^{j} + \mathcal{O}\left( \frac{1}{N} \right) .
  \label{eq_alternating_shrinking_r_pmf_polynomial}
\end{align}
See \ref{APPENDIX_ALTERNATING_SHRINKING_MODEL_PDF} for the derivation.
We use this form of the polynomial (Eq.~\eqref{eq_alternating_shrinking_r_pmf_polynomial} to formally show how a widespread sparse PDF emerges, which follows next.

In the limit of $N \rightarrow \infty$, our population rate PMF  becomes the continuous density given by (see 
\ref{APPENDIX_ALTERNATING_SHRINKING_MODEL_LIMIT})
\begin{align}
   \lim_{N \rightarrow \infty} \mathcal{P}\left( r_N | \boldsymbol{\theta}_N \right) & = p\left( r | \boldsymbol{\lambda} \right) dr  \nonumber \\
   & = \frac{1}{Z} \exp\left[   -\mathpzc{f} \sum_{j=1}^{\infty} \left(-1 \right)^{j+1} C_j r^{j} \right] dr,
    \label{eq_alternating_shrinking_r_pmf_to_pdf}
\end{align}
\noindent where $\boldsymbol{\lambda}=\left\{ \mathpzc{f}, \left\{ C_j \right\}_{j \in \mathbb{N}^{+}}\right\}$. Depending on the choice of each $C_j$, we obtain different types of densities. Here we provide two examples where the polynomial term $Q_N\left( r_N; \boldsymbol{\theta}_N \right)$ converges to a non-positive decreasing function with respect to $r$, and therefore the corresponding densities result in widespread distributions with a non-concentrated maximum at 0 (Corollary \ref{corollary_h_function_g_polynomial}). 

\vspace{1em} 
\textbf{Polylogarithmic exponential distribution} If we define $C_j = \frac{1}{j^{m}} \;\; \forall j\;\;$ then the probability density function in Eq.~\eqref{eq_alternating_shrinking_r_pmf_to_pdf} is
\begin{equation}
    \label{eq_alternating_shrinking_r_pdf_polylogarithm}
    p\left( r | \mathpzc{f}, m \right)  =
    \frac{ 1 }{ Z } \exp\left[    \mathpzc{f} \text{Li}_m\left[ - r \right]  \right ],
\end{equation}
where $\text{Li}_m[ \cdot ]$ is the polylogarithm function of order $m=1,2,3,\ldots$ (See 
\ref{APPENDIX_ALTERNATING_SHRINKING_MODEL_LIMIT}). We call the density in Eq.~\eqref{eq_alternating_shrinking_r_pdf_polylogarithm} the polylogarithmic exponential density, where the function $\mathpzc{f}\text{Li}_m\left[-r\right]$ is non-positive (see 
\ref{APPENDIX_NONPOSITIVE_PROPERTY}) and strictly decreasing (see 
\ref{APPENDIX_DECREASING_PROPERTY}) for $r \in [0,1]$ with a maximum at $r=0$. See Fig.~\ref{fig:POLYLOGARITHMIC_PDF_VARY_F_m} for the density functions for different $\mathpzc{f}$ and $m$. Note that for $m=1$, we obtain the natural logarithm, i.e.,
\begin{equation}
  \label{eq_polylogarithm_to_log}
  \text{Li}_1\left[ - r \right] = -\log\left[ 1 + r \right].
\end{equation}
We provide the mean and variance of the distribution for $m=1$ in \ref{APPENDIX_MEAN_AND_VARIANCE}. We note that, albeit on finite support, the resulting distribution with the logarithmic nonlinearity resembles the deformed exponential family distribution obtained under the R\'enyi entropy maximization principle with the deformation parameter $\gamma=-1$ \citep{morales2021generalization, aguilera2024explosive}.

\begin{center}
  \includegraphics[width=1.0 \textwidth]{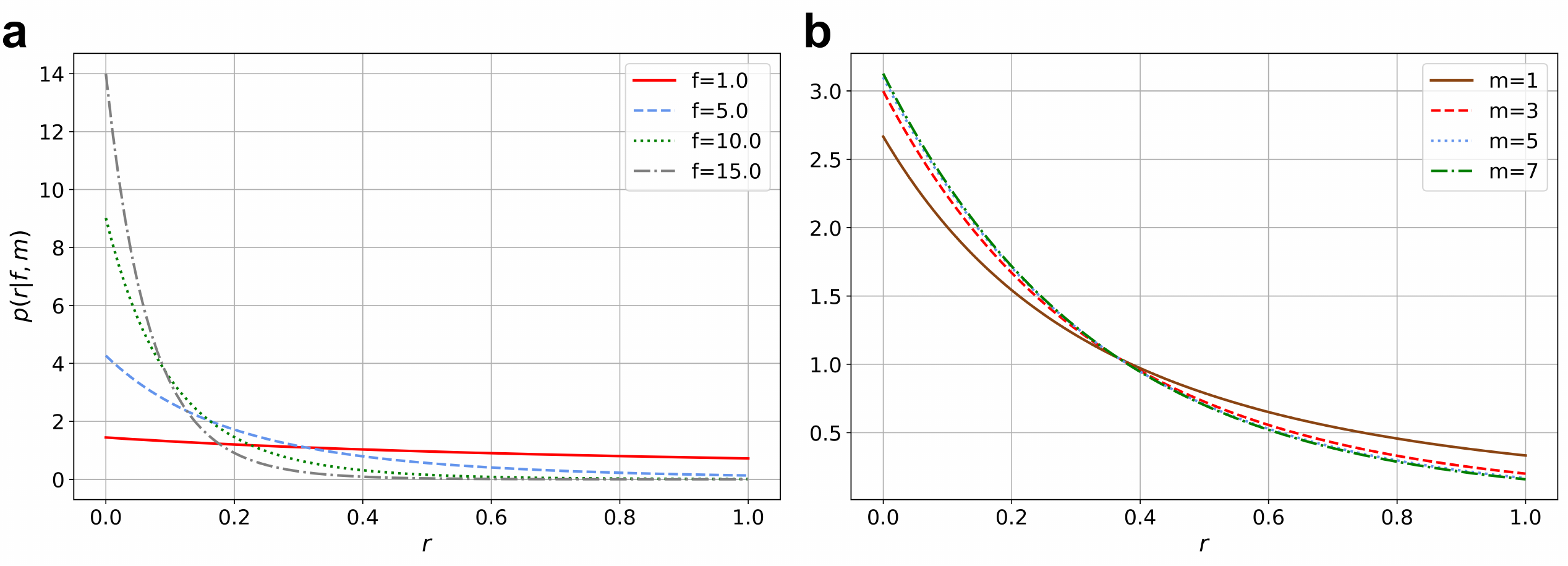}
\end{center}
\captionof{figure}{The polylogarithmic exponential PDF. \textbf{a} The PDFs for different values of $\mathpzc{f}$ ($m=1$). \textbf{b} The PDFs for different values of $m$  ($\mathpzc{f}=3$).}
\label{fig:POLYLOGARITHMIC_PDF_VARY_F_m}

The distribution function of the polylogarithmic exponential density corresponding to the PDF in Eq.~\eqref{eq_alternating_shrinking_r_pdf_polylogarithm} is as follows
\begin{align}
    \label{eq_F_r_polylogarithmic_form}
    F\left( u |  \mathpzc{f}, m \right) = \frac{1}{Z} \int_{0} ^{u} \exp\left( \mathpzc{f} \text{Li}_m\left[-r\right] \right) dr,
\end{align}
\noindent where $u \in [0,1]$. For $m=1$, we obtain the distribution function
\begin{align}
    F\left(u | \mathpzc{f}, m=1 \right) &= \frac{1}{Z}\int_{0}^{u} \exp\left( - \mathpzc{f} \log\left( 1 + r \right) \right) dr \nonumber \\
    &=
    \left \{ \begin{array}{cc}
         \frac{ 1 -\left(1 + u\right)^{-\mathpzc{f}+1}}{1 - 2^{-\mathpzc{f} + 1 }}  & \text{for} \;\; \mathpzc{f} \neq 1 \\
         \; & \; \\
         \frac{\log\left(1 + u\right)}{\log 2 } & \text{for} \;\; \mathpzc{f} = 1.
    \end{array} \right .
    \label{eq_F_r_polylogarithmic_form_m1}
\end{align}
For $m = 2, 3, 4, \hdots$, one may employ a numerical integration method to approximate Eq.~\eqref{eq_F_r_polylogarithmic_form}. \\  

\vspace{1em}
\textbf{Shifted-geometric exponential distribution} If we instead define $C_j = \left( \tau \right)^{j}$, with $\;0<\tau < 1, \; \forall j\;\;$ so that $\tau r < 1$, then the probability density function in Eq.~\eqref{eq_alternating_shrinking_r_pmf_to_pdf} is 
\begin{align}
    p\left( r | \mathpzc{f}, \tau \right)  =&
    \frac{1 }{ Z } \exp\left[   -\frac{\mathpzc{f}}{1 + \frac{1}{\tau r}} \right] \nonumber \\
   = & \frac{ 1 }{ Z } \exp\left[ 
   \mathpzc{f} \left( \frac{1}{1 + \tau r} - 1 \right) \right ],
   \label{eq_alternating_shrinking_r_pdf_shifted_geometric}
\end{align}
where the last exponential argument corresponds to a shifted-geometric series. See 
\ref{APPENDIX_ALTERNATING_SHRINKING_MODEL_LIMIT} for the details. Therefore, we call the density in Eq.~\eqref{eq_alternating_shrinking_r_pdf_shifted_geometric} the shifted-geometric exponential density. See Fig.~\ref{fig:SHIFTED_GEOMETRIC_PDF_VARY_F_tau} for the density functions for different $\mathpzc{f}$ and $\tau$. In addition, the function $\mathpzc{f}\left( \frac{1}{1 + \tau r} - 1 \right)$ in Eq.~\eqref{eq_alternating_shrinking_r_pdf_shifted_geometric} is non-positive (see 
\ref{APPENDIX_NONPOSITIVE_PROPERTY}) and strictly decreasing (see 
\ref{APPENDIX_DECREASING_PROPERTY}) for $r \in [0,1]$ with a maximum at $r=0$. See \ref{APPENDIX_MEAN_AND_VARIANCE} for the mean and variance of this distribution.

The distribution function corresponding to the shifted-geometric exponential density in Eq.~\eqref{eq_alternating_shrinking_r_pdf_shifted_geometric} is calculated as
\begin{align}
    F\left( u |  \mathpzc{f}, \tau  \right) &= \frac{1}{Z} \int_{0} ^{u} \exp\left[ \mathpzc{f} \left( \frac{1}{1 + \tau r} - 1 \right) \right] dr \nonumber \\
    &=\frac{ \left( 1 + \tau u \right) \exp\left[ \mathpzc{f}\left( \frac{1}{1 + \tau u} - 1\right) \right] - 1 + \mathpzc{f} e^{-\mathpzc{f}}\left\{ \text{Ei}\left( \mathpzc{f}\right) - \text{Ei}\left( \frac{\mathpzc{f}}{1 + \tau u}\right) \right\} }
    { \left( 1 + \tau \right) \exp\left[ \mathpzc{f}\left(\frac{1}{1 + \tau} - 1\right)\right] - 1 + \mathpzc{f} e^{-\mathpzc{f}} \left\{ \text{Ei}\left(\mathpzc{f}\right) - \text{Ei}\left( \frac{\mathpzc{f}}{1 + \tau } \right) \right\}  }. 
    \label{eq_F_r_shifted_geometric_form}
\end{align}
Here, the special exponential integral function $\text{Ei}\left(x\right)$ is defined as follows for $x \;\in \mathbb{R}$ \citep{Masina_2019_exponential_integral_func}
\begin{equation}
      \label{eq_exponential_integral_function_real_series_representation}
      \text{Ei}\left(x\right) = \gamma + \log\left(x\right) + \sum\limits_{k=1}^{\infty} \frac{x^{k}}{k\; k!},
  \end{equation}
\noindent where $\gamma$ is the Euler-Mascheroni constant ($\gamma \approx 0.5772156649$). See \ref{APPENDIX_SHIFTED_GEOMETRIC_DISTRIBUTION_FUNCTION} for verification of Eq.~\eqref{eq_F_r_shifted_geometric_form}.  \\

\begin{center}
  \includegraphics[width=1.0 \textwidth]{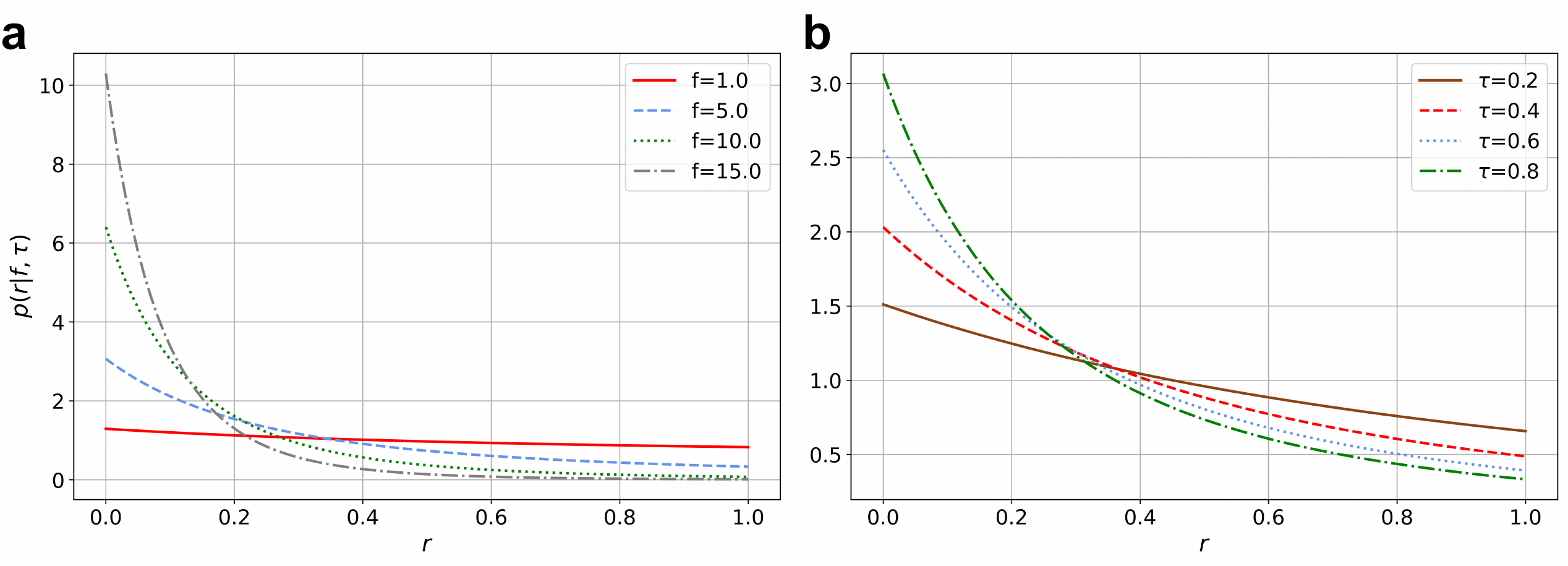}
\end{center}
\captionof{figure}{The shifted-geometric exponential PDF. \textbf{a} The PDFs for different values of $\mathpzc{f}$ ($\tau = 0.8$). \textbf{b} The PDFs for different values of $\tau$ ($\mathpzc{f} = 5$).}
\label{fig:SHIFTED_GEOMETRIC_PDF_VARY_F_tau}

\section{Experimental verification, neural interpretation, and statistical properties}
\label{SECTION_EXP_NEURAL_STAT}

In this section, we evaluate the proposed models incorporating alternating and shrinking HOIs by testing their ability to predict experimentally observed neuronal activities, interpreting the resulting distributions in the context of recurrent neural networks, and summarizing their statistical properties.

\subsection{Experimental data fitting}
\label{SUBSECTION_DATA_FITTING}

Here, we assess the goodness of fit of the proposed models to the empirical data of parallel spike sequences. We used data sets \citep{Loback_etal_2017,marre2017multi} consisting of preprocessed multi-electrode array recordings from salamander retinal ganglion cells responding to either natural movies or white noise checkerboards.

The fitting procedure is as follows. We constructed binarized salamander datasets using either a 100 millisecond (ms) time window \citep{Loback_etal_2017} or a 20 ms time window \citep{marre2017multi} under different stimuli. We then extracted $N_s$ samples $\mathbf{x}^{(i)}, i=1,2, \ldots, N_s$, each composed of population activity of $N$ retinal ganglion cells under a particular visual stimuli condition. Next we transformed them into population rate samples. That is, we obtained $r_N ^{(i)} \in [0,1]$ where $r_N ^{(i)} = \frac{1}{N} \sum_{k=1}^{N} x_k ^{(i)}$ for each sample. Then, using the following log-likelihood function for the PDFs from the models
\begin{align}
\ell \left(\boldsymbol{\lambda} ; \left\{r_N ^{(i)}\right\}_{i=1}^{N_s}\right) &=\sum_{i=1}^{N_s} \log p\left(r_N^{(i)} \mid \boldsymbol{\lambda} \right),
\end{align}
\noindent we obtained the parameters fitted to the data under the maximum likelihood principle, i.e.,
\begin{align}
    \hat{\boldsymbol{\lambda}} &=\underset{\boldsymbol{\lambda}}{\operatorname{argmax}}\left\{ \ell \left(\boldsymbol{\lambda} ;\left\{r_N ^{(i)}\right\}_{i=1}^{N_s}\right)\right\}.
    \label{eq_maximum_likelihood_homogeneous_population_rate_models}
\end{align}

In addition to the polylogarithmic and shifted-geometric models, we fit the first and second-order homogeneous population rate models (with our $h(\mathbf{x})$ function). For the first-order model, we used the PDF from Eq.~\eqref{eq_homogeneous_first_order_negative_interactions_r_pdf}, which corresponds to a truncated exponential distribution, the simplest sparse homogeneous model. It serves as a reference baseline to determine if other distributions are heavy-tailed or not. See Subsection \ref{SUBSECTION_HEAVY_TAILEDNESS} 
for our precise definition of a heavy-tailed distribution. For the second-order model, we used the PDF from Eq.~\eqref{eq_homogeneous_second_order_sparse_r_pdf}. The parameters optimized under Eq.~\eqref{eq_maximum_likelihood_homogeneous_population_rate_models} for each model are 
\begin{align}
\boldsymbol{\lambda}^* & \approx 
    \left \{ \begin{array}{ll}
     \hat{\mathpzc{f}} & \text{for the first-order exp. distr.}  \\
     \hat{\mathpzc{f}}_1, \hat{\mathpzc{f}}_2 & \text{for the second-order exp. distr.}  \\
     \hat{\mathpzc{f}}, \hat{m} & \text{for the polylogarithmic exp. distr.}  \\
     \hat{\mathpzc{f}}, \hat{\tau} & \text{for the shifted-geometric exp. distr.}
\end{array} \right .
\end{align}

\begin{center}
  \includegraphics[width=1.0 \textwidth]{NECO-24-001-00R2-Figure.4.pdf}
\end{center}
\captionof{figure}{Models fitted to a binary dataset from salamander retinal ganglion cells ($N=160$) responding to visual stimuli from a natural movie. \textbf{a} The fitted polylogarithmic (red) and shifted-geometric (blue) models to the data (green) in the logarithmic scale. In comparison, the plot shows the fitted first (gray) and second-order (yellow) models with the entropy-cancelling base measure function. Each plot shows the best-fit curve in terms of average log-likelihood. \textbf{b} The same as in (a) in a linear scale for the vertical axis and the (original) dataset histogram.}
\label{fig:MODELS_ALL_FIT_160_MARRE_ETAL}

\noindent
\begin{center}
\begin{tabular}{|c|c|c|c|c|}
    \hline
    \rowcolor{gray!20} 
       & \textbf{First-order} & \textbf{Second-order} & \textbf{Poly-} & \textbf{Shifted-} \\
      \rowcolor{gray!20}
      &  &  & \textbf{logarithmic} & \textbf{geometric} \\
    \hline
    \cite{marre2017multi} & 1.9390 & 2.3329 & 1.9849 &  \cellcolor{green!20} 2.3400 \\
    natural movie & $\mathpzc{f}=61.37$ &  $\mathpzc{f}_1=-25.86,$ & $\mathpzc{f}=62.28,$ & \cellcolor{green!20} $\mathpzc{f}_1=63.27,$ \\
    stimuli dataset &  & $\mathpzc{f}_2=2.52,$ & $m=1$ & \cellcolor{green!20} $\tau=0.5$ \\
    \hline 
    \cite{Loback_etal_2017} & 2.8543 & 2.8891 & 2.8713 &  \cellcolor{green!20} 2.9083  \\
    white noise checkerboard & $\mathpzc{f}=64.55$ & $\mathpzc{f}_1=-54.58,$ & $\mathpzc{f}=65.51$ & \cellcolor{green!20} $\mathpzc{f}=67.87$ \\
     stimuli dataset    &  &  $\mathpzc{f}_2 =4.85$  & $m=1$ & \cellcolor{green!20} $\tau=0.8$ \\
    \hline
    \cite{Loback_etal_2017} & 2.8542 & 2.8895 & 2.8738 &  \cellcolor{green!20} 2.9116  \\
    natural movie & $\mathpzc{f}=64.55$ & $\mathpzc{f}_1=-54.58,$ & $\mathpzc{f}=65.52$ & \cellcolor{green!20} $\mathpzc{f}=67.88$ \\
     stimuli dataset    &  &  $\mathpzc{f}_2 =4.85$  & $m=1$ & \cellcolor{green!20} $\tau=0.8$ \\
    \hline
\end{tabular}
\end{center}
\vspace{1ex}
\noindent\captionof{table}{Cross-validated log-likelihood per sample of each fitted model to three datasets 
  from salamander ganglion cells' (binary) responses to specific visual stimuli (natural movie or white noise checkerboard stimuli). The binary spikes were obtained with a time window of 100 ms for Loback et al's  
  \citeyearpar{Loback_etal_2017} datasets and with a time window of 20 ms for Marre et al's \citeyearpar{marre2017multi} 
  dataset. For Loback et al's \citeyearpar{Loback_etal_2017} 
  datasets, we selected the $N=40$ most active cells 
  and for Marre et al's \citeyearpar{marre2017multi} 
  dataset we kept 160 neurons. We trained the models using a training set of size $N_s=80,000$ for each dataset. We then computed the cross-validated log-likelihood using a test sample of size $N_s=20,000$ for each dataset. The value of the cross-validated log-likelihood per a test sample is shown in the table. To learn the sparsity-inducing parameter $\mathpzc{f}$ (or $\mathpzc{f}_1$ and $\mathpzc{f}_2$ for the second-order model), we used a gradient ascent method with the best learning rate among the values $\left\{ 0.0001, 0.0005, 0.001, 0.005, 0.01, 0.05, 0.1  \right\}$. A grid search was performed for the $C_j$ values of the alternating series of the corresponding model. The search for $m$ in the polylogarithmic model was performed in $\left\{ 1 + 2 t \right\}_{t=0}^{5}$  and for $\tau$ in the shifted-geometric model in $\left\{ 0.1 + 0.05t \right\}_{t=0}^{18}$.}
\label{tab:AVG_ML_RESULTS}

Figure \ref{fig:MODELS_ALL_FIT_160_MARRE_ETAL} shows the fit to exemplary data from $N=160$ neurons (green dots) by the proposed models as well as the independent and pairwise homogeneous models with the entropy-cancelling base measure function. The data set shown here is the same as in Fig.~\ref{fig:MODELS_MARRE_MAX_ENT}, where we confirmed the failure of independent and pairwise homogeneous MaxEnt models ($h(\mathbf{x})=1$). To examine the predictive ability of the models, we divided the data into training and testing datasets. After fitting the models to the training data sets, we compared the polylogarithmic and shifted-geometric models against the first and second-order homogeneous population rate models with the entropy-cancelling base measure function, using the likelihoods of the unseen test data sets. Table \ref{tab:AVG_ML_RESULTS} presents the results of the optimal fit for each type of model. 

Overall, the shifted-geometric model showed superior predictive performance in all datasets over other models (Table \ref{tab:AVG_ML_RESULTS}). However, the improvement of the shifted-geometric model over the pairwise model in modeling the tail of the observed distribution was marginal, where both models captured the heavy tail well (Fig.~\ref{fig:MODELS_ALL_FIT_160_MARRE_ETAL}a). Since the HOIs of the second-order model are derived solely from the entropy-cancelling base measure function $h\left( \mathbf{x} \right)$ (Eq.~\eqref{eq_pmf_h_function}), this result indicates that the HOIs induced by $h\left(\mathbf{x}\right)$ facilitated capturing the heavy tail 
whereas the contribution by the structured HOIs introduced by the polynomial term $Q_{N}\left(r_N ; \boldsymbol{\theta}_N \right)$ (Eq.~\eqref{eq_polynomial_r}) was minor. Thus, the introduction of the entropy-cancelling base measure function was critical for modeling the heavy-tailed distribution of neural activities. The improvement in the shifted-geometric distribution over the second-order model is likely due to the improved fitting at the silent states of the distribution ($r \ll 1$) (Fig.~\ref{fig:MODELS_ALL_FIT_160_MARRE_ETAL}b). In contrast, the polylogarithmic model shows only marginal improvements over the first-order model in all datasets, indicating significantly greater flexibility of the shifted-geometric model in capturing sparse, heavy-tailed distributions of neuronal spike-rate distributions. 
This is because, although the polylogarithmic model is also heavy-tailed, it is constrained to stay close to the first-order model. 
In particular, the $q(r; \boldsymbol{\lambda})$ function in the polylogarithmic model is bounded below by $-\mathpzc{f}r$, which corresponds to the first-order model, and above by $-\mathpzc{f}\log(1+r)$, which is a sublinear function close to the linear case in the support of $[0,1]$ (see \ref{APPENDIX_HEAVY_TAILEDNESS}). 

Finally, we also evaluated the effect of the neural population size $N$ on the estimated sparsity-inducing parameters, $\mathpzc{\hat{f}}, \mathpzc{\hat{f}}_1$, and $\mathpzc{\hat{f}}_2$, depending on the specific model (Fig.~\ref{fig:PARAMETERS_FIT_AS_N_VARIES}). The results show logarithmic growth for each sparsity-inducing parameter as $N$ grows, with the exception of the $\mathpzc{\hat{f}}_2$ parameter for the second-order model that stays roughly constant. These results indicate that the population spike-rate distribution has not converged within the 
size of neurons analyzed. Since the divergence of $\mathpzc{\hat{f}}$ results in a concentrated distribution at $0$, this result indicates that the successful fitting by the shifted-geometric model for the finite data does not imply that larger neuronal networks exhibit sparse, widespread population activity distributions. It remains a future challenge to determine if larger neural population activity exhibits scale-invariant widespread distributions.

\begin{center}
  \includegraphics[width=0.8 \textwidth]{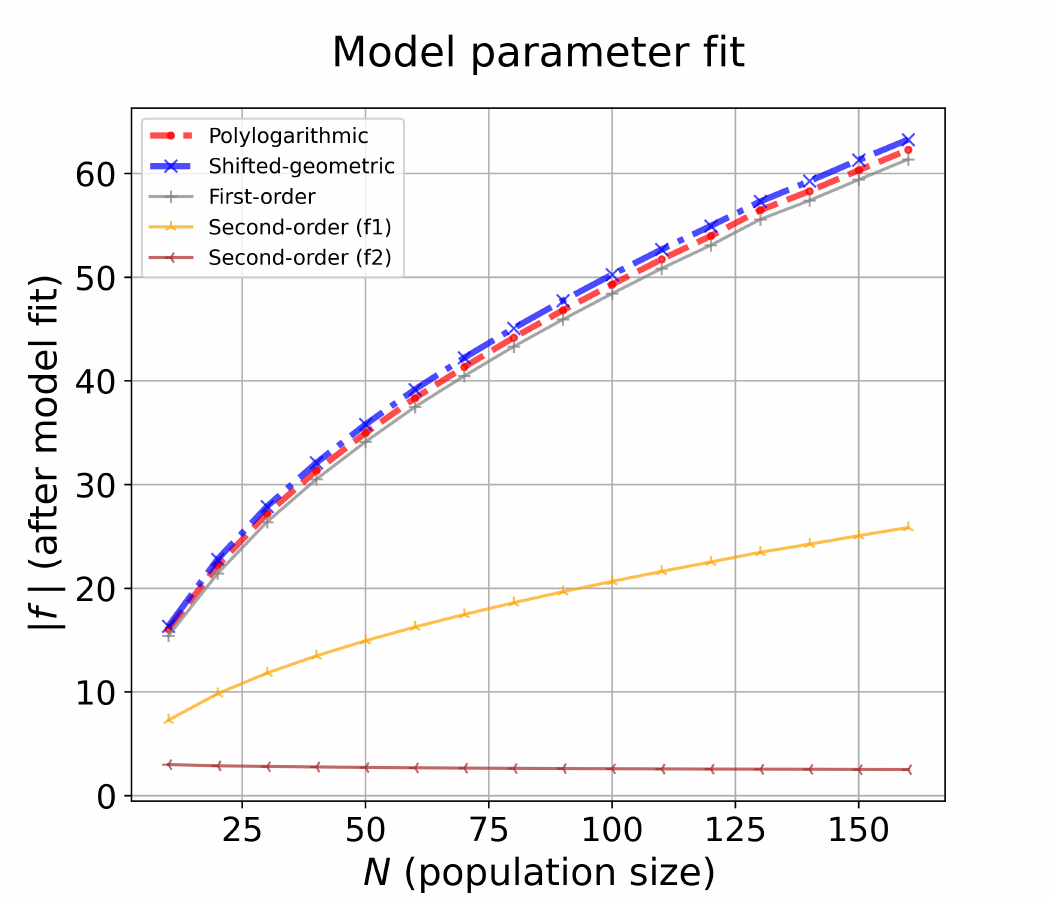}
\end{center}
\captionof{figure}{Absolute value of estimated sparsity-inducing parameters as the neural population size $(N)$ varies. A binary dataset from salamander retinal ganglion cells responding to natural movie stimuli 
 \citep{marre2017multi} was used for the data fitting procedure. Each point represents a different fit to the data. The estimation of each parameter is as ($N$) varies with the polylogarithmic (red), shifted-geometric (blue), first-order (gray) and second-order (brown) models, all with our $h(\mathbf{x})$ function.}
\label{fig:PARAMETERS_FIT_AS_N_VARIES}

\subsection{Neural Interpretation}
\label{SUBSECTION_NEURAL_INTERPRETATION}

\begin{center}
  \includegraphics[width=0.95 \textwidth]{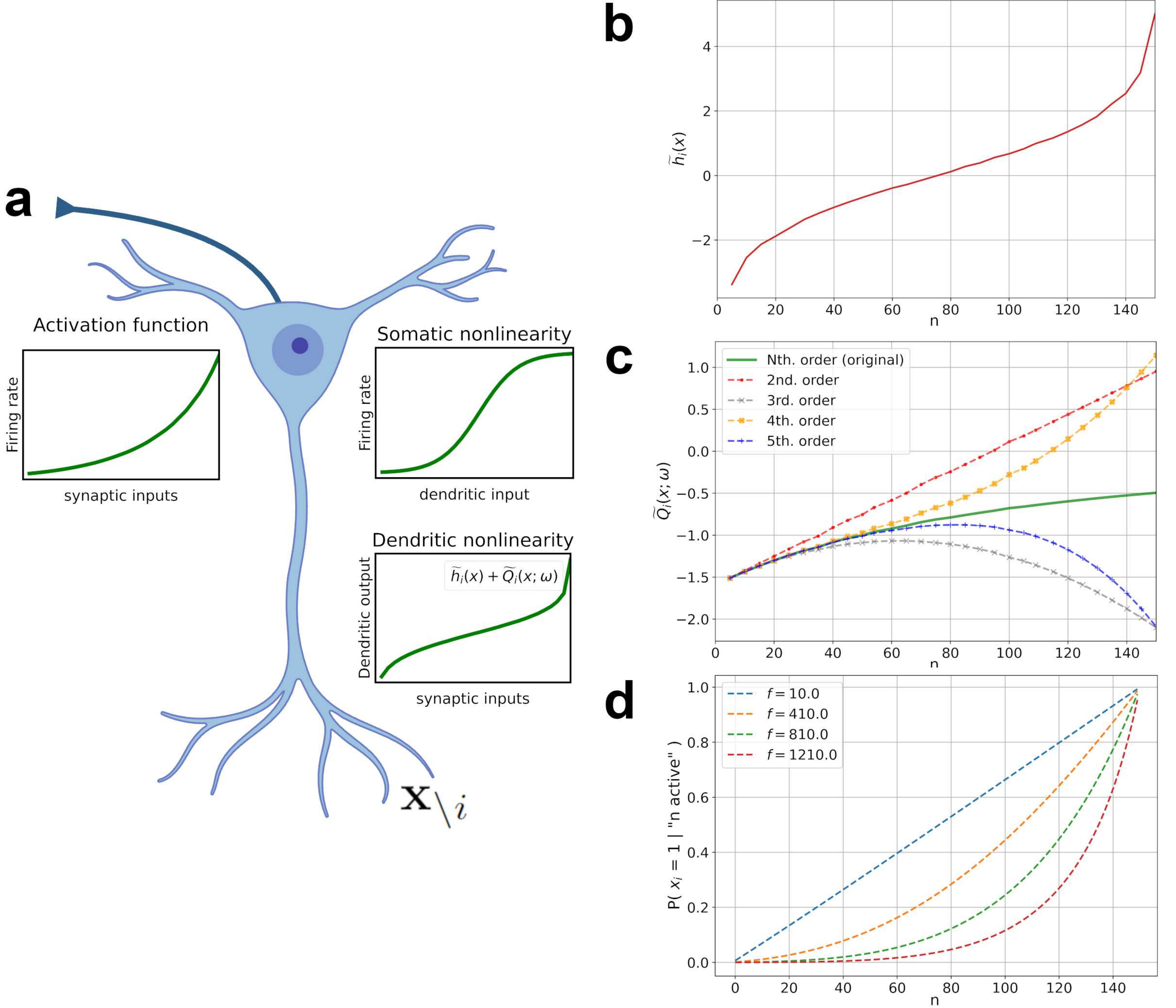}
\end{center}
\captionof{figure}{Neural interpretation of the alternating and shrinking HOIs models. \textbf{a} A schematic of a neuron $i$  with somatic and dendritic nonlinearity of the shifted-geometric exponential distribution. (Top left) A nonlinear activation function of the neuron $i$ that relates the output firing rate and synaptic inputs from the other neurons ($\textbf{x}_{\backslash i}$). (Top right) A somatic activation function (logistic function), receiving the output of the dendritic activity as input. (Bottom right) Dendritic nonlinearity receiving inputs from presynaptic neurons $\mathbf{x}_{\backslash i}$. Created with \textit{BioRender.com}. \textbf{b} The component of dendritic nonlinearity $\widetilde{h}_i$ for neuron $i$ of the shifted-geometric exponential distribution for $\tau=0.8$ as a function of the number of active input neurons, $n$, in a population of $N=150$ neurons for a fixed value of $\mathpzc{f}=2N$. \textbf{c} The same plot as in (b) but for the dendritic nonlinearity component $\widetilde{Q}_{i}$ 
 for a fixed value of $\mathpzc{f}=2N$ (solid green). The dashed lines are the functions truncated at the second, third, fourth, and fifth-order interactions. \textbf{d} The activation functions of a neuron in a recurrent network composed of $N=150$ neurons for different sparsity parameters. The activation function is the probability of activating the $i$th neuron given $n$ active input neurons.}
\label{fig:neural_nonlinearities}

The sampling dynamics used to realize the joint distribution of the binary population activity, such as the Gibbs sampler (also known as Glauber dynamics), can be interpreted as the dynamics of recurrent neural networks. In such dynamics, each neuron responds to the inputs from the rest of the neurons through a specific nonlinear activation function. In this approach, the activity of an individual neuron is sampled from the conditional probability given the activities of the other neurons, either sequentially or in random order. 
The activity rate of the $i$th neuron is obtained as the following logistic function (see \ref{APPENDIX_NEURAL_INTERPRETATION} for the details)
\begin{align}
    \mathcal{P}(x_i = 1 | \mathbf{x}_{\backslash i}, \boldsymbol{\omega} ) 
    &= \frac{1}{1 + \exp \left(  - \log
              \frac{  \mathcal{P}(x_i = 1, \mathbf{x}_{\backslash i},\boldsymbol{\omega}) }
              { \mathcal{P}(x_i = 0, \mathbf{x}_{\backslash i},\boldsymbol{\omega}) }
              \right) },
    \label{eq_conditional_distrib_single_neuron}
\end{align}
where the log ratio is computed as
\begin{align}
   \log \frac{ \mathcal{P}\left(x_i = 1, \mathbf{x}_{\backslash i}, \boldsymbol{\omega} \right) }{  \mathcal{P}\left( x_i = 0, \mathbf{x}_{\backslash i} , \boldsymbol{\omega} \right) }   
   = \widetilde{h}_{i}\left( \mathbf{x} \right) + \widetilde{Q}_{i} \left( \mathbf{x} ; \boldsymbol{\omega} \right),
       \label{eq_log_ratio_single_neuron}
\end{align}
Here, for the alternating and shrinking HOIs model, $\widetilde{h}_{i}\left( \cdot \right)$ and $\widetilde{Q}_{i} \left( \cdot \right)$ are given as
\begin{align}
    \label{eq_h_tilde}
    \widetilde{h}_{i}\left( \mathbf{x} \right) & =
    \log\left( \frac{ 1 + \sum_{j \neq i} x_j }{ N - \sum_{j \neq i} x_j } \right)
\end{align}
and 
\begin{align}
    \label{eq_Q_tilde}
    & \widetilde{Q}_{i} \left( \mathbf{x} ; \boldsymbol{\omega} \right) \nonumber \\
    & =
    -\mathpzc{f} \left( \sum_{j=1}^{N} \left( -1 \right)^{j+1} C_j \left( \frac{1 + \sum_{k \neq i} x_k }{ N } \right)^j   - \sum_{j=1}^{N} \left( -1 \right)^{j+1} C_j \left( \frac{ \sum_{k \neq i} x_k }{ N } \right)^j \right)
\end{align}

One can regard the logistic function as a somatic activation function of a neuron and the log ratio as a nonlinearity over synaptic inputs from other neurons at the dendrites (Fig.~\ref{fig:neural_nonlinearities}a), though other explanations could also apply. Note that $\widetilde{h}_{i}\left( \mathbf{x} \right)$ and $\widetilde{Q}_{i} \left( \mathbf{x} ; \boldsymbol{\omega} \right)$ include the higher-order terms with respect to $\mathbf{x}$, indicating that the neuron is sensitive to higher-order statistics of input activities. This result contrasts with the Gibbs sampler of the pairwise MaxEnt model, resulting in the log-ratio given by the first-order terms with respect to $\mathbf{x}$. In this case, the neurons are sensitive only to the individual activities of the presynaptic neurons, leading to the linear integration of the presynaptic inputs.

Both $\widetilde{h}_{i}\left( \mathbf{x} \right)$ and $\widetilde{Q}_{i} \left( \mathbf{x}; \boldsymbol{\omega} \right)$ are increasing functions with respect to the number of active presynaptic neurons, $n=\sum_{j \neq i} x_j$ (see the sum of their profile at the diagram in Fig.~\ref{fig:neural_nonlinearities}a). $\widetilde{h}_{i}\left( \mathbf{x} \right)$ is an increasing function that diverges as $n$ approaches to $N$ (Fig.~\ref{fig:neural_nonlinearities}b). In contrast, $\widetilde{Q}_{i} \left( \mathbf{x}; \boldsymbol{\omega} \right)$ is a sublinear function with respect to the number of active presynaptic neurons, in contrast to the linear function expected for the neurons under the pairwise (second-order) model (Fig.~\ref{fig:neural_nonlinearities}c). The even and odd order attenuate and promote the effect of sublinearity, respectively, resulting in the alternating convergence to $\widetilde{Q}_{i} \left( \mathbf{x}; \boldsymbol{\omega} \right)$ by adding higher-order terms. Overall, $\widetilde{h}_{i}\left( \mathbf{x} \right)$ facilitates supra-linear integration of the presynapatic inputs in the high-input regime, whereas the polynomial term $\widetilde{Q}_{i} \left( \mathbf{x}; \boldsymbol{\omega} \right)$ contributes to sublinear integration in the low-input regime, although contribution by $\widetilde{h}_{i}\left( \mathbf{x} \right)$ is not negligible.

The final input-output relation of the neurons caused by both the dendritic nonlinearity and the somatic logistic activation function can be calculated as follows. Let $n$ be the number of active presynaptic neurons for a neuron $i$. Then the probability that the postsynaptic neuron $i$ is activated is given as (see \ref{APPENDIX_NEURAL_INTERPRETATION} for the derivation)
\begin{align}
    \label{eq_conditional_prob_x_k_1_given_n_active}
    \mathcal{P}\left( x_i = 1 \left | \sum_{j\neq i} x_j = n \right. \right) 
& = \frac{ 1 }{ 1  +  \exp\left[ \log \frac{N-n}{n+1} + \theta_{n+1}  -  \sum_{k=1}^{n} \theta_k  
   \left( \begin{array}{c}
   n \\
   k-1 
\end{array} \right) 
\right]  }.
\end{align}
Figure \ref{fig:neural_nonlinearities}d shows the conditional probability (Eq.~\eqref{eq_conditional_prob_x_k_1_given_n_active}) as a function of $n$ for the shifted-geometric exponential distribution. The spiking probability is facilitated by a larger number of active presynaptic neurons in a strictly increasing manner. However, the sparsity parameter $\mathpzc{f}$ modifies this activation function nonlinearly. With larger $\mathpzc{f}$, the conditional probability of the activation is significantly attenuated for small $n$ due to the sublinear property of $\widetilde{Q}_{i} \left( \cdot \right)$ at small $n(\ll N)$. In contrast, this probability is kept high for larger $n$ due to the diverging property of $\widetilde{h}_{i}\left( \cdot \right)$ at large $n(\sim N)$. The combined effect of these two functions gives rise to an activation function with properties akin to a threshold nonlinearity followed by a supra-linear function. Such an activation function, especially with large $\mathpzc{f}$, is expected to induce sparse population activity because the hyperactive state of the population becomes highly improbable, keeping neurons silent most of the time. However, once the population activity surpasses the threshold, the network facilitates its activity, increasing the probability of highly synchronous patterns that shape the tail of the population spike rate distribution. We note that the form of the activation function induced by the alternating HOIs and the entropy-canceling base measure function closely resembles the activation function obtained from the rectified polynomial energy introduced in the original dense associative memories \citep{krotov2016dense}, which supports sparse representations.

\subsection{Properties of the distributions} \label{SUBSUBSECTION_PROPERTIES_OF_DISTRIBUTIONS}

Finally, in this section, we summarize key properties of the proposed models, including their sparsity, heavy-tailedness, entropy, and heat capacity.

\vskip 1em
\noindent \textbf{Sparsity} 
Figures \ref{fig:POLYLOGARITHMIC_PDF_VARY_F_m} and \ref{fig:SHIFTED_GEOMETRIC_PDF_VARY_F_tau} show that the sparsity for both the polylogarithmic exponential and the shifted-geometric exponential densities is controlled non-linearly by the $\mathpzc{f}$ parameter. The densities with small values of $\mathpzc{f}$ approach a uniform distribution, while the densities with large $\mathpzc{f}$ values become very sparse. We can confirm this in the following limits for both distributions
\begin{align}
    \lim_{\mathpzc{f} \to 0} p\left( r | \boldsymbol{\lambda} \right)
 = \frac{ e^{0} }{\int_{0}^{1} e^{0}  dr'} = 1.
    \label{eq_alternating_PDF_f_limit_zero}
\end{align}

On the other extreme, as $\mathpzc{f} \to \infty$ both distributions tend to a Dirac delta distribution centered at $0$ (see \ref{APPENDIX_LIMIT_DISTRIBUTIONS} for the proof):
\begin{align}
    \lim_{\mathpzc{f} \to \infty} p\left( r | \boldsymbol{\lambda} \right) 
 = \delta\left( r \right).
    \label{eq_alternating_PDF_f_limit_infinity}
\end{align}
Compared to the polylogarithmic exponential distribution, the shifted-geometric exponential distribution exhibits fatter tails due to a slower decay in probability for increasing values of the population rate (See $\mathpzc{f} \in \left\{ 10, 15\right\}$ in Figs.~\ref{fig:POLYLOGARITHMIC_PDF_VARY_F_m}a and \ref{fig:SHIFTED_GEOMETRIC_PDF_VARY_F_tau}a).

The $m$ parameter also modulates sparsity for the polylogarithmic exponential distribution (Fig.~\ref{fig:POLYLOGARITHMIC_PDF_VARY_F_m}b) but to a much lesser extent than the $\mathpzc{f}$ parameter, i.e., the distribution is less sensitive to changes in the $m$ parameter. Because of this, $m=1$ is the simplest choice of the polylogarithmic exponential distribution, for which we provide the complete analytical PDF and distribution function. Similarly, for the shifted-geometric exponential distribution, the $\tau$ parameter is less relevant for inducing sparsity (Fig.~\ref{fig:SHIFTED_GEOMETRIC_PDF_VARY_F_tau}b) compared to the $\mathpzc{f}$ parameter, but more when compared to the $m$ polylogarithmic parameter. \\

\noindent \textbf{Heavy-tailedness} 

Our alternating shrinking models share a common sparsity-inducing parameter $\mathpzc{f}$. We introduced a novel definition for a heavy-tailed distribution within the compact support of $[0,1]$ and at a sparseness level induced by a particular value of $\mathpzc{f}$ in Subsection \ref{SUBSECTION_HEAVY_TAILEDNESS}. 
Based on these definitions, we summarize the following results. See \ref{APPENDIX_HEAVY_TAILEDNESS_PROOF} for the proof of these results. 

\begin{itemize}
    \item Polylogarithmic exponential distributions are heavy-tailed for $\mathpzc{f} \in (0,\infty)$ and for all  $m=1,2,3,\cdots$. The exception is at the special case taking the limit of $m \to \infty$, which corresponds to the reference truncated exponential distribution.
    \item Shifted-geometric exponential distributions are heavy-tailed for $\mathpzc{f} \in (0,\infty)$ and for $\tau \in (0,1)$.
\end{itemize}

\vskip 1em
\noindent \textbf{Entropy and heat capacity} 
Here we assess the entropy and heat capacity of the distributions. The entropy quantifies the average uncertainty of a probability distribution. The heat capacity quantifies the variability (fluctuation) of that uncertainty, reflecting the sensitivity of energy to a global scaling of the parameters (i.e., temperature). Notably, peaks in heat capacity are often associated with critical points, where the system exhibits maximized responsiveness. These are important quantities for assessing the informational capacity and potential criticality of the system \citep{Tkacik_etal_2015_thermodynamics}.

First, the entropy of both distributions is computed as
\begin{align}
\mathbb{E}_{R}\left[ -\log\left( p\left( r | \boldsymbol{\lambda} \right) \right) \right] 
   = -\int_{0}^{1} p\left( r | \boldsymbol{\lambda} \right) \log\left( p\left( r | \boldsymbol{\lambda} \right) \right) dr
    \label{eq_entropy_alternating_shrinking_densities}.
\end{align}
For the explicit entropy of the polylogarithmic exponential and the shifted-geometric exponential distributions, see 
\ref{APPENDIX_ENTROPY}.

Second, the heat capacity is computed as follows. Let $\mathpzc{f} = \frac{1}{T}$ and $d \mathpzc{f} = -\frac{1}{T^2} d T$, where $T$ denotes a temperature parameter. Then, the heat capacity of both the polylogarithmic exponential distribution (with $m=1$) and the shifted-geometric distribution is computed  as
\begin{align}
C \left( \mathpzc{f} \right) &= \frac{\partial }{\partial  T} \left( -\frac{1}{Z} \frac{d Z}{d \mathpzc{f} } \right) 
= \mathpzc{f}^{2} \frac{ d^2 Z }{ d \mathpzc{f}^2 } \frac{1}{Z} - \frac{ \mathpzc{f}^{2} }{ Z^{2} } \left( \frac{ d Z }{ d \mathpzc{f} } \right)^2.
    \label{eq_heat_capacity_definition}
\end{align}
See \ref{APPENDIX_HEAT_CAPACITY} for the specific values of the normalization constant and its derivatives for the heat capacity of the polylogarithmic exponential and the shifted-geometric exponential distributions, as well as some limits with respect to $\mathpzc{f}$.

\begin{center}
  \includegraphics[width=1.0 \textwidth]{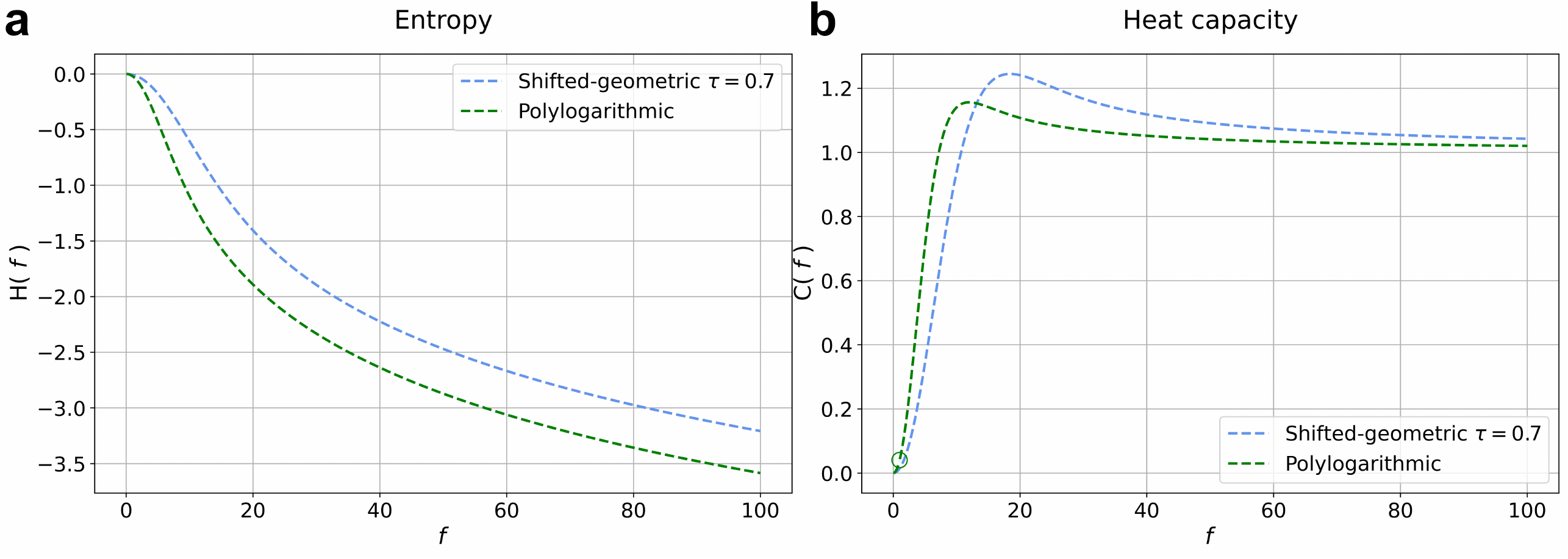}
\end{center}
\captionof{figure}{Entropy and heat capacity of the polylogarithmic exponential and the shifted-geometric exponential distributions. \textbf{a} Entropy of both distributions as $\mathpzc{f}$ varies ($m=1$, $\tau=0.7$).\textbf{b} Heat capacity of both distributions as $\mathpzc{f}$ varies ($m=1$, $\tau=0.7$). Notice the empty point at $\mathpzc{f}=1$ for the heat capacity of the polylogarithmic model, where it is undetermined.}
\label{fig:ENTROPY_HEAT_CAPACITY}

The entropy of the polylogarithmic exponential distribution ($m=1$) is non-positive and a decreasing function of $\mathpzc{f}$ as can be seen in Fig.~\ref{fig:ENTROPY_HEAT_CAPACITY}a. Such negative entropy is compatible with a neural system that promotes a high level of organization. On the other hand, the heat capacity increases with $\mathpzc{f}$ until a numerically found maximum at $\mathpzc{f}\approx 11.96$, after which it decreases until $\lim_{\mathpzc{f} \to \infty} C\left( \mathpzc{f} \right) = 1 $ (see 
\ref{APPENDIX_HEAT_CAPACITY}). Such a limit can be intuitively observed in Fig.~\ref{fig:ENTROPY_HEAT_CAPACITY}b. At $\mathpzc{f}=1$, the heat capacity is 
undetermined (represented by an open circle in Fig.~\ref{fig:ENTROPY_HEAT_CAPACITY}b for $m=1$). The result that the heat capacity is maximized at the finite $\mathpzc{f}$ indicates that the fluctuating regime can coexist with a certain level of sparseness.

The entropy for the shifted-geometric exponential distribution is also non-positive and a decreasing function of $\mathpzc{f}$, compatible with a high level of organization, as can be seen in Fig.~\ref{fig:ENTROPY_HEAT_CAPACITY}a for $\tau=0.7$. The heat capacity (for $\tau=0.7$) has a numerical maximum at $\mathpzc{f} \approx 18.44$ (Fig.~\ref{fig:ENTROPY_HEAT_CAPACITY}b), after which it decreases until $C\left(\mathpzc{f}\right) \approx 1$. However, unlike the polylogarithmic exponential distribution ($m=1$), we obtain that $\lim_{\mathpzc{f} \to \infty} C\left( \mathpzc{f} \right) $ is undetermined. \\

\noindent \textbf{Effect of HOIs by the polynomial term} 
We assess the effect of HOIs by the polynomial term in shaping the distributions by comparing the model with the argument of the exponent in the right-hand side of Eq.~\eqref{eq_alternating_shrinking_r_pmf_to_pdf} truncated at the $k$-th order term against the original model. Figure \ref{fig:log_k_approx_PDFs} compares the polylogarithmic exponential PDF and the shifted-geometric exponential PDF in the log-log plot. In both PDFs, all of the $k$-th order approximations overall approximate better the ground truth values for larger $r$ (i.e., smaller probability density) compared to the region near $r=0$, where most neurons in the underlying population remain silent. In addition, the $k$-th order approximation alternates around the diagonal line (baseline) depending on whether $k$ is even or odd, indicating overestimation and underestimation of the low probability densities, respectively, while this relation reverses for the approximation of the high probability densities. The approximated PDF becomes closer to the original model with increasing $k$ (Fig.~\ref{fig:log_k_approx_PDFs}). Overall, the shifted-geometric exponential PDF exhibts more significant deviation even at the region of large probability density (small $r$, silent states) if we consider only the low order interactions (Fig.~\ref{fig:log_k_approx_PDFs}b), compared to the polylogarithmic exponential PDF (Fig.~\ref{fig:log_k_approx_PDFs}a). Thus, for the shifted-geometric exponential PDF, the HOIs shape not only the tail of the distribution but also the probabilities of silence. 

\begin{center}
  \includegraphics[width=1.0 \textwidth]{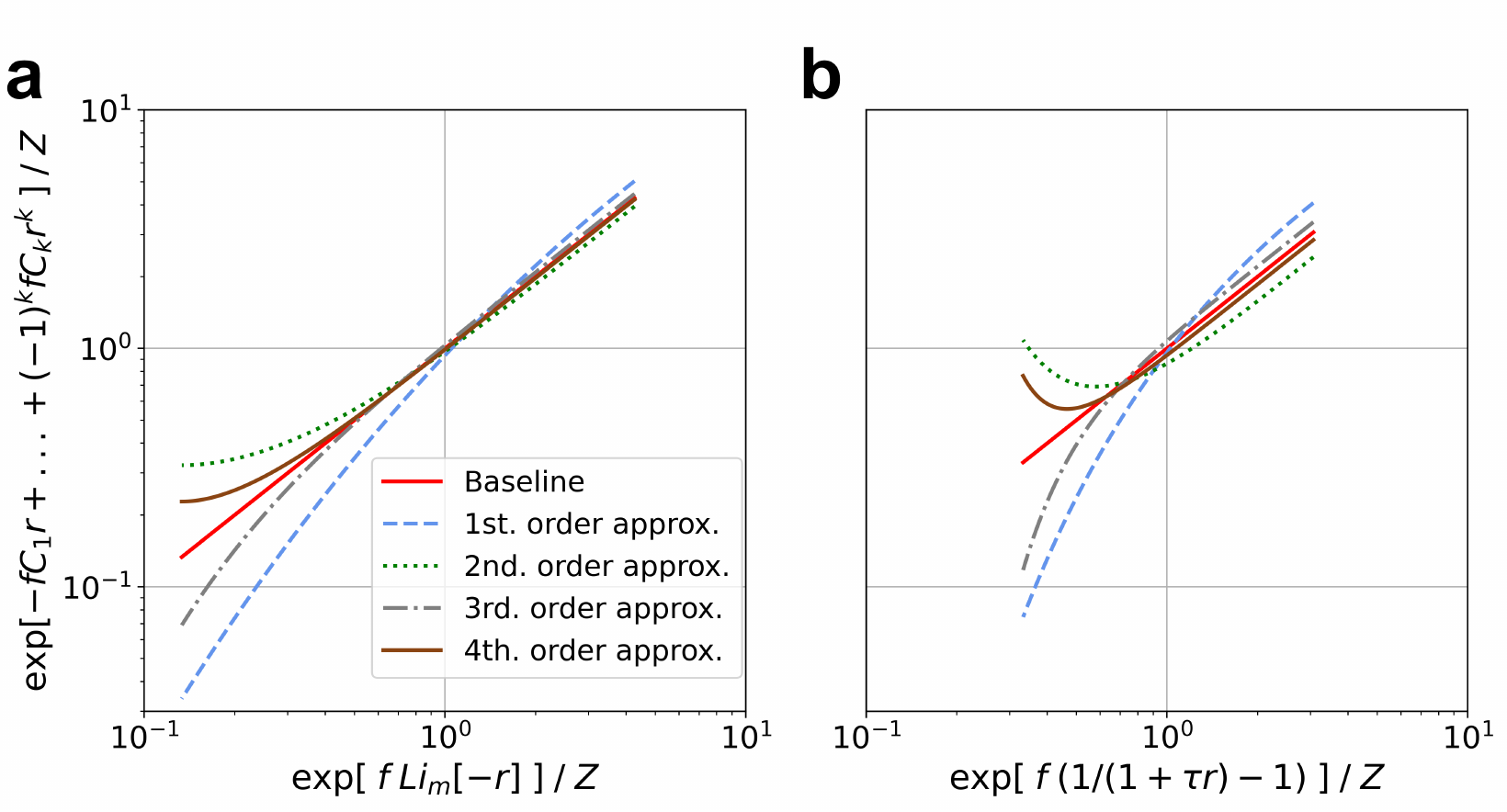}
\end{center}
\captionof{figure}{Comparison of the PDFs of the proposed models vs their approximation, including up to the $k$-th order HOIs.
\textbf{a} The polylogarithmic exponential PDF approximations with $m=1$. The baseline is a red diagonal line. 
\textbf{b} The shifted-geometric exponential PDF approximations with $\tau=0.8$. In all plots, $\mathpzc{f}=5$.}
\label{fig:log_k_approx_PDFs}

\section{Discussion} \label{SECTION_DISCUSSION}

This study presented parametric models for distributions of the sparse collective activity of homogeneous binary neurons, inspired by the experimentally observed alternating HOIs. These distributions remain widespread with parameter-dependent sparsity and exhibit heavy-tails in the limit of a large number of neurons. We derived these models by establishing sufficient conditions that ensure the PMF of the homogeneous binary population results in a widespread continuous-valued population rate distribution in an infinitely large network.

The proposed models explain how a sparse, widespread distribution arises from specific HOIs with alternating shrinking structures, features that the pairwise MaxEnt model fails to explain \citep{Ganmor_etal_2011,Tkacik_etal_2014}. The models are consistent with a previous theoretical prediction by \cite{Amari_etal_2003}, 
which states that all orders of interactions are required to produce a widespread activity distribution in a large population of correlated neurons. However, here we showed that the base measure function, independently of the (canonical) interactions, must be chosen carefully to avoid the dominance of the entropy term in the homogeneous distribution. We also showed that the proposed models outperformed the first- and second-order interaction models when fitting to spiking data of salamander retinal ganglion cells responding to visual stimuli. In particular, the shifted-geometric exponential distribution improved the likelihood of observing the data, underscoring the importance of both the entropy-canceling base measure function and the polynomial term to account for the highly synchronized and sparse states, respectively.


Despite the prevalence of sparse, widespread distributions in neural systems, the mechanisms driving these patterns are still not fully understood and continue to be a subject of active investigation. One of the simplest yet insightful models that can reproduce these key features is the dichotomized Gaussian (DG) model \citep{Amari_etal_2003, Yu_etal_2011} and its extensions \citep{MONTANI_etal_2013,montangie2015quantifying,montangie2017higher,montangie2018common,LeenSheaBrown_2015}. The DG model consists of threshold-neurons that become active if inputs sampled from a correlated multivariate Gaussian distribution exceed a threshold. The outputs of the DG neurons, represented as $(0,1)$ patterns, exhibit sparse population activity \citep{Macke_etal_2011, Shimazaki_etal_2015} with characteristic HOIs. Specifically, they display alternating signs in the interactions at successive orders, such as negative triplewise and positive quadruple-wise interactions and so on, with a shrinking magnitude \citep{Shimazaki_etal_2015,montangie2018common}. The structured HOIs contribute to the sparse activity and create the widespread distribution \citep{Amari_etal_2003}. Supporting this theoretical prediction, the specific alternating structure of HOIs of up to the fifth order was found in population activity of hippocampal neurons \citep{Shimazaki_etal_2015}. Furthermore, using a more biologically plausible integrate-and-fire neuron model under the balanced inputs of excitatory and inhibitory neurons near the spiking threshold, it was discovered that the network architecture with excitatory shared (and hence correlated) inputs to pairs of neurons can explain the positive pairwise and negative triple-wise interactions observed in monkey and mouse visual neurons while neurons receiving common inhibition can not \citep{shomali_etal_2023_uncovering}. In this study, we demonstrated that recurrent networks with neurons exhibiting a threshold-like activation function produce sparse population activity that aligns well with empirical distributions. Altogether, these findings suggest that physiological pairwise connections, either from hidden neurons or among observed neurons, combined with a threshold-like activation function — rather than a winner-take-all network architecture — are sufficient to explain the observed sparse neuronal population activity.

 
The nonlinear functions derived from the convergence of the entropy-cancelling base measure function and alternating shrinking HOI series are at the core of our models and
fundamental for inducing the sparse and heavy-tailed
population profile. We consider that the nonlinearity driving the HOIs in neurons could originate from the combination of spiking nonlinearity at the soma and nonlinear dynamics at the dendrites.  Directional selectivity, coincidence detection in auditory neurons, temporal integration, image denoising, forward masking \citep{LondonHausser_2005_dendritic_comp}, and nonlinear integration of spatial cortical feedback in V1 neurons \citep{Fisek_etal_2023} exemplify the nonlinearity of dendritic computation. While modeling these nonlinearities opens up new avenues, sparsity is one of the main features induced by the HOIs in this study. The structured HOIs introduced by the polynomial term contribute to sparsity by inducing sublinear integration of synaptic inputs (see also \citep{ZylberbergSheaBrown_2015} for the sublinear integration induced by the negative triple-wise interactions and its optimality in coding natural scenes). The polynomial terms nonlinearly modify the input-output relationship of the neurons, endowing it with the threshold-like nonlinearity that induces sparsity in the population activity. We proposed two nonlinear functions: a logarithmic function and a function based on a shifted-geometric series, both of which are strictly decreasing functions of the population rate. Supporting the former, a modeling study on collision-sensitive locust neurons suggests that their dendritic trees implement logarithmic transformations \citep{JonesGabbiani_2012}, a finding reinforced by the fan-like dendritic structures identified in these neurons.

We note that the induced nonlinearity in a single neuron is further shaped by the base measure function. Specifically, the entropy-canceling base measure function alone can transform the logistic activation function into a linear one, promoting heavy-tailed population spike rate distributions. When combined with sparsity-inducing polynomial terms, it produces a threshold-like nonlinearity followed by supralinear activation for highly synchronous inputs. This activation profile closely resembles those observed in dense associative memories \citep{krotov2016dense, demircigil2017model}, which demonstrate significantly higher memory capacity compared to classical associative memories without HOIs. This similarity seems to suggest that both sparsity and the heavy-tailed nature of widespread distributions are involved in accommodating a large number of memories, offering insights into memory-storage capacity in biological networks.

The current model of HOIs is limited by its assumptions of homogeneity and stationarity in neural activity distributions. These assumptions simplify analysis and enable the derivation of key properties, and are legitimate depending on the questions being asked. However, they fail to capture the heterogeneity and dynamic nature observed in biological neural networks. 
Introducing heterogeneity entails allowing parameters at each order to vary across different combinations of neurons, resulting in a combinatorial increase in complexity and necessitating reliable estimation methods \citep{cayco2018moment}. One possible approach to mitigate this challenge is to reduce the number of parameters by incorporating nonlinearities into the energy function, as demonstrated in dense associative memories \citep{krotov2016dense, demircigil2017model} and curved neural networks \citep{aguilera2024explosive, barra2018new}, which exhibit enhanced memory capacity. The present study provides key insights into such nonlinearities resulting in biologically plausible neuronal activities. However, the connection between memory capacity, sparsity, and heavy-tailedness of heterogeneous models remains to be thoroughly investigated. Additionally, heterogeneity in reciprocal connections disrupts detailed balance, leading to nonequilibrium dynamics \citep{schnakenberg_network_1976, seifert2012stochastic}. In the case of the classical pairwise model, this results in the kinetic Ising model with asymmetric connections \citep{crisanti1987dynamics, aguilera2021unifying, aguilera2023nonequilibrium}, which has been applied to analyze neuronal spiking data \citep{tyrcha2013effect,hertz_ising_2013,dunn2015correlations,roudi_multi-neuronal_2015}. Such systems can exhibit oscillatory, rotational, and temporal patterns with a definitive past-future order. However, the role of HOIs in governing such dynamics remains an open question requiring further exploration.

Observed neural systems also exhibit non-stationary activity, characterized by time-varying firing rates and correlations. While fitting a stationary model can be entirely appropriate — particularly when the goal is to investigate interactions arising from the marginalization over dynamic statistical features \citep{schwab2014zipf} —, it becomes essential to introduce time-dependent models when the objective is to disentangle interactions induced by shared inputs from those stemming from firing rate dynamics \citep{tyrcha2013effect, granot2013stimulus}. 
In this context, state-space Ising models have been developed to simultaneously estimate the firing-rate and correlation dynamics \citep{shimazaki2009state, shimazaki2012state, shimazaki2013single, donner2017approximate, gaudreault2018state, gaudreault2019online}, revealing dynamical decorrelations of visual neurons within oscillatory activities during visual stimulation \citep{donner2017approximate}, or time-dependent triple-wise interaction of M1 neurons linked to a specific behavioral paradigm \citep{shimazaki2012state}. These models also facilitate the study of time-dependent thermodynamic properties like entropy and heat capacity of neural systems \citep{donner2017approximate, gaudreault2018state}. Developing a tractable state-space framework for models with HOIs and analyzing their properties — such as the dynamics of sparsity and heavy-tailed behavior — remains a key challenge. Addressing this requires advanced computational approaches accounting for non-stationarity and heterogeneity while preserving analytical tractability.

In light of recent advances in high-throughput recording technologies for investigating HOIs of biological neurons and in the enrichment of artificial neural networks through specific nonlinearities incorporating HOIs, we anticipate significant progress in constructing high-performance, energy-efficient computational architectures. Such progress can be achieved by deepening statistical and mechanistic insights into biological HOIs that induce sparsity and heavy-tailedness in neural dynamics.

\section{Acknowledgements}
We thank Miguel Aguilera and Keito Wakatsuki for discussions and valuable comments on this manuscript. This work was supported by JSPS KAKENHI Grant Number JP 20K11709, 21H05246, 25K03085.

\bibliography{references}

\begin{thebibliography}{75}
\expandafter\ifx\csname natexlab\endcsname\relax\def\natexlab#1{#1}\fi
\providecommand{\url}[1]{\texttt{#1}}
\providecommand{\href}[2]{#2}
\providecommand{\path}[1]{#1}
\providecommand{\DOIprefix}{doi:}
\providecommand{\ArXivprefix}{arXiv:}
\providecommand{\URLprefix}{URL: }
\providecommand{\Pubmedprefix}{pmid:}
\providecommand{\doi}[1]{\href{http://dx.doi.org/#1}{\path{#1}}}
\providecommand{\Pubmed}[1]{\href{pmid:#1}{\path{#1}}}
\providecommand{\bibinfo}[2]{#2}
\ifx\xfnm\relax \def\xfnm[#1]{\unskip,\space#1}\fi
\bibitem[{Aguilera et~al.(2023)Aguilera, Igarashi \& Shimazaki}]{aguilera2023nonequilibrium}
\bibinfo{author}{Aguilera, M.}, \bibinfo{author}{Igarashi, M.}, \& \bibinfo{author}{Shimazaki, H.} (\bibinfo{year}{2023}).
\newblock \bibinfo{title}{Nonequilibrium thermodynamics of the asymmetric sherrington-kirkpatrick model}.
\newblock {\it \bibinfo{journal}{Nature Communications}\/},  {\it \bibinfo{volume}{14}\/}, \bibinfo{pages}{3685}.
\bibitem[{Aguilera et~al.(2021)Aguilera, Moosavi \& Shimazaki}]{aguilera2021unifying}
\bibinfo{author}{Aguilera, M.}, \bibinfo{author}{Moosavi, S.~A.}, \& \bibinfo{author}{Shimazaki, H.} (\bibinfo{year}{2021}).
\newblock \bibinfo{title}{A unifying framework for mean-field theories of asymmetric kinetic ising systems}.
\newblock {\it \bibinfo{journal}{Nature communications}\/},  {\it \bibinfo{volume}{12}\/}, \bibinfo{pages}{1197}.
\bibitem[{Aguilera et~al.(2024)Aguilera, Morales, Rosas \& Shimazaki}]{aguilera2024explosive}
\bibinfo{author}{Aguilera, M.}, \bibinfo{author}{Morales, P.~A.}, \bibinfo{author}{Rosas, F.~E.}, \& \bibinfo{author}{Shimazaki, H.} (\bibinfo{year}{2024}).
\newblock \bibinfo{title}{Explosive neural networks via higher-order interactions in curved statistical manifolds}.
\newblock {\it \bibinfo{journal}{arXiv preprint arXiv:2408.02326}\/}, .
\bibitem[{Amari et~al.(2003)Amari, Nakahara, Wu \& Sakai}]{Amari_etal_2003}
\bibinfo{author}{Amari, S.-I.}, \bibinfo{author}{Nakahara, H.}, \bibinfo{author}{Wu, S.}, \& \bibinfo{author}{Sakai, Y.} (\bibinfo{year}{2003}).
\newblock \bibinfo{title}{{Synchronous firing and higher-order interactions in neuron pool}}.
\newblock {\it \bibinfo{journal}{Neural computation}\/},  {\it \bibinfo{volume}{15}\/}, \bibinfo{pages}{127--42}.
\bibitem[{Ambrogioni(2023)}]{ambrogioni2023search}
\bibinfo{author}{Ambrogioni, L.} (\bibinfo{year}{2023}).
\newblock \bibinfo{title}{In search of dispersed memories: Generative diffusion models are associative memory networks}.
\newblock {\it \bibinfo{journal}{arXiv preprint arXiv:2309.17290}\/}, .
\bibitem[{Attneave(1954)}]{attneave1954some}
\bibinfo{author}{Attneave, F.} (\bibinfo{year}{1954}).
\newblock \bibinfo{title}{Some informational aspects of visual perception.}
\newblock {\it \bibinfo{journal}{Psychological review}\/},  {\it \bibinfo{volume}{61}\/}, \bibinfo{pages}{183}.
\bibitem[{Barlow et~al.(1961)}]{barlow1961possible}
\bibinfo{author}{Barlow, H.~B.} et~al. (\bibinfo{year}{1961}).
\newblock \bibinfo{title}{Possible principles underlying the transformation of sensory messages}.
\newblock {\it \bibinfo{journal}{Sensory communication}\/},  {\it \bibinfo{volume}{1}\/}, \bibinfo{pages}{217--233}.
\bibitem[{Barra et~al.(2018)Barra, Beccaria \& Fachechi}]{barra2018new}
\bibinfo{author}{Barra, A.}, \bibinfo{author}{Beccaria, M.}, \& \bibinfo{author}{Fachechi, A.} (\bibinfo{year}{2018}).
\newblock \bibinfo{title}{A new mechanical approach to handle generalized hopfield neural networks}.
\newblock {\it \bibinfo{journal}{Neural Networks}\/},  {\it \bibinfo{volume}{106}\/}, \bibinfo{pages}{205--222}.
\bibitem[{Barreiro et~al.(2014)Barreiro, Gjorgjieva, Rieke \& Shea-Brown}]{Barreiro_etal_2014}
\bibinfo{author}{Barreiro, A.}, \bibinfo{author}{Gjorgjieva, J.}, \bibinfo{author}{Rieke, F.}, \& \bibinfo{author}{Shea-Brown, E.} (\bibinfo{year}{2014}).
\newblock \bibinfo{title}{When do microcircuits produce beyond-pairwise correlations?}
\newblock {\it \bibinfo{journal}{Frontiers in Computational Neuroscience}\/},  {\it \bibinfo{volume}{8}\/}.
\bibitem[{Berry et~al.(1997)Berry, Warland \& Meister}]{Berry_etal_1997_retinal_ganglion}
\bibinfo{author}{Berry, M.}, \bibinfo{author}{Warland, D.}, \& \bibinfo{author}{Meister, M.} (\bibinfo{year}{1997}).
\newblock \bibinfo{title}{The structure and precision of retinal spike trains}.
\newblock {\it \bibinfo{journal}{Proceedings of the National Academy of Sciences of the United States of America}\/},  {\it \bibinfo{volume}{94}\/}, \bibinfo{pages}{5411--6}.
\bibitem[{Cayco-Gajic et~al.(2015)Cayco-Gajic, Zylberberg \& Shea-Brown}]{CaycoGajic_etal_2015}
\bibinfo{author}{Cayco-Gajic, N.~A.}, \bibinfo{author}{Zylberberg, J.}, \& \bibinfo{author}{Shea-Brown, E.} (\bibinfo{year}{2015}).
\newblock \bibinfo{title}{Triplet correlations among similarly tuned cells impact population coding}.
\newblock {\it \bibinfo{journal}{Frontiers in Computational Neuroscience}\/},  {\it \bibinfo{volume}{9}\/}.
\bibitem[{Cayco-Gajic et~al.(2018)Cayco-Gajic, Zylberberg \& Shea-Brown}]{cayco2018moment}
\bibinfo{author}{Cayco-Gajic, N.~A.}, \bibinfo{author}{Zylberberg, J.}, \& \bibinfo{author}{Shea-Brown, E.} (\bibinfo{year}{2018}).
\newblock \bibinfo{title}{A moment-based maximum entropy model for fitting higher-order interactions in neural data}.
\newblock {\it \bibinfo{journal}{Entropy}\/},  {\it \bibinfo{volume}{20}\/}, \bibinfo{pages}{489}.
\bibitem[{Chettih \& Harvey(2019)}]{Chettih_and_Harvey_2019}
\bibinfo{author}{Chettih, S.}, \& \bibinfo{author}{Harvey, C.} (\bibinfo{year}{2019}).
\newblock \bibinfo{title}{Single-neuron perturbations reveal feature-specific competition in v1}.
\newblock {\it \bibinfo{journal}{Nature}\/},  {\it \bibinfo{volume}{567}\/}, \bibinfo{pages}{334–340}.
\bibitem[{Crisanti \& Sompolinsky(1987)}]{crisanti1987dynamics}
\bibinfo{author}{Crisanti, A.}, \& \bibinfo{author}{Sompolinsky, H.} (\bibinfo{year}{1987}).
\newblock \bibinfo{title}{Dynamics of spin systems with randomly asymmetric bonds: Langevin dynamics and a spherical model}.
\newblock {\it \bibinfo{journal}{Physical Review A}\/},  {\it \bibinfo{volume}{36}\/}, \bibinfo{pages}{4922}.
\bibitem[{Demircigil et~al.(2017)Demircigil, Heusel, L{\"o}we, Upgang \& Vermet}]{demircigil2017model}
\bibinfo{author}{Demircigil, M.}, \bibinfo{author}{Heusel, J.}, \bibinfo{author}{L{\"o}we, M.}, \bibinfo{author}{Upgang, S.}, \& \bibinfo{author}{Vermet, F.} (\bibinfo{year}{2017}).
\newblock \bibinfo{title}{On a model of associative memory with huge storage capacity}.
\newblock {\it \bibinfo{journal}{Journal of Statistical Physics}\/},  {\it \bibinfo{volume}{168}\/}, \bibinfo{pages}{288--299}.
\bibitem[{Donner et~al.(2017)Donner, Obermayer \& Shimazaki}]{donner2017approximate}
\bibinfo{author}{Donner, C.}, \bibinfo{author}{Obermayer, K.}, \& \bibinfo{author}{Shimazaki, H.} (\bibinfo{year}{2017}).
\newblock \bibinfo{title}{Approximate inference for time-varying interactions and macroscopic dynamics of neural populations}.
\newblock {\it \bibinfo{journal}{PLoS computational biology}\/},  {\it \bibinfo{volume}{13}\/}, \bibinfo{pages}{e1005309}.
\bibitem[{Dunn et~al.(2015)Dunn, M{\o}rreaunet \& Roudi}]{dunn2015correlations}
\bibinfo{author}{Dunn, B.}, \bibinfo{author}{M{\o}rreaunet, M.}, \& \bibinfo{author}{Roudi, Y.} (\bibinfo{year}{2015}).
\newblock \bibinfo{title}{Correlations and functional connections in a population of grid cells}.
\newblock {\it \bibinfo{journal}{PLoS computational biology}\/},  {\it \bibinfo{volume}{11}\/}, \bibinfo{pages}{e1004052}.
\bibitem[{Field(1994)}]{field1994goal}
\bibinfo{author}{Field, D.~J.} (\bibinfo{year}{1994}).
\newblock \bibinfo{title}{What is the goal of sensory coding?}
\newblock {\it \bibinfo{journal}{Neural computation}\/},  {\it \bibinfo{volume}{6}\/}, \bibinfo{pages}{559--601}.
\bibitem[{Fişek et~al.(2023)Fişek, Herrmann, Egea-Weiss, Cloves, Bauer, Lee, Russell \& Häusser}]{Fisek_etal_2023}
\bibinfo{author}{Fişek, M.}, \bibinfo{author}{Herrmann, D.}, \bibinfo{author}{Egea-Weiss, A.}, \bibinfo{author}{Cloves, M.}, \bibinfo{author}{Bauer, L.}, \bibinfo{author}{Lee, T.-Y.}, \bibinfo{author}{Russell, L.~E.}, \& \bibinfo{author}{Häusser, M.} (\bibinfo{year}{2023}).
\newblock \bibinfo{title}{Cortico-cortical feedback engages active dendrites in visual cortex}.
\newblock {\it \bibinfo{journal}{Nature}\/}, .
\bibitem[{Froudarakis et~al.(2014)Froudarakis, Berens, Ecker, Cotton, Sinz, Yatsenko, Saggau, Bethge \& Tolias}]{Froudakaris_etal_2014}
\bibinfo{author}{Froudarakis, E.}, \bibinfo{author}{Berens, P.}, \bibinfo{author}{Ecker, A.}, \bibinfo{author}{Cotton, R.}, \bibinfo{author}{Sinz, F.}, \bibinfo{author}{Yatsenko, D.}, \bibinfo{author}{Saggau, P.}, \bibinfo{author}{Bethge, M.}, \& \bibinfo{author}{Tolias, A.} (\bibinfo{year}{2014}).
\newblock \bibinfo{title}{Population code in mouse v1 facilitates readout of natural scenes through increased sparseness}.
\newblock {\it \bibinfo{journal}{Nature neuroscience}\/},  {\it \bibinfo{volume}{17}\/}, \bibinfo{pages}{851--7}.
\bibitem[{Ganmor et~al.(2011)Ganmor, Segev \& Schneidman}]{Ganmor_etal_2011}
\bibinfo{author}{Ganmor, E.}, \bibinfo{author}{Segev, R.}, \& \bibinfo{author}{Schneidman, E.} (\bibinfo{year}{2011}).
\newblock \bibinfo{title}{Sparse low-order interaction network underlies a highly correlated and learnable neural population code}.
\newblock {\it \bibinfo{journal}{Proceedings of the National Academy of Sciences}\/},  {\it \bibinfo{volume}{108}\/}, \bibinfo{pages}{9679--9684}.
\bibitem[{Gaudreault et~al.(2019)Gaudreault, Saxena \& Shimazaki}]{gaudreault2019online}
\bibinfo{author}{Gaudreault, J.}, \bibinfo{author}{Saxena, A.}, \& \bibinfo{author}{Shimazaki, H.} (\bibinfo{year}{2019}).
\newblock \bibinfo{title}{Online estimation of multiple dynamic graphs in pattern sequences}.
\newblock In {\it \bibinfo{booktitle}{2019 International Joint Conference on Neural Networks (IJCNN)}\/} (pp. \bibinfo{pages}{1--8}).
\newblock \bibinfo{organization}{IEEE}.
\bibitem[{Gaudreault \& Shimazaki(2018)}]{gaudreault2018state}
\bibinfo{author}{Gaudreault, J.}, \& \bibinfo{author}{Shimazaki, H.} (\bibinfo{year}{2018}).
\newblock \bibinfo{title}{State-space analysis of an ising model reveals contributions of pairwise interactions to sparseness, fluctuation, and stimulus coding of monkey v1 neurons}.
\newblock In {\it \bibinfo{booktitle}{Artificial Neural Networks and Machine Learning--ICANN 2018: 27th International Conference on Artificial Neural Networks, Rhodes, Greece, October 4-7, 2018, Proceedings, Part III 27}\/} (pp. \bibinfo{pages}{641--651}).
\newblock \bibinfo{organization}{Springer}.
\bibitem[{Granot-Atedgi et~al.(2013)Granot-Atedgi, Tka{\v{c}}ik, Segev \& Schneidman}]{granot2013stimulus}
\bibinfo{author}{Granot-Atedgi, E.}, \bibinfo{author}{Tka{\v{c}}ik, G.}, \bibinfo{author}{Segev, R.}, \& \bibinfo{author}{Schneidman, E.} (\bibinfo{year}{2013}).
\newblock \bibinfo{title}{Stimulus-dependent maximum entropy models of neural population codes}.
\newblock {\it \bibinfo{journal}{PLoS computational biology}\/},  {\it \bibinfo{volume}{9}\/}, \bibinfo{pages}{e1002922}.
\bibitem[{Hertz et~al.(2013)Hertz, Roudi \& Tyrcha}]{hertz_ising_2013}
\bibinfo{author}{Hertz, J.}, \bibinfo{author}{Roudi, Y.}, \& \bibinfo{author}{Tyrcha, J.} (\bibinfo{year}{2013}).
\newblock \bibinfo{title}{Ising model for inferring network structure from spike data: {In} {Principal} of {Neural} {Coding}}.
\newblock In {\it \bibinfo{booktitle}{Principles of {Neural} {Coding}}\/} (pp. \bibinfo{pages}{527--546}).
\newblock \bibinfo{publisher}{CRC Press}.
\bibitem[{Jaynes(1957)}]{jaynes1957information}
\bibinfo{author}{Jaynes, E.~T.} (\bibinfo{year}{1957}).
\newblock \bibinfo{title}{Information theory and statistical mechanics}.
\newblock {\it \bibinfo{journal}{Physical review}\/},  {\it \bibinfo{volume}{106}\/}, \bibinfo{pages}{620}.
\bibitem[{Jones \& Gabbiani(2012)}]{JonesGabbiani_2012}
\bibinfo{author}{Jones, P.~W.}, \& \bibinfo{author}{Gabbiani, F.} (\bibinfo{year}{2012}).
\newblock \bibinfo{title}{Logarithmic compression of sensory signals within the dendritic tree of a collision-sensitive neuron}.
\newblock {\it \bibinfo{journal}{The Journal of neuroscience: the official journal of the Society for Neuroscience}\/},  {\it \bibinfo{volume}{32}\/}, \bibinfo{pages}{4923–4934}.
\bibitem[{Krotov(2023)}]{krotov2023new}
\bibinfo{author}{Krotov, D.} (\bibinfo{year}{2023}).
\newblock \bibinfo{title}{A new frontier for hopfield networks}.
\newblock {\it \bibinfo{journal}{Nature Reviews Physics}\/},  {\it \bibinfo{volume}{5}\/}, \bibinfo{pages}{366--367}.
\bibitem[{Krotov \& Hopfield(2016)}]{krotov2016dense}
\bibinfo{author}{Krotov, D.}, \& \bibinfo{author}{Hopfield, J.~J.} (\bibinfo{year}{2016}).
\newblock \bibinfo{title}{Dense associative memory for pattern recognition}.
\newblock {\it \bibinfo{journal}{Advances in neural information processing systems}\/},  {\it \bibinfo{volume}{29}\/}.
\bibitem[{Leen \& Shea-Brown(2015)}]{LeenSheaBrown_2015}
\bibinfo{author}{Leen, D.}, \& \bibinfo{author}{Shea-Brown, E.} (\bibinfo{year}{2015}).
\newblock \bibinfo{title}{A simple mechanism for beyond-pairwise correlations in integrate-and-fire neurons}.
\newblock {\it \bibinfo{journal}{Journal of mathematical neuroscience}\/},  {\it \bibinfo{volume}{5}\/}.
\bibitem[{Loback et~al.(2017)Loback, Tkačik, Prentice, Ioffe, Berry~II, Marre \& Berry}]{Loback_etal_2017}
\bibinfo{author}{Loback, A.~R.}, \bibinfo{author}{Tkačik, G.}, \bibinfo{author}{Prentice, J.~S.}, \bibinfo{author}{Ioffe, M.~L.}, \bibinfo{author}{Berry~II, M.~J.}, \bibinfo{author}{Marre, O.}, \& \bibinfo{author}{Berry, M.~J.} (\bibinfo{year}{2017}).
\newblock \bibinfo{title}{Data from: Error-robust modes of the retinal population code [dataset]}.
\newblock \bibinfo{note}{Dryad}.
\bibitem[{London \& Häusser(2005)}]{LondonHausser_2005_dendritic_comp}
\bibinfo{author}{London, M.}, \& \bibinfo{author}{Häusser, M.} (\bibinfo{year}{2005}).
\newblock \bibinfo{title}{Dendritic computation}.
\newblock {\it \bibinfo{journal}{Annual Review of Neuroscience}\/},  {\it \bibinfo{volume}{28}\/}, \bibinfo{pages}{503--532}.
\newblock \bibinfo{note}{PMID: 16033324}.
\bibitem[{Macke et~al.(2011)Macke, Opper \& Bethge}]{Macke_etal_2011}
\bibinfo{author}{Macke, J.~H.}, \bibinfo{author}{Opper, M.}, \& \bibinfo{author}{Bethge, M.} (\bibinfo{year}{2011}).
\newblock \bibinfo{title}{Common input explains higher-order correlations and entropy in a simple model of neural population activity}.
\newblock {\it \bibinfo{journal}{Physical review letters}\/},  {\it \bibinfo{volume}{106}\/}, \bibinfo{pages}{208102}.
\bibitem[{Marre et~al.(2017)Marre, Tka{\v{c}}ik, Amodei, Schneidman, Bialek \& Berry}]{marre2017multi}
\bibinfo{author}{Marre, O.}, \bibinfo{author}{Tka{\v{c}}ik, G.}, \bibinfo{author}{Amodei, D.}, \bibinfo{author}{Schneidman, E.}, \bibinfo{author}{Bialek, W.}, \& \bibinfo{author}{Berry, M.} (\bibinfo{year}{2017}).
\newblock {\it \bibinfo{title}{Multi-electrode array recording from salamander retinal ganglion cells}\/}.
\newblock \bibinfo{type}{Technical Report} Institute of Science and Technology Austria.
\bibitem[{Martignon et~al.(2000)Martignon, Deco, Laskey, Diamond, Freiwald \& Vaadia}]{Martignol_etal_2000}
\bibinfo{author}{Martignon, L.}, \bibinfo{author}{Deco, G.}, \bibinfo{author}{Laskey, K.}, \bibinfo{author}{Diamond, M.}, \bibinfo{author}{Freiwald, W.}, \& \bibinfo{author}{Vaadia, E.} (\bibinfo{year}{2000}).
\newblock \bibinfo{title}{{Neural coding: higher-order temporal patterns in the neurostatistics of cell assemblies}}.
\newblock {\it \bibinfo{journal}{Neural computation}\/},  {\it \bibinfo{volume}{12}\/}, \bibinfo{pages}{2621--2653}.
\bibitem[{Masina(2019)}]{Masina_2019_exponential_integral_func}
\bibinfo{author}{Masina, E.} (\bibinfo{year}{2019}).
\newblock \bibinfo{title}{Useful review on the exponential-integral special function}.
\newblock \href{http://arxiv.org/abs/1907.12373}{\tt arXiv:1907.12373}.
\bibitem[{Montangie \& Montani(2015)}]{montangie2015quantifying}
\bibinfo{author}{Montangie, L.}, \& \bibinfo{author}{Montani, F.} (\bibinfo{year}{2015}).
\newblock \bibinfo{title}{Quantifying higher-order correlations in a neuronal pool}.
\newblock {\it \bibinfo{journal}{Physica A: Statistical Mechanics and its Applications}\/},  {\it \bibinfo{volume}{421}\/}, \bibinfo{pages}{388--400}.
\bibitem[{Montangie \& Montani(2017)}]{montangie2017higher}
\bibinfo{author}{Montangie, L.}, \& \bibinfo{author}{Montani, F.} (\bibinfo{year}{2017}).
\newblock \bibinfo{title}{Higher-order correlations in common input shapes the output spiking activity of a neural population}.
\newblock {\it \bibinfo{journal}{Physica A: Statistical Mechanics and its Applications}\/},  {\it \bibinfo{volume}{471}\/}, \bibinfo{pages}{845--861}.
\bibitem[{Montangie \& Montani(2018)}]{montangie2018common}
\bibinfo{author}{Montangie, L.}, \& \bibinfo{author}{Montani, F.} (\bibinfo{year}{2018}).
\newblock \bibinfo{title}{Common inputs in subthreshold membrane potential: The role of quiescent states in neuronal activity}.
\newblock {\it \bibinfo{journal}{Physical Review E}\/},  {\it \bibinfo{volume}{97}\/}, \bibinfo{pages}{060302}.
\bibitem[{Montani et~al.(2009)Montani, Ince, Senatore, Arabzadeh, Diamond \& Panzeri}]{Montani_etal_2009}
\bibinfo{author}{Montani, F.}, \bibinfo{author}{Ince, R.~A.}, \bibinfo{author}{Senatore, R.}, \bibinfo{author}{Arabzadeh, E.}, \bibinfo{author}{Diamond, M.~E.}, \& \bibinfo{author}{Panzeri, S.} (\bibinfo{year}{2009}).
\newblock \bibinfo{title}{{The impact of high-order interactions on the rate of synchronous discharge and information transmission in somatosensory cortex}}.
\newblock {\it \bibinfo{journal}{Philosophical transactions. Series A, Mathematical, physical, and engineering sciences}\/},  {\it \bibinfo{volume}{367}\/}, \bibinfo{pages}{3297--3310}.
\bibitem[{Montani et~al.(2013)Montani, Phoka, Portesi \& Schultz}]{MONTANI_etal_2013}
\bibinfo{author}{Montani, F.}, \bibinfo{author}{Phoka, E.}, \bibinfo{author}{Portesi, M.}, \& \bibinfo{author}{Schultz, S.~R.} (\bibinfo{year}{2013}).
\newblock \bibinfo{title}{Statistical modelling of higher-order correlations in pools of neural activity}.
\newblock {\it \bibinfo{journal}{Physica A: Statistical Mechanics and its Applications}\/},  {\it \bibinfo{volume}{392}\/}, \bibinfo{pages}{3066--3086}.
\bibitem[{Morales \& Rosas(2021)}]{morales2021generalization}
\bibinfo{author}{Morales, P.~A.}, \& \bibinfo{author}{Rosas, F.~E.} (\bibinfo{year}{2021}).
\newblock \bibinfo{title}{Generalization of the maximum entropy principle for curved statistical manifolds}.
\newblock {\it \bibinfo{journal}{Phys. Rev. Res.}\/},  {\it \bibinfo{volume}{3}\/}, \bibinfo{pages}{033216}.
\bibitem[{Moshitch \& Nelken(2014)}]{MOSHITCH_NELKEN_2014}
\bibinfo{author}{Moshitch, D.}, \& \bibinfo{author}{Nelken, I.} (\bibinfo{year}{2014}).
\newblock \bibinfo{title}{Using tweedie distributions for fitting spike count data}.
\newblock {\it \bibinfo{journal}{Journal of Neuroscience Methods}\/},  {\it \bibinfo{volume}{225}\/}, \bibinfo{pages}{13--28}.
\bibitem[{Mulero et~al.(2017)Mulero, Sordo, de Souza \& Suárez-LLorens}]{Mulero_etal_2017_stochastic_dom}
\bibinfo{author}{Mulero, J.}, \bibinfo{author}{Sordo, M.~A.}, \bibinfo{author}{de Souza, M.~C.}, \& \bibinfo{author}{Suárez-LLorens, A.} (\bibinfo{year}{2017}).
\newblock \bibinfo{title}{Two stochastic dominance criteria based on tail comparisons}.
\newblock {\it \bibinfo{journal}{Applied Stochastic Models in Business and Industry}\/},  {\it \bibinfo{volume}{33}\/}, \bibinfo{pages}{575--589}.
\bibitem[{Nair et~al.(2022)Nair, Wierman \& Zwart}]{Nair_Wierman_Zwart_2022}
\bibinfo{author}{Nair, J.}, \bibinfo{author}{Wierman, A.}, \& \bibinfo{author}{Zwart, B.} (\bibinfo{year}{2022}).
\newblock {\it \bibinfo{title}{The Fundamentals of Heavy Tails: Properties, Emergence, and Estimation}\/}.
\newblock Cambridge Series in Statistical and Probabilistic Mathematics.
\newblock \bibinfo{publisher}{Cambridge University Press}.
\bibitem[{Nakahara \& Amari(2002)}]{NakaharaAmari_2002}
\bibinfo{author}{Nakahara, H.}, \& \bibinfo{author}{Amari, S.} (\bibinfo{year}{2002}).
\newblock \bibinfo{title}{{Information-geometric measure for neural spikes}}.
\newblock {\it \bibinfo{journal}{Neural computation}\/},  {\it \bibinfo{volume}{14}\/}, \bibinfo{pages}{2269--2316}.
\bibitem[{Ohiorhenuan et~al.(2010)Ohiorhenuan, Mechler, Purpura, Schmid, Hu \& Victor}]{Ohiorhenuan_etal_2010}
\bibinfo{author}{Ohiorhenuan, I.~E.}, \bibinfo{author}{Mechler, F.}, \bibinfo{author}{Purpura, K.~P.}, \bibinfo{author}{Schmid, A.~M.}, \bibinfo{author}{Hu, Q.}, \& \bibinfo{author}{Victor, J.~D.} (\bibinfo{year}{2010}).
\newblock \bibinfo{title}{{Sparse coding and high-order correlations in fine-scale cortical networks}}.
\newblock {\it \bibinfo{journal}{Nature}\/},  {\it \bibinfo{volume}{466}\/}, \bibinfo{pages}{617--621}.
\bibitem[{Olshausen \& Field(1997)}]{OLSHAUSEN_FIELD_1997}
\bibinfo{author}{Olshausen, B.~A.}, \& \bibinfo{author}{Field, D.~J.} (\bibinfo{year}{1997}).
\newblock \bibinfo{title}{Sparse coding with an overcomplete basis set: A strategy employed by v1?}
\newblock {\it \bibinfo{journal}{Vision Research}\/},  {\it \bibinfo{volume}{37}\/}, \bibinfo{pages}{3311 -- 3325}.
\bibitem[{Ramsauer et~al.(2020)Ramsauer, Sch{\"a}fl, Lehner, Seidl, Widrich, Adler, Gruber, Holzleitner, Pavlovi{\'c}, Sandve et~al.}]{ramsauer2020hopfield}
\bibinfo{author}{Ramsauer, H.}, \bibinfo{author}{Sch{\"a}fl, B.}, \bibinfo{author}{Lehner, J.}, \bibinfo{author}{Seidl, P.}, \bibinfo{author}{Widrich, M.}, \bibinfo{author}{Adler, T.}, \bibinfo{author}{Gruber, L.}, \bibinfo{author}{Holzleitner, M.}, \bibinfo{author}{Pavlovi{\'c}, M.}, \bibinfo{author}{Sandve, G.~K.} et~al. (\bibinfo{year}{2020}).
\newblock \bibinfo{title}{Hopfield networks is all you need}.
\newblock {\it \bibinfo{journal}{arXiv preprint arXiv:2008.02217}\/}, .
\bibitem[{Rolls \& Tovee(1995)}]{rolls1995sparseness}
\bibinfo{author}{Rolls, E.~T.}, \& \bibinfo{author}{Tovee, M.~J.} (\bibinfo{year}{1995}).
\newblock \bibinfo{title}{Sparseness of the neuronal representation of stimuli in the primate temporal visual cortex}.
\newblock {\it \bibinfo{journal}{Journal of neurophysiology}\/},  {\it \bibinfo{volume}{73}\/}, \bibinfo{pages}{713--726}.
\bibitem[{Roudi et~al.(2015)Roudi, Dunn \& Hertz}]{roudi_multi-neuronal_2015}
\bibinfo{author}{Roudi, Y.}, \bibinfo{author}{Dunn, B.}, \& \bibinfo{author}{Hertz, J.} (\bibinfo{year}{2015}).
\newblock \bibinfo{title}{Multi-neuronal activity and functional connectivity in cell assemblies}.
\newblock {\it \bibinfo{journal}{Current opinion in neurobiology}\/},  {\it \bibinfo{volume}{32}\/}, \bibinfo{pages}{38--44}. \DOIprefix\doi{10.1016/j.conb.2014.10.011}.
\bibitem[{Roudi et~al.(2009)Roudi, Nirenberg \& Latham}]{roudi2009pairwise}
\bibinfo{author}{Roudi, Y.}, \bibinfo{author}{Nirenberg, S.}, \& \bibinfo{author}{Latham, P.~E.} (\bibinfo{year}{2009}).
\newblock \bibinfo{title}{Pairwise maximum entropy models for studying large biological systems: when they can work and when they can't}.
\newblock {\it \bibinfo{journal}{PLoS computational biology}\/},  {\it \bibinfo{volume}{5}\/}, \bibinfo{pages}{e1000380}.
\bibitem[{Schnakenberg(1976)}]{schnakenberg_network_1976}
\bibinfo{author}{Schnakenberg, J.} (\bibinfo{year}{1976}).
\newblock \bibinfo{title}{Network theory of microscopic and macroscopic behavior of master equation systems}.
\newblock {\it \bibinfo{journal}{Reviews of Modern Physics}\/},  {\it \bibinfo{volume}{48}\/}, \bibinfo{pages}{571--585}.
\bibitem[{Schneidman et~al.(2006)Schneidman, Berry, Segev \& Bialek}]{Schneidmain_etal_2006}
\bibinfo{author}{Schneidman, E.}, \bibinfo{author}{Berry, M.~J.}, \bibinfo{author}{Segev, R.}, \& \bibinfo{author}{Bialek, W.} (\bibinfo{year}{2006}).
\newblock \bibinfo{title}{{Weak pairwise correlations imply strongly correlated network states in a neural population}}.
\newblock {\it \bibinfo{journal}{Nature}\/},  {\it \bibinfo{volume}{440}\/}, \bibinfo{pages}{1007--1012}.
\bibitem[{Schwab et~al.(2014)Schwab, Nemenman \& Mehta}]{schwab2014zipf}
\bibinfo{author}{Schwab, D.~J.}, \bibinfo{author}{Nemenman, I.}, \& \bibinfo{author}{Mehta, P.} (\bibinfo{year}{2014}).
\newblock \bibinfo{title}{Zipf’s law and criticality in multivariate data without fine-tuning}.
\newblock {\it \bibinfo{journal}{Physical review letters}\/},  {\it \bibinfo{volume}{113}\/}, \bibinfo{pages}{068102}.
\bibitem[{Schwartz et~al.(2004)Schwartz, Movellan, Wachtler, Albright \& Sejnowski}]{Schwartz_etal_2004}
\bibinfo{author}{Schwartz, O.}, \bibinfo{author}{Movellan, J.}, \bibinfo{author}{Wachtler, T.}, \bibinfo{author}{Albright, T.}, \& \bibinfo{author}{Sejnowski, T.} (\bibinfo{year}{2004}).
\newblock \bibinfo{title}{Spike count distributions, factorizability, and contextual effects in area v1}.
\newblock {\it \bibinfo{journal}{Neurocomputing}\/},  {\it \bibinfo{volume}{58-60}\/}, \bibinfo{pages}{893--900}.
\bibitem[{Seifert(2012)}]{seifert2012stochastic}
\bibinfo{author}{Seifert, U.} (\bibinfo{year}{2012}).
\newblock \bibinfo{title}{Stochastic thermodynamics, fluctuation theorems and molecular machines}.
\newblock {\it \bibinfo{journal}{Reports on progress in physics}\/},  {\it \bibinfo{volume}{75}\/}, \bibinfo{pages}{126001}.
\bibitem[{Shimazaki(2013)}]{shimazaki2013single}
\bibinfo{author}{Shimazaki, H.} (\bibinfo{year}{2013}).
\newblock \bibinfo{title}{Single-trial estimation of stimulus and spike-history effects on time-varying ensemble spiking activity of multiple neurons: a simulation study}.
\newblock In {\it \bibinfo{booktitle}{Journal of Physics: Conference Series}\/} (p. \bibinfo{pages}{012009}).
\newblock \bibinfo{organization}{IOP Publishing} volume \bibinfo{volume}{473}.
\bibitem[{Shimazaki et~al.(2009)Shimazaki, Amari, Brown \& Grun}]{shimazaki2009state}
\bibinfo{author}{Shimazaki, H.}, \bibinfo{author}{Amari, S.-i.}, \bibinfo{author}{Brown, E.~N.}, \& \bibinfo{author}{Grun, S.} (\bibinfo{year}{2009}).
\newblock \bibinfo{title}{State-space analysis on time-varying correlations in parallel spike sequences}.
\newblock In {\it \bibinfo{booktitle}{2009 IEEE International Conference on Acoustics, Speech and Signal Processing}\/} (pp. \bibinfo{pages}{3501--3504}).
\newblock \bibinfo{organization}{IEEE}.
\bibitem[{Shimazaki et~al.(2012{\natexlab{a}})Shimazaki, Amari, Brown \& Gr{\"u}n}]{shimazaki2012state}
\bibinfo{author}{Shimazaki, H.}, \bibinfo{author}{Amari, S.-i.}, \bibinfo{author}{Brown, E.~N.}, \& \bibinfo{author}{Gr{\"u}n, S.} (\bibinfo{year}{2012}{\natexlab{a}}).
\newblock \bibinfo{title}{State-space analysis of time-varying higher-order spike correlation for multiple neural spike train data}.
\newblock {\it \bibinfo{journal}{PLoS computational biology}\/},  {\it \bibinfo{volume}{8}\/}, \bibinfo{pages}{e1002385}.
\bibitem[{Shimazaki et~al.(2012{\natexlab{b}})Shimazaki, Amari, Brown \& Grün}]{Shimazaki_etal_2012}
\bibinfo{author}{Shimazaki, H.}, \bibinfo{author}{Amari, S.-i.}, \bibinfo{author}{Brown, E.~N.}, \& \bibinfo{author}{Grün, S.} (\bibinfo{year}{2012}{\natexlab{b}}).
\newblock \bibinfo{title}{State-space analysis of time-varying higher-order spike correlation for multiple neural spike train data}.
\newblock {\it \bibinfo{journal}{PLOS Computational Biology}\/},  {\it \bibinfo{volume}{8}\/}, \bibinfo{pages}{1--27}.
\bibitem[{Shimazaki et~al.(2015)Shimazaki, Sadeghi, Ishikawa, Ikegaya \& Toyoizumi}]{Shimazaki_etal_2015}
\bibinfo{author}{Shimazaki, H.}, \bibinfo{author}{Sadeghi, K.}, \bibinfo{author}{Ishikawa, T.}, \bibinfo{author}{Ikegaya, Y.}, \& \bibinfo{author}{Toyoizumi, T.} (\bibinfo{year}{2015}).
\newblock \bibinfo{title}{Simultaneous silence organizes structured higher-order interactions in neural populations}.
\newblock {\it \bibinfo{journal}{Scientific Reports}\/},  {\it \bibinfo{volume}{5}\/}.
\bibitem[{Shlens et~al.(2006)Shlens, Field, Gauthier, Grivich, Petrusca, Sher, Litke \& Chichilnisky}]{Shlens_etal_2006}
\bibinfo{author}{Shlens, J.}, \bibinfo{author}{Field, G.~D.}, \bibinfo{author}{Gauthier, J.~L.}, \bibinfo{author}{Grivich, M.~I.}, \bibinfo{author}{Petrusca, D.}, \bibinfo{author}{Sher, A.}, \bibinfo{author}{Litke, A.~M.}, \& \bibinfo{author}{Chichilnisky, E.~J.} (\bibinfo{year}{2006}).
\newblock \bibinfo{title}{The structure of multi-neuron firing patterns in primate retina}.
\newblock {\it \bibinfo{journal}{Journal of Neuroscience}\/},  {\it \bibinfo{volume}{26}\/}, \bibinfo{pages}{8254--8266}.
\bibitem[{Shomali et~al.(2023)Shomali, Rasuli, Ahmadabadi \& Shimazaki}]{shomali_etal_2023_uncovering}
\bibinfo{author}{Shomali, S.~R.}, \bibinfo{author}{Rasuli, S.~N.}, \bibinfo{author}{Ahmadabadi, M.~N.}, \& \bibinfo{author}{Shimazaki, H.} (\bibinfo{year}{2023}).
\newblock \bibinfo{title}{Uncovering hidden network architecture from spiking activities using an exact statistical input-output relation of neurons}.
\newblock {\it \bibinfo{journal}{Communications Biology}\/},  {\it \bibinfo{volume}{6}\/}, \bibinfo{pages}{169}.
\bibitem[{Tkačik et~al.(2014)Tkačik, Marre, Amodei, Schneidman, Bialek \& Berry}]{Tkacik_etal_2014}
\bibinfo{author}{Tkačik, G.}, \bibinfo{author}{Marre, O.}, \bibinfo{author}{Amodei, D.}, \bibinfo{author}{Schneidman, E.}, \bibinfo{author}{Bialek, W.}, \& \bibinfo{author}{Berry, M.~J., II} (\bibinfo{year}{2014}).
\newblock \bibinfo{title}{Searching for collective behavior in a large network of sensory neurons}.
\newblock {\it \bibinfo{journal}{PLOS Computational Biology}\/},  {\it \bibinfo{volume}{10}\/}, \bibinfo{pages}{1--23}.
\bibitem[{Tkačik et~al.(2015)Tkačik, Mora, Marre, Amodei, Palmer, Berry \& Bialek}]{Tkacik_etal_2015_thermodynamics}
\bibinfo{author}{Tkačik, G.}, \bibinfo{author}{Mora, T.}, \bibinfo{author}{Marre, O.}, \bibinfo{author}{Amodei, D.}, \bibinfo{author}{Palmer, S.~E.}, \bibinfo{author}{Berry, M.~J.}, \& \bibinfo{author}{Bialek, W.} (\bibinfo{year}{2015}).
\newblock \bibinfo{title}{Thermodynamics and signatures of criticality in a network of neurons}.
\newblock {\it \bibinfo{journal}{Proceedings of the National Academy of Sciences}\/},  {\it \bibinfo{volume}{112}\/}, \bibinfo{pages}{11508--11513}.
\bibitem[{Tyrcha et~al.(2013)Tyrcha, Roudi, Marsili \& Hertz}]{tyrcha2013effect}
\bibinfo{author}{Tyrcha, J.}, \bibinfo{author}{Roudi, Y.}, \bibinfo{author}{Marsili, M.}, \& \bibinfo{author}{Hertz, J.} (\bibinfo{year}{2013}).
\newblock \bibinfo{title}{The effect of nonstationarity on models inferred from neural data}.
\newblock {\it \bibinfo{journal}{Journal of Statistical Mechanics: Theory and Experiment}\/},  {\it \bibinfo{volume}{2013}\/}, \bibinfo{pages}{P03005}.
\bibitem[{Vinje \& Gallant(2000)}]{VinjeGallant_2000}
\bibinfo{author}{Vinje, W.~E.}, \& \bibinfo{author}{Gallant, J.~L.} (\bibinfo{year}{2000}).
\newblock \bibinfo{title}{Sparse coding and decorrelation in primary visual cortex during natural vision}.
\newblock {\it \bibinfo{journal}{Science}\/},  {\it \bibinfo{volume}{287}\/}, \bibinfo{pages}{1273–1276}.
\bibitem[{Willmore et~al.(2011)Willmore, Mazer \& Gallant}]{WillmoreMazerGallant_2011}
\bibinfo{author}{Willmore, B.}, \bibinfo{author}{Mazer, J.}, \& \bibinfo{author}{Gallant, J.} (\bibinfo{year}{2011}).
\newblock \bibinfo{title}{Sparse coding in striate and extrastriate visual cortex}.
\newblock {\it \bibinfo{journal}{Journal of neurophysiology}\/},  {\it \bibinfo{volume}{105}\/}, \bibinfo{pages}{2907–2919}.
\bibitem[{Willmore \& Tolhurst(2001)}]{willmore2001characterizing}
\bibinfo{author}{Willmore, B.}, \& \bibinfo{author}{Tolhurst, D.~J.} (\bibinfo{year}{2001}).
\newblock \bibinfo{title}{Characterizing the sparseness of neural codes}.
\newblock {\it \bibinfo{journal}{Network: Computation in Neural Systems}\/},  {\it \bibinfo{volume}{12}\/}, \bibinfo{pages}{255}.
\bibitem[{Wohrer et~al.(2013)Wohrer, Humphries \& Machens}]{Wohrer_etal_2013_PopWideDistr}
\bibinfo{author}{Wohrer, A.}, \bibinfo{author}{Humphries, M.}, \& \bibinfo{author}{Machens, C.} (\bibinfo{year}{2013}).
\newblock \bibinfo{title}{Population-wide distributions of neural activity during perceptual decision-making}.
\newblock {\it \bibinfo{journal}{Progress in neurobiology}\/},  (pp. \bibinfo{pages}{156--93}).
\bibitem[{Yen et~al.(2007)Yen, Baker \& Gray}]{YenBakerGray_2007}
\bibinfo{author}{Yen, S.~C.}, \bibinfo{author}{Baker, J.}, \& \bibinfo{author}{Gray, C.~M.} (\bibinfo{year}{2007}).
\newblock \bibinfo{title}{Heterogeneity in the responses of adjacent neurons to natural stimuli in cat striate cortex}.
\newblock {\it \bibinfo{journal}{Journal of neurophysiology}\/},  {\it \bibinfo{volume}{97}\/}, \bibinfo{pages}{1326–1341}.
\bibitem[{Yildiz(2015)}]{Yildiz_2015_mit}
\bibinfo{author}{Yildiz, M.} (\bibinfo{year}{2015}).
\newblock \bibinfo{title}{Chapter 4: Stochastic dominance lecture notes}.
\newblock In {\it \bibinfo{booktitle}{Microeconomic Theory III---MIT Course 14.123}\/}.
\newblock \bibinfo{address}{Cambridge~MA}.
\newblock \bibinfo{note}{{MIT OpenCourseWare}}.
\bibitem[{Yu et~al.(2011)Yu, Yang, Nakahara, Santos, Nikoli{\'c} \& Plenz}]{Yu_etal_2011}
\bibinfo{author}{Yu, S.}, \bibinfo{author}{Yang, H.}, \bibinfo{author}{Nakahara, H.}, \bibinfo{author}{Santos, G.~S.}, \bibinfo{author}{Nikoli{\'c}, D.}, \& \bibinfo{author}{Plenz, D.} (\bibinfo{year}{2011}).
\newblock \bibinfo{title}{Higher-order interactions characterized in cortical activity}.
\newblock {\it \bibinfo{journal}{Journal of Neuroscience}\/},  {\it \bibinfo{volume}{31}\/}, \bibinfo{pages}{17514--17526}.
\bibitem[{Zylberberg \& Shea-Brown(2015)}]{ZylberbergSheaBrown_2015}
\bibinfo{author}{Zylberberg, J.}, \& \bibinfo{author}{Shea-Brown, E.} (\bibinfo{year}{2015}).
\newblock \bibinfo{title}{Input nonlinearities can shape beyond-pairwise correlations and improve information transmission by neural populations}.
\newblock {\it \bibinfo{journal}{Phys. Rev. E}\/},  {\it \bibinfo{volume}{92}\/}, \bibinfo{pages}{062707}.

\end{thebibliography}

\newpage
\appendix

\setcounter{figure}{0}
\setcounter{page}{1}

\section{Homogeneous sparse population of neurons}

\subsection{Proof of Theorem \ref{theorem_widespread_zero_peaked}} \label{APPENDIX_THEOREM_PROOF}

We define the distribution to be concentrated at $r^{0}_N$ in the limit $N \to \infty$ if it converges to a delta function centered at $r^{0}_N$. For the PMF Eq.~\eqref{eq_homogeneous_r_pmf_g_function} to remain a non-concentrated or widespread distribution in the limit, the following condition must be satisfied \citep{Amari_etal_2003}
\begin{equation}
    \label{eq_theorem_widespread_zero_peak_widespread_proof_01}
    \lim_{N \to \infty} N \left[ G_N\left(r_N ; \boldsymbol{\theta}_N \right) - G_N\left(r^{0}_N ; \boldsymbol{\theta}_N \right)   \right] < \infty
\end{equation}
$$
\forall \; r^{0} _N \neq r_N \; \in S_r.
$$
This condition ensures that the difference in the exponent values between any two distinct points remains finite for all 
$\; r^{0} _N \neq r_N$ within the support $S_r$, thereby preventing the distribution from concentrating at a single point as $N \to \infty$. 

Because $N G_N\left( r_N ; \boldsymbol{\theta}_N \right)$ is of the constant order (see Eq.~\eqref{eq_theorem_widespread_zero_peaked_order}), we have
\begin{align}
    \mathcal{O}\left( N G_N \left(r_N ; \boldsymbol{\theta}_N \right) - N G_N\left(r^{0} _N; \boldsymbol{\theta}_N \right)\right) = \mathcal{O}\left(1\right),
    \label{eq_theorem_widespread_zero_peak_widespread_proof_02}
\end{align}
which implies that the inequality in Eq.~\eqref{eq_theorem_widespread_zero_peak_widespread_proof_01} holds for all $r_N \in S_r$ with $G_N\left( r_N ; \boldsymbol{\theta}_N \right) < \infty$. Therefore, the PMF Eq.~\eqref{eq_homogeneous_r_pmf_g_function} does not concentrate but instead remains widespread in the limit of $N \to \infty$.
 
Next, we prove that the maximum of the PMF is located at zero in the limit as $N \to \infty$. Recall that $G_N\left( \cdot \right)$ is a non-positive, strictly decreasing function with finite values over the support $S_r$. This implies that the maximum of $G_N\left(r_N ; \boldsymbol{\theta}_N \right)$, and consequently the maximum of the PMF Eq.~\eqref{eq_homogeneous_r_pmf_g_function}, occurs at 
\begin{align}
    \inf \left\{ S_r \right\} = \inf \left\{  0, \frac{1}{N}, \frac{2}{N}, \hdots, 1 \right\} = 0.
\end{align}
Furthermore, the maximum remains at $r=0$ in the limit, \begin{align}
    \lim_{N \to \infty} \mathcal{P}\left( r_N | \boldsymbol{\theta}_N \right)  = p\left( r | \boldsymbol{\lambda} \right) dr, 
\end{align}
since $S_r \to [0,1]$ as $N \to \infty$ and by continuity of the strictly decreasing property of $G_N\left( \cdot \right)$.

\newpage
\subsection{First-order homogeneous model}\label{APPENDIX_SUBSECTION_INDEPENDENT_HOM_MODEL}

The distribution function corresponding to the PDF from Eq.~\eqref{eq_homogeneous_first_order_negative_interactions_r_pdf} is
\begin{align}
\label{eq_homogeneous_first_order_negative_interactions_r_F}
    F\left( u | f \right)
    &=\frac{1 - e^{-\mathpzc{f} u} }{ 1 - e^{-\mathpzc{f}}  }.
\end{align}
The mean is given by
\begin{align}
\mu_{R} &= \frac{1}{Z}\int_{0}^{1}  e^{-\mathpzc{f}r} r dr = \frac{1}{\mathpzc{f}\left( 1 -e^{-\mathpzc{f}} \right)} - \frac{e^{-\mathpzc{f}}}{1-e^{-\mathpzc{f}}}\left( 1 + \frac{1}{\mathpzc{f}} \right)    \label{eq_independent_dist_mean_r},
\end{align}
and the variance is
\begin{align}
\sigma_{R} ^{2} &= \frac{1}{Z}\int_{0}^{1} e^{-\mathpzc{f}r} r^2 dr - \mu_{R} ^{2} \nonumber \\
&= \frac{2}{\mathpzc{f}^2 \left( 1 - e^{-\mathpzc{f}} \right)} - \frac{e^{-\mathpzc{f}}}{ 1 - e^{-\mathpzc{f}} } \left( \frac{2}{\mathpzc{f}^2} + \frac{2}{\mathpzc{f}}  \right) 
 -\frac{1}{\mathpzc{f}} e^{-\mathpzc{f}} - \mu_{R} ^{2}.
    \label{eq_independent_dist_variance_r}
\end{align}

\newpage
\subsection{Second-order homogeneous model}\label{APPENDIX_SUBSECTION_SECOND_HOM_MODEL}
The normalization constant of the second-order homogeneous model is obtained as
\begin{align}
Z & = \int_{0}^{1} e^{ \mathpzc{f}_1 r + \mathpzc{f}_2 r^2 }  dr  \nonumber \\
& = \left \{ \begin{array}{ccc}
     \frac{ \sqrt{\pi} }{2 \sqrt{\mathpzc{f}_2}} \exp\left[ -\left( \frac{ \mathpzc{f_1} }{ 2 \sqrt{\mathpzc{f_2}} } \right)^2 \right] \text{erfi}\left( \frac{ \mathpzc{f_1} }{ 2 \sqrt{\mathpzc{f_2}} } + \sqrt{\mathpzc{f}_2} r \right) \Bigg |_0 ^{1} & & \text{if} \;\mathpzc{f}_2 > 0  \\
     \frac{ \sqrt{\pi} }{2 \sqrt{ | \mathpzc{f}_2} | } \exp\left[ \left( \frac{ \mathpzc{f_1} }{ 2 \sqrt{ | \mathpzc{f_2} | } } \right)^2 \right] \text{erf} \left( - \frac{ \mathpzc{f_1} }{ 2 \sqrt{ | \mathpzc{f_2} | } } + \sqrt{ | \mathpzc{f}_2 | } r \right) \Bigg |_0 ^{1} &  & \text{if}\;\mathpzc{f}_2 < 0
\end{array} \right . ,
\label{eq_homogeneous_second_order_sparse_r_pdf_Z}
\end{align}
where the imaginary error function $\text{erfi}\left(\cdot\right)$ is defined as
\begin{equation}
    \label{eq_special_function_erfi}
    \text{erfi}\left( x \right) = \sum\limits_{k=0}^{\infty} \frac{ x^{2k + 1} }{(2k + 1) k!},
\end{equation}
and the error function $\text{erf}\left(\cdot\right)$ as
\begin{equation}
    \label{eq_special_function_erf}
    \text{erf}\left( x \right) = \sum\limits_{k=0}^{\infty} \frac{ \left(-1\right)^{k} x^{2k + 1} }{(2k + 1) k!}.
\end{equation}

\newpage
\subsection{The first-order stochastic dominance} \label{APPENDIX_HEAVY_TAILEDNESS}

The inequality condition given by Eq.~\eqref{eq_definition_heavy_tail_distr_in_r_02} represents a  case of strict first-order stochastic dominance. Let $X$ and $Y$ be two random variables with distribution $F_X\left(x\right)$ and $F_Y\left(y\right)$, respectively. Then $Y$ is said to be stochastically dominated by $X$ in the first-order sense if and only if either of the following equivalent conditions hold \citep{Mulero_etal_2017_stochastic_dom,Yildiz_2015_mit} 
\begin{align}
    \label{eq_first_order_stochastic_dominance_01}
    F_Y\left(x\right) & \geq F_X\left(x\right) \;\; \forall x \in \mathbb{R}, \\
    \label{eq_first_order_stochastic_dominance_02}
   \mathbb{E}_{Y}\left[ u\left( Y \right)\right] 
   &\leq \mathbb{E}_{X}\left[ u\left( X \right) \right],
\end{align}
for any non-decreasing function $u\left( \cdot \right)$.

It follows that the heavy-tail definition in Eq.~\eqref{eq_definition_heavy_tail_distr_in_r_02} constitutes a case of first-order stochastic dominance with strict inequality, wherein the two distribution functions do not overlap for $r \in \left(0,1\right)$. Hence, here we provide a proof of the equivalence of the two conditions stated above, following the approach in \citep{Yildiz_2015_mit}, adapted to the strict inequality case by assuming that $u\left( \cdot \right)$ is a non-decreasing and \textit{non-constant} function.

For any two continuous and invertible distribution functions $F_X\left( x\right)$ and $F_Y \left( y \right)$, suppose that the following strict inequality holds
\begin{align}
    \label{eq_first_order_stochastic_dominance_strict_01}
    F _{Y}\left(x\right) > F_{X}\left(x\right)  \;\;\;\forall x \in \mathbb{R}.
\end{align}
By taking the inverse with respect to $F_Y ( \cdot)$ on both sides, we have
\begin{align}
    \label{eq_first_order_stochastic_dominance_strict_relative_ordering_points_x_y}
    x > F_{Y} ^{-1} \left( F_{X}\left( x \right) \right) \equiv y(x).
\end{align}
This inequality reflects the relative ordering of the points $x$ and $y$ that correspond to the same cumulative probability under the two distributions. We note that $F_{X}\left( x \right) = F_{Y}(y(x))$.  

Then, for any non-decreasing and non-constant function $u(x)$, the strict inequality $ x > y(x) $ implies $ u(x) > u(y(x)) $. Hence,
\begin{align}
\label{eq_first_order_stochastic_dominance_strict_02}
    \int u(x) \, dF_X(x) &> \int u(y(x)) \, dF_X(x).
\end{align}
Now, since $F_X(x) = F_Y(y(x))$, and assuming $ F_Y $ is continuous and strictly increasing (so \( y(x) \) is invertible and differentiable), the change of variables $y = y(x)$ implies
\begin{align}
    \int u(y(x)) \, dF_X(x) = \int u(y) \, dF_Y(y).
\end{align}
Therefore, we have
\begin{align}
    \int u(x) \, dF_X(x) > \int u(y) \, dF_Y(y),
\end{align}
which proves the strict form of Eq.~\eqref{eq_first_order_stochastic_dominance_02}.

On the other hand, departing from the strict inequality in the expectations,
\begin{align}
    \label{eq_first_order_stochastic_dominance_strict_expectation_01}
    \mathbb{E}_{Y}\left[ u\left( Y \right) \right] < \mathbb{E}_{X}\left[ u\left( X \right) \right],
\end{align}

\noindent and assuming $F_X\left( x \right) < F_Y \left( x \right)$ does not hold for all $x$, then there exists some $x_0$ such that $F_X(x_0) \geq F_Y(x_0)$. In particular, taking $u(x) = \mathbbm{1}_{[x\geq x_0]}$ the following holds
\begin{align}
    \label{eq_first_order_stochastic_dominance_strict_expectation_01_proof_by_contradiction}
    \int \mathbbm{1}_{[x\geq x_0]} d F_{X}\left(x\right) & = 1 - F_X \left(x_0\right) \nonumber \\
    & \leq 1 - F_Y \left( x_0 \right) \nonumber \\
    &= \int \mathbbm{1}_{[ y \geq x_0 ]} d F_Y \left( y \right),
\end{align}

\noindent which contradicts the initial strict inequality in \eqref{eq_first_order_stochastic_dominance_strict_expectation_01} and therefore if \eqref{eq_first_order_stochastic_dominance_strict_expectation_01} holds the following strict inequality must hold
\begin{align}
    F_X\left( x \right) < F_Y \left( x \right) \;\;\;\forall x \in \mathbb{R}.
\end{align}
This completes the proof of the equivalence between Eqs.~\eqref{eq_first_order_stochastic_dominance_strict_01} and \eqref{eq_first_order_stochastic_dominance_strict_02} under the assumption of a non-decreasing, non-constant function $u(x)$.

Returning to our original sparse distributions, we identify $F_{X}$ with $F_{\boldsymbol{\lambda}}$ and $F_Y$ with $F_{\mathpzc{f}}^{exp}$. 
Then, the equivalence of Eqs.~\eqref{eq_first_order_stochastic_dominance_strict_01} and \eqref{eq_first_order_stochastic_dominance_strict_02}
enables us to prove the heavy-tailedness of the distributions $F_{\boldsymbol{\lambda}}$  directly from Eq.~\eqref{eq_first_order_stochastic_dominance_strict_expectation_01}.

Note that $p\left( r ; \boldsymbol{\lambda} \right)$ is unimodal with a peak at $r=0$ and that the exponent of the exponential model, $q\left( r ; \boldsymbol{\lambda} \right)$, is a strictly decreasing function of $r$.  Increasing the sparsity parameter $\mathpzc{f}$ makes $q\left( r ; \boldsymbol{\lambda} \right)$ decrease faster along $r$, which is preserved for $\exp\left( q\left( r ; \boldsymbol{\lambda} \right) \right)$. Thus, as $\mathpzc{f}$ increases,  the density decreases faster along $r$, while the peak value of the PDF at $r=0$ increases monotonically. Conversely, as $\mathpzc{f}$ decreases, $q\left( r ; \boldsymbol{\lambda} \right)$ approaches to zero, and the density becomes flatter. These behaviours ensure that $\lim_{\mathpzc{f} \to \infty} p\left( r | \boldsymbol{\lambda} \right) = \delta(0)$ and $\lim_{\mathpzc{f} \to 0} p\left( r | \boldsymbol{\lambda} \right) = 1$. 




For our purposes, we additionally assume $u\left(r\right)$ is non-negative and finite on $[0,1]$, ensuring that the expectation remains well-defined and strictly ordered. This yields the following inequalities:
\begin{align}
 \int_{0} ^{1} \left(\lim_{\mathpzc{f} \to 0} p\left( r | \boldsymbol{\lambda} \right)  \right) u\left(r\right) dr 
     & = \int_{0} ^{1}  u\left(r\right) dr \nonumber \\
     &> \int_0 ^{1} p\left( r | \boldsymbol{\lambda}_k \right) u\left(r\right) dr \nonumber \\
     &> \int_{0} ^{1} \left(\lim_{\mathpzc{f} \to \infty} p\left( r | \boldsymbol{\lambda} \right)  \right) u\left(r\right) dr 
     = u\left(0\right),
     \label{eq_alternating_shrinking_expectation_bounds}
\end{align}
where $\mathpzc{f}_k \in \boldsymbol{\lambda}_k $ denotes sparsity level with $0 < \mathpzc{f}_k < +\infty$. This inequality indicates that the expectation strictly decreases as the distribution becomes more concentrated near $r=0$, reflecting the increasing sparsity induced by larger values of $\mathpzc{f}$. We now formalize this monotonic behavior in the following lemma.

\begin{lemma}
    \label{lemma_expectation_decreasing_property}
    The expectation of a non-negative, non-decreasing, and non-constant function $u\left(r\right)$ over a distribution $F_{\boldsymbol{\lambda}}\left( r \right)$ whose density $p\left(r | \boldsymbol{\lambda}\right)$ satisfies Corollary \ref{corollary_h_function_g_polynomial} 
    \begin{equation}
        \label{eq_lemma_expectation_decreasing_property}
            \mathbb{E}_{R | \boldsymbol{\lambda}}\left[ u\left(r\right)\right] = \int_0 ^{1} p\left( r | \boldsymbol{\lambda} \right) u\left(r\right) dr
    \end{equation}
    is a strictly decreasing function with respect to the parameter $\mathpzc{f}$.
\end{lemma}
We note that the bounds in Eq.~\eqref{eq_alternating_shrinking_expectation_bounds} and Lemma \ref{lemma_expectation_decreasing_property} also apply to the (truncated) exponential distribution (our baseline reference distribution). See Appendix \ref{APPENDIX_HEAVY_TAILEDNESS_PROOF} for proofs that our proposed distributions are heavy-tailed.

\newpage
\section{Entropy-dominated homogeneous population}
\label{APPENDIX_ENTROPY_DOMINATED_CASE}

We now obtain Eq.~\eqref{eq_G_r_function_02}. First, by using the Stirling formula with order notation (Eq.~\eqref{eq_Stirling_identity_factorials}), the logarithm of the binomial coefficient becomes
\begin{align}
\log 
\left( \begin{array}{c}
     N\\
     N r_N 
\end{array} \right) = & \log N! - \log \left( N r_N \right) ! - \log \left( N \left( 1 - r_N\right) \right)! \nonumber \\
=& \log\left( \sqrt{2 \pi N} N^{N} e^{-N} \right) + \log\left(  1 + \mathcal{O}\left(\frac{1}{N} \right) \right) \nonumber \\
& - \log\left( \sqrt{2 \pi N r_N} \left(N r_N\right)^{N r_N} e^{-N r_N} \right) - \log \left( 1 + \mathcal{O}\left(\frac{1}{N r_N}\right)\right) \nonumber \\
& - \log\left( \sqrt{2 \pi N \left(1 - r_N\right)} \left( N \left(1 - r_N\right) \right)^{N\left(1 - r_N\right)} e^{-N\left(1 - r_N\right)} \right)  \nonumber \\
& -\log\left( 1 + \mathcal{O}\left(\frac{1}{N\left(1-r_N\right)}\right)\right) \nonumber \\
=& -\log\left( \sqrt{2 \pi N r_N \left(1 - r_N\right)} \right) + N H\left(r_N\right) + \mathcal{O}\left( \frac{1}{N} \left( 1 - \frac{1}{1-r_N} - \frac{1}{r_N} \right) \right),
\label{eq_binomial_coefficient_large_N_equivalence}
\end{align}
where the entropy term $H\left( r_N \right)$ is defined in Eq.~\eqref{eq_entropy_r}. Since we consider $h\left(N r_N\right) = 1$ for this case and given Eq.~\eqref{eq_binomial_coefficient_large_N_equivalence} the function $G_N\left( r_N ; \theta_N \right)$ becomes
\begin{align}
G_N\left( r_N ; \boldsymbol{\theta}_N \right) = & \frac{1}{N} \log 
\left( \begin{array}{c}
     N\\
     N r_N 
\end{array} \right)  + \frac{1}{N} Q_N\left( r_N ; \boldsymbol{\theta}_N \right)  \nonumber \\
=& -\frac{1}{N}\log\left( \sqrt{2 \pi N r_N \left(1 - r_N\right)} \right) + H\left(r_N\right) + \frac{1}{N}Q_N\left( r_N ; \boldsymbol{\theta}_N \right) \nonumber \\
& +\frac{1}{N} \mathcal{O}\left( \frac{1}{N} \left( 1 - \frac{1}{1-r_N} - \frac{1}{r_N} \right) \right).
    \label{eq_G_r_function_03}
\end{align}

The discrete support for the population rate can be partitioned into two sets where the $G_N\left( \cdot; \boldsymbol{\theta}_N \right)$ function is either positive or non-positive  as follows
\begin{equation}
    \label{eq_r_support_positive_negative}
    r_N \in S_{r}^{G_N +} \cup S_{r}^{G_N -},
\end{equation}
\noindent where
\begin{equation}
    \label{eq_Sr_G_positive}
    S_r ^{G_N +} \equiv \left\{ z_N \in S_r \; | \;G_N\left( r_N ; \boldsymbol{\theta}_N \right) > 0  \right\},
\end{equation}
and
\begin{equation}
    \label{eq_Sr_G_negative}
    S_r ^{G_N -} \equiv \left\{ z_N \in S_r\; | \;G_N\left( r_N ; \boldsymbol{\theta}_N \right) \leq 0  \right\}.
\end{equation}
The dominance of the entropy term seen in Eq.~\eqref{eq_entropy_dominating_case_NG_order} for large values of $N$ happens for $r_N \in S_r ^{G_N +}$. The fact that the entropy is non-negative and that it dominates over any non-positive term at a region allows the existence of the following maximizer
\begin{align}
   r^{*} _N &= \underset{r_N  \in S_r ^{G_N +} \cup S_r ^{G_N -}}{\text{argmax}}\left\{ G_N\left( r_N ; \boldsymbol{\theta}_N \right) \right\} \nonumber \\
   &= \underset{r_N \in S_r ^{G_N +}}{\text{argmax}}\left\{ G_N\left( r_N;  \boldsymbol{\theta}_N \right) \right\},
    \label{eq_r_maximizer_positive_negative_support}
\end{align}
\noindent with $0 < r^{*}_N < 1$. Such a maximizer is strictly not at the extremes because the entropy vanishes there.

We now provide the limiting distribution for this entropy-dominated case. Recall that $r$ denotes the continuous value that the random variable $R$ takes.

We note that Eq.~\eqref{eq_homogeneous_r_pmf_g_function} can be rewritten by sending the numerator to the denominator and using the Kronecker notation as indicator function for each term in  the PMF as
\begin{align}
    \mathcal{P}\left( r_N | \boldsymbol{\theta}_N \right) 
    &=\sum\limits_{r ^{j} \in S_r}
    \frac{1}{\sum\limits_{r' \in S_r} \exp\left[ N \left( G_N\left(r'_N ; \boldsymbol{\theta}_N\right) - G_N \left(r ^{j} ; \boldsymbol{\theta}_N\right)\right) \right]} \delta_{r_N , r ^{j} } \nonumber \\
    &= 
    \sum\limits_{r ^{j} \in S_r} \frac{1}{ \Xi\left( r ^{j} ; \boldsymbol{\theta}_N \right) } \delta_{r_N , r ^{j} } .
    \label{eq_homogeneous_r_pmf_entropy_dominating_case_02}
\end{align}
 
Here, we can express the summation in each denominator (i.e., $\Xi\left( r_N  ; \boldsymbol{\theta}_N \right)$) by the disjoint parts as 
\begin{align}
    \Xi\left( r_N ; \boldsymbol{\theta}_N \right) =  z ^{0} + z_N ^{+}\left(r_N ;  \boldsymbol{\theta}_N \right) + z_N ^{-}\left(r_N ; \boldsymbol{\theta}_N \right),
\end{align}
where
\begin{align}
    z ^{0}  &= \exp\left[ N \left( 0 \right) \right] = 1 , \nonumber\\
    z_N ^{+} \left( r_N ; \boldsymbol{\theta}_N \right) &= \sum_{r'_N \in B_{r'}\left( r_N \right) ^{+} } \exp\left[ N \left[( G_N\left(r'_N ; \boldsymbol{\theta}_N\right) - G_N\left( r_N ; \boldsymbol{\theta}_N\right)  \right) \right] , \nonumber\\
    z_N ^{-} \left( r_N ; \boldsymbol{\theta}_N \right) &= \sum_{ r'_N  \in B_{r'}\left( r_N \right) ^{-} } \exp\left[ N \left( G_N\left(r'_N ; \boldsymbol{\theta}_N\right) - G_N\left( r_N ; \boldsymbol{\theta}_N\right)  \right)  \right].
    \label{eq_z_disjoint_parts}
\end{align}
\noindent with
\begin{equation}
    \label{eq_rr_G_diff_pos}
    B_{r'}\left( r_N \right) ^{+} \equiv \left\{ r'_N \; | G_N\left( r'_N ; \boldsymbol{\theta}_N \right) - \;G_N\left( r_N ; \boldsymbol{\theta}_N \right) > 0  \right\}.
\end{equation}
\begin{equation}
    \label{eq_rr_G_diff_neg}
    B_{r'}\left( r_N \right) ^{-} \equiv \left\{ r'_N \; | G_N\left( r'_N ; \boldsymbol{\theta}_N \right) - \;G_N\left( r_N ; \boldsymbol{\theta}_N \right) < 0  \right\}.
\end{equation}

For the evaluation at the extremes of the support, we note that $G_N\left(r^{*}_N; \boldsymbol{\theta}_N \right) > G_N \left( r_N ; \boldsymbol{\theta}_N \right)  \;\;\;\forall r_N \neq r^{*}_N$. In the limit we obtain for each denominator
\begin{align}
   & \lim_{N \rightarrow \infty } \Xi\left( r_N ^{j} ; \boldsymbol{\theta}_N \right) \\
   & = \left\{
   \begin{array}{cccc}
        z ^{0} + z_N ^{+}\left(r_N ;  \boldsymbol{\theta}_N \right) + z_N ^{-}\left(r_N ; \boldsymbol{\theta}_N \right)  & = & 1 + \infty + 0  &  \text{if\;} r_N = 0 \\
        z ^{0} + z_N ^{+}\left(r_N ;  \boldsymbol{\theta}_N \right) + z_N ^{-}\left(r_N ; \boldsymbol{\theta}_N \right)  & = & 1 + \infty + 0  &  \text{if\;} r_N = 1 \\
        z ^{0} + z_N ^{+}\left(r_N ;  \boldsymbol{\theta}_N \right) + z_N ^{-}\left(r_N ; \boldsymbol{\theta}_N \right)  & = & 1 + \infty + 0  &  \text{if\;} r_N \neq 0 , r_N \neq 1,\\
        \;  & \; & \;  & \phantom{;;;}  r_N \neq r_N ^{*} \\
       z ^{0} + z_N ^{-}\left(r_N ; \boldsymbol{\theta}_N \right)  & = & 1 + 0  &  \text{if\;} r_N = r_N ^{*} ,
   \end{array}
   \right .
    \label{eq_Xi_limit}
\end{align}
\noindent where $r_N ^{*} \rightarrow r ^{*}$ for sufficiently large $N$. Then, the evaluation of the PMF at each of those points in the limit becomes
\begin{align}
    \label{eq_P_r_limits}
    \lim_{N \rightarrow \infty } \mathcal{P}\left( r_N | \boldsymbol{\theta}_N \right) =& \left\{ 
    \begin{array}{cccc}
        \frac{1}{\infty } \delta\left( r  \right)  & = & 0  & \text{if\;} r_N = 0 \\
          \frac{1}{\infty } \delta\left( r  - 1 \right)  & = & 0  & \text{if\;} r_N = 1 \\
         \frac{1}{\infty } \delta\left( 0 \right)  & = & 0  & \text{if\;} r_N \neq 0 , r_N \neq 1,\\
         \;  & \; & \;  & \phantom{;;;}  r_N \neq r_N ^{*} \\
          \delta\left( r  - r^{*}  \right)  & = & +\infty  & \text{if\;} r_N = r_N ^{*} 
    \end{array}
    \right .
\end{align}
Hence, by combining the four cases for $r_N$, we obtain in the limit that the PDF is zero everywhere except at $r^{*}$ where it tends to infinity, which corresponds to the delta PDF in Eq.~\eqref{eq_P_r_limit_delta_entropy_peak} and the distribution function in Eq.~\eqref{eq_F_r_limit_heavyside_entropy_peak}.

\newpage
\section{The model with alternating and shrinking higher-order interactions}

\subsection{Canonical coordinates} \label{APPENDIX_CANONICAL_COORDINATES}

The sum of total binary activity elevated to a given power $k > 0$ and $k \leq N$ can be expanded as follows (using the Multinomial Theorem), 
\begin{align}
\left( \sum_{i=1}^{N} x_i \right) ^{k}
        =&
        \sum_{i=1}^{N} x_i ^{k} +
        \sum_{ i_1 < i_2 } 
        \sum_{\substack{
                  k_{i_1} + k_{i_2} = k\\
                  }} \left( \begin{array}{c}
                       k\\
                       k_{i_1}, k_{i_2} 
                  \end{array} \right)
                  x_{i_1} ^{k_{i_1}} x_{i_2} ^{k_{i_2}} 
                  \nonumber \\
        & + \sum_{ i_1 < i_2 <i_3} 
        \sum_{\substack{
                  k_{i_1} + k_{i_2} + k_{i_3}= k\\
                  }} \left( \begin{array}{c}
                       k\\
                       k_{i_1}, k_{i_2}, k_{i_3} 
                  \end{array} \right)
                  x_{i_1} ^{k_{i_1}} x_{i_2} ^{k_{i_2}}
                  x_{i_3} ^{k_{i_3}}
                   \nonumber \\
        & + \cdots + \sum_{ i_1 < i_2 < \hdots < i_k  } 
        k! \, 
                  x_{i_1} x_{i_2} \hdots x_{i_k},
    \label{eq_multinomial_theorem_expansion}
\end{align}
where 
$\left( \begin{array}{c}
    k\\
    k_{i_1}, k_{i_2}, \hdots, k_{N} 
    \end{array} \right) = \frac{k!} {k_{i_1}! k_{i_2}! \cdots k_{i_N}!}$ is a multinomial coefficient with $k_{i_1}>0, k_{i_2}>0,\ldots,k_{_{i_N}}>0$. Using the fact that any $x_i$ is binary, the powers of $x_i$ in Eq.~\eqref{eq_multinomial_theorem_expansion} reduce as follows
\begin{align}
\left( \sum_{i=1}^{N} x_i \right) ^{k} =&
        \sum_{i=1}^{N} x_i +
        \sum_{ i_1 < i_2 } 
        \sum_{\substack{
                  k_{i_1} + k_{i_2} = k\\
                  }} \left( \begin{array}{c}
                       k\\
                       k_{i_1}, k_{i_2} 
                  \end{array} \right) x_{i_1}  x_{i_2} 
                  + \cdots + \nonumber \\
        & + \sum_{ i_1 < i_2 <i_3}
        \sum_{\substack{
                  k_{i_1} + k_{i_2} + k_{i_3}= k\\
                  }} \left( \begin{array}{c}
                       k\\
                       k_{i_1}, k_{i_2}, k_{i_3} 
                  \end{array} \right)
                  x_{i_1}  x_{i_2} 
                  x_{i_3} 
                   \nonumber \\
        & + \cdots  + \sum_{ i_1 < i_2 < \hdots < i_k  } 
                  k! \, x_{i_1}  x_{i_2}  \hdots x_{i_k} . 
    \label{eq_multinomial_theorem_expansion_02}
\end{align}

Following the multinomial expansion from Eq.~\eqref{eq_multinomial_theorem_expansion_02}, the canonical form of our binary PMF can be obtained as
\begin{align}
    \mathcal{P}\left( \mathbf{x} | \boldsymbol{\omega} \right) 
  &= \frac{h\left(\mathbf{x}\right)}{Z} \exp\left[ -\mathpzc{f} \sum\limits_{j=1}^{N} \left(-1\right)^{j+1} C_j \left( \frac{\sum_{i=1}^{N} x_i}{N}\right)^{j}  \right] \nonumber \\
  &=
  \frac{h\left( \mathbf{x}\right)}{Z}\exp\left[ -\mathpzc{f}\left( C_1 \frac{\sum_{i=1}^{N} x_i }{N} - C_2 \frac{\left( \sum_{i=1}^{N} x_i\right)^2}{N^2}  \right . \right. \nonumber \\
  &
  \;\;\;\;\;\;\;\;\;\; \;\;\;\;\;\;\;\;\;\; \;\;\;\;\;\;\;\;\;\; \left . \left. + \hdots + \left(-1\right)^{N+1} C_N \frac{\left( \sum_{i=1}^{N} x_i  \right)^{N}}{N^{N}} \right)  \right] \nonumber \\
  &= \frac{h\left( \mathbf{x} \right)}{Z}\exp\left[ - \frac{\mathpzc{f} C_1}{N} \sum\limits_{i=1}^{N} x_i    \right . \nonumber \\
  & \phantom{\frac{h\left( \mathbf{x} \right)}{Z}\exp=}
  + \frac{\mathpzc{f} C_2}{N^{2}} \left( \sum\limits_{i=1}^{N} x_i  \right)^2 \nonumber \\
  & \phantom{\frac{h\left( \mathbf{x} \right)}{Z}\exp=}
  - \frac{\mathpzc{f} C_3}{N^{3}} \left( \sum\limits_{i=1}^{N} x_i  \right)^3 \nonumber \\
  &\phantom{\frac{h\left( \mathbf{x} \right)}{Z}\exp==} \vdots  \nonumber \\
  &\phantom{\frac{h\left( \mathbf{x} \right)}{Z}\exp=}
  \left. + \left(-1\right)^{N+1} \frac{\mathpzc{f} C_N}{N^{N}} \left( \sum\limits_{i=1}^{N} x_i  \right)^N \right] \nonumber \\
  & =
  \frac{h\left( \mathbf{x}\right)}{Z}\exp\left[- \frac{\mathpzc{f} C_1}{N} \sum\limits_{i=1}^{N} x_i  \right . \nonumber \\
  & \phantom{\frac{h\left( \mathbf{x} \right)}{Z}\exp=} 
  + \frac{\mathpzc{f} C_2}{N^{2}} \left( \sum_{i=1}^{N} x_{i} + 2 \sum_{ i_1 < i_2  } x_{i_1} x_{i_2}  \right) \nonumber \\
  & \phantom{\frac{h\left( \mathbf{x} \right)}{Z}\exp=} 
  - \frac{\mathpzc{f} C_3}{N^3} \left( \sum_{i=1}^{N} x_{i} + 6 \sum_{ i_1 <  i_2  } x_{i_1} x_{i_2} + 6 \sum_{ i_1 < i_2 < i_3 } x_{i_1} x_{i_2} x_{i_3}   \right) \nonumber \\
  & \phantom{\frac{h\left( \mathbf{x} \right)}{Z}\exp==} \vdots \nonumber\\
  & \phantom{\frac{h\left( \mathbf{x} \right)}{Z}\exp=}
    + \left. \left(-1\right)^{N+1} \frac{\mathpzc{f} C_N}{N^{N}} \left( \sum\limits_{i=1}^{N} x_{i} +
  \sum_{ i_1 < i_2  }  \sum_{\substack{
                  k_{1} + k_{2} = N
                  }} \left( \begin{array}{c}
                       N\\
                       k_{1}, k_{2} 
                  \end{array} \right) x_{i_1}  x_{i_2}  \right. \right. \nonumber \\
  & \;\;\;\;\; \;\;\;\;\; \;\;\;\;\; \;\;\;\;\; \;\;\;\;\; \;\;\;\;\; \;\;\;\;\; \;\;\;\;\; \;\;\;\;\; \;\;\;\;\; \;\;\;\;\; \;\;\;\;\; \;\;\;\;\;   + \hdots +  N! \,  x_{i_1} x_{i_2} \hdots x_{i_N}  \Bigg ) \Bigg ] \nonumber \\
    \label{eq_alternating_shrinking_binary_to_canonical_mapping}
\end{align}

Here, we compare the above equation with the canonical form of the exponential family distribution:
\begin{align}
  \mathcal{P}\left( \mathbf{x} | \boldsymbol{\theta}_N \right) = \frac{h\left( \mathbf{x}\right)}{Z}\exp\left[ \theta_1 \sum\limits_{i=1}^{N} x_i + \theta_2 \sum_{i_1 < i_2 } x_{i_1} x_{i_2} + \hdots + \theta_N x_{i_1} x_{i_2} \hdots  x_{i_N} \right],
\end{align}
where $h\left(\mathbf{x}\right)$ is defined as in Eq.~\eqref{eq_pmf_h_function}. The canonical parameter $\boldsymbol{\theta}_N$ is obtained by grouping the coefficients that multiply every power function of the coordinates in $\mathbf{x}$, i.e.,
\begin{align}
    \theta_1 =& \sum_{l=1}^{N} \left( -1 \right)^{l} \frac{\mathpzc{f} C_l }{ N^{l} } , \nonumber \\
  \theta_2 =& \sum_{l=2}^{N} \left(-1\right)^{l} \frac{ \mathpzc{f} C_l }{ N^{l} } \sum\limits_{\substack{
                  k_{1} + k_{2} = l
                  }} \left( \begin{array}{c}
       l\\
       k_1 , k_2 
  \end{array} \right), \nonumber \\
  & \vdots \nonumber \\
  \theta_N = &
  \left(-1\right)^{N} \frac{\mathpzc{f} C_N }{ N^{N} } N!.
  \label{eq_alternating_shrinking_canonical_vector2}
\end{align}

\newpage
\subsection{Population rate probability density function}
\label{APPENDIX_ALTERNATING_SHRINKING_MODEL_PDF}
We now show the proof to obtain our limiting alternating PDF from Eq.~\eqref{eq_alternating_shrinking_r_pmf_to_pdf}. For the remainder of the homogeneous analysis we make use of Stirling numbers of the first kind, which are defined as follows for $m>0$
\begin{align}
  s\left(k,m\right) = \left(-1\right)^{k-m} \left[ \begin{array}{c}
       k\\
       m 
  \end{array} \right],
      \label{eq_Stirling_first_kind_definition}
\end{align}
\noindent where
\begin{align}
  \left[ \begin{array}{c}
       k\\
       m 
  \end{array} \right] = \left( k - 1 \right) \left[ \begin{array}{c}
       k-1\\
       m
  \end{array} \right] + \left[ \begin{array}{c}
       k-1\\
       m-1
  \end{array}\right],
      \label{eq_Stirling_first_kind_coefficient}
\end{align}
\noindent with the initial conditions (for $k>0$)  
\begin{align}
  \left[ \begin{array}{c}
       0\\
       0
  \end{array}\right] = 1, \;\;\;\;\;
  \left[ \begin{array}{c}
       k\\
       0
  \end{array}\right] = \left[ \begin{array}{c}
       0\\
       k
  \end{array}\right] = 0,
      \label{eq_Stirling_first_kind_initial_cond_01}
\end{align}
\noindent and the identity  
\begin{align}
  \left[ \begin{array}{c}
       k\\
       k
  \end{array}\right] = 1.
      \label{eq_Stirling_first_kind_identity_01}
\end{align}
\noindent Stirling numbers of the first kind are useful for rewriting the binomial coefficients in an expanded polynomial form as follows  
\begin{equation}
      \label{eq_combinations_using_Stirling_S1}
      \left( \begin{array}{c}
           n\\
           k 
      \end{array} \right) = \frac{1}{k!} \sum_{m=0}^{k} s\left( k, m \right) n^{m} .
\end{equation}

The polynomial term in the exponential argument of our alternating PMF is  
\begin{align}
      Q_N \left( r_N ; \boldsymbol{\theta}_N \right) &= \sum\limits_{k=1}^{N r_N }\left( \begin{array}{c}
           N r_N\\
           k
      \end{array} \right) \theta_k \nonumber \\
      &= \sum\limits_{k=1}^{N r_N } \frac{1}{k!} \sum\limits_{m=0}^{k}s\left( k,m \right)\left( N r_N \right)^m \theta_k \nonumber \\
      &= \sum\limits_{k=1}^{N r_N} \theta_k \sum\limits_{m=0}^{k} \left(-1\right)^{k-m} \left[ \begin{array}{c}
           k\\
           m 
      \end{array} \right] \frac{\left( N r_N \right)^{m}}{ k! }  \nonumber \\
       &= \sum\limits_{k=1}^{N r_N} \theta_k \left[ \left(-1\right)^{k}\left[ \begin{array}{c}
            k\\
            0 
       \end{array} \right] \frac{\left( N r_N \right)^{0}}{k!} + \left(-1\right)^{0} \left[ \begin{array}{c}
            k\\
            k 
       \end{array} \right] \frac{\left( N r_N \right)^{k}}{ k! }  + \right . \nonumber \\
       & \phantom{======} \left . \sum\limits_{m=1}^{k-1} \left(-1\right)^{k-m} \left[ \begin{array}{c}
            k\\
            m 
       \end{array} \right] \frac{\left( N r_N \right)^{m}}{ k! }  \right] \nonumber \\
       & = \sum\limits_{k=1}^{N r_N} \theta_k \left[ \frac{\left( N r_N \right)^{k}}{ k! } \right] + \sum\limits_{k=1}^{N r_N} \theta_k \left[ \sum\limits_{m=1}^{k-1} \left(-1\right)^{k-m} \left[ \begin{array}{c}
            k\\
            m 
       \end{array} \right] \frac{\left( N r_N \right)^{m}}{ k! } \right],
      \label{eq_alternating_shrinking_prior_polynomial_S1_01}
\end{align}
where the later was obtained using the identities Eqs.~\eqref{eq_Stirling_first_kind_initial_cond_01} and \eqref{eq_Stirling_first_kind_identity_01}. The first term in the last right-hand side of Eq.~\eqref{eq_alternating_shrinking_prior_polynomial_S1_01} can be expanded as
\begin{align}
  \sum\limits_{k=1}^{N r_N} \theta_k \left[ \frac{\left( N r_N \right)^{k}}{ k! } \right] &= 
   \sum\limits_{l=1}^{N} \left(-1\right)^{l} \mathpzc{f} \frac{ C_l}{N^{l}} \left( N r_N \right)  \nonumber \\
   &\;\;\;\;\; +  \sum\limits_{l=2}^{N}\left(-1\right)^{l} \mathpzc{f} \frac{C_l}{ N^{l} } \sum\limits_{\substack{
                  k_{1} + k_{2} = l 
                  }}  \left( \begin{array}{c}
        l\\
        k_1, k_2 
   \end{array} \right) \frac{ \left( N r_N \right)^{2} }{ 2! } \nonumber \\
   &\;\;\;\;\; + \hdots \nonumber \\
   & \;\;\;\;\; +  \sum\limits_{l=N r_N} ^{N} \left( -1 \right)^{l} \mathpzc{f} \frac{C_l}{N^{l}} \sum\limits_{\substack{
                  k_{1} + \hdots + k_{N r_N} = l 
                  }}  \left( \begin{array}{c}
        l\\
        k_1, \hdots, k_{N r_N} 
   \end{array} \right) \frac{\left( N r_N \right)^{N r_N }}{ \left( N r_N \right)! }   \nonumber \\
   &=  \left(-1\right)^{1} \mathpzc{f} \frac{ C_1}{N^{1}}  N r_N + \sum\limits_{l=2}^{N} \left(-1\right)^{l} \mathpzc{f} \frac{ C_l}{N^{l}} N r_N   \nonumber \\
   & \;\;\;\;\; + \rule{0cm}{1cm} \left(-1\right)^{2} \mathpzc{f} \frac{C_2 2! }{N^{2}} \frac{N^2 \left(r_N\right)^2 }{2!}  
   + \sum\limits_{l=3}^{N}\left(-1\right)^{l} \mathpzc{f} \frac{C_l}{ N^{l} } \sum\limits_{\substack{
                  k_{1} + k_{2} = l
                  }}  \left( \begin{array}{c}
        l\\
        k_1, k_2 
   \end{array} \right) \frac{ \left( N r_N \right)^{2} }{ 2! }   \nonumber \\
   & \;\;\;\;\; + \hdots \nonumber \\
   & \;\;\;\;\; + \rule{0cm}{1cm}  \left(-1\right)^{N r_N} \mathpzc{f} \frac{C_{N r_N} \left( N r_N \right)! }{ N^{N r_N} } \frac{N^{N r_N} r^{N r_N} }{ \left( N r_N \right)! }   \nonumber \\
   & \;\;\;\;\; \;\;\;\;\;\;\;\;   +\sum\limits_{l=N r_N + 1} ^{N} \left( -1 \right)^{l} \mathpzc{f} \frac{C_l}{N^{l}} \sum\limits_{\substack{
                  k_{1} + \hdots + k_{N r_N} = l 
                  }}  \left( \begin{array}{c}
        l\\
        k_1, \hdots, k_{N r_N} 
   \end{array} \right) \frac{ N ^{N r_N } \left(r_N\right)^{N r_N } }{ \left( N r_N \right)! }   \nonumber \\
   &= \sum\limits_{j=1}^{N r_N} \left( -1 \right)^{j} \mathpzc{f} \frac{C_j}{N^{j}} N^{j} \left(r_N\right)^{j} + \mathcal{O}\left( \frac{1}{N} \right) \nonumber \\
   &= \sum\limits_{j=1}^{N r_N} \left( -1 \right)^{j} \mathpzc{f} C_j \left( r_N \right)^{j} + \mathcal{O}\left( \frac{1}{N} \right) \nonumber \\
   &= - \mathpzc{f} \sum\limits_{j=1}^{N r_N} \left( -1 \right)^{j+1}  C_j \left( r_N \right)^{j} + \mathcal{O}\left( \frac{1}{N} \right) .
      \label{eq_alternating_shrinking_prior_polynomial_S1_01_first_term_01}
\end{align}
On the other hand, we can obtain the order for the second term in the last right-hand side of Eq.~\eqref{eq_alternating_shrinking_prior_polynomial_S1_01} as follows
\begin{align}
  & \mathcal{O}\left( \sum\limits_{k=1}^{N r_N} \theta_k \left[ \sum\limits_{m=1}^{k-1} \left(-1\right)^{k-m} \left[ \begin{array}{c}
       k\\
       m 
  \end{array} \right] \frac{N^{m} \left(r_N\right)^{m} }{ k! }  \right]  \right) \nonumber \\
  \;\; &= 
  \mathcal{O}\left( \theta_1 + \theta_2  \left(-1 \right)\left[ \begin{array}{c}
       2\\
       1 
  \end{array} \right] \frac{N^{1} \left( r_N\right)^{1} }{ 2! }  + \right . \nonumber \\
  \;\; & \;\;\;\;\; \;\;\;\;\; \left . + \hdots + \theta_{N r_N}  \sum\limits_{m=1}^{N r_N - 1} \left(-1\right)^{N r_N - m} \left[ \begin{array}{c}
       N r_N \\
       m
  \end{array} \right] \frac{ N^{m} \left( r_N \right)^{m} }{ \left( N r_N \right)! }  \right),
      \label{eq_alternating_shrinking_prior_polynomial_S1_01_second_term_order_01}
\end{align}
where the highest order for each $\theta_k$ corresponds to its first term (since its terms grow inversely in $N$). The order is then
\begin{align}
  &   \mathcal{O}\left( \sum\limits_{k=1}^{N r_N} \theta_k \left[ \sum\limits_{m=1}^{k-1} \left(-1\right)^{k-m} \left[ \begin{array}{c}
       k\\
       m 
  \end{array} \right] \frac{N^{m} \left(r_N\right)^{m} }{ k! }  \right]  \right) \nonumber \\
  \;\; &= \mathcal{O}\left( \frac{1}{N} + \frac{1}{N^2} 2!  \frac{N}{2!}  + \hdots + \frac{1}{N^{N r_N}} \left( N r_N\right)!  \frac{N^{N r_N - 1}}{ \left( N r_N \right)! }  \right) \nonumber \\
  \;\; &= \mathcal{O}\left( \frac{1}{N} \right).
      \label{eq_alternating_shrinking_prior_polynomial_S1_01_second_term_order_02}
  \end{align}
  
  Considering Eqs.~\eqref{eq_alternating_shrinking_prior_polynomial_S1_01}, \eqref{eq_alternating_shrinking_prior_polynomial_S1_01_first_term_01} and \eqref{eq_alternating_shrinking_prior_polynomial_S1_01_second_term_order_02}, the polynomial term can be written as in Eq.~\eqref{eq_alternating_shrinking_r_pmf_polynomial} and our PMF becomes
  
  \begin{align}
  \mathcal{P}\left( r_N | \boldsymbol{\theta}_N \right) &=   \frac{1}{Z} \exp\left[ -\mathpzc{f} \sum\limits_{j=1}^{N r_N} \left(-1\right)^{j+1} C_j \left(r_N\right)^{j} + \mathcal{O}\left( \frac{1}{N} \right) \right] .
      \label{eq_alternating_shrinking_prior_discrete_order_form}
\end{align}
  
\newpage
\subsection{Widespread probability density function limit}
  \label{APPENDIX_ALTERNATING_SHRINKING_MODEL_LIMIT}
Using Eq.~\eqref{eq_alternating_shrinking_prior_discrete_order_form} in the limit as $N \to \infty$ our PMF becomes
\begin{align}
\lim_{N \rightarrow \infty}
   \mathcal{P}\left( r_N | \boldsymbol{\theta}_N \right)  &= \frac{1}{Z} \exp\left[  - \mathpzc{f} \sum\limits_{j=1}^{\infty} \left(-1\right)^{j+1} C_j r^{j}  \right] dr \nonumber \\
   &= p\left(r | \boldsymbol{\lambda}\right) dr,
   \label{eq_alternating_prior_r_convergence_limit_02}
\end{align}
\noindent where $\boldsymbol{\lambda} = \left\{ \mathpzc{f}, \left\{ C_j \right\}_{j \in \mathbb{N}^{+}} \right\}$. The specific form of the distribution Eq.~\eqref{eq_alternating_prior_r_convergence_limit_02} and its convergence depend on the $C_j$ constants. By the Leibniz criterion for alternating series, the following series in our alternating density
\begin{align}
-\mathpzc{f} \sum_{j=1}^{\infty} \left( -1 \right)^{j+1}  C_j r ^{j} =&
\mathpzc{f} \sum_{j=1}^{\infty} \left(-1\right)^{j} C_j r^j
    \label{eq_alternating_series_theorem_01}
\end{align}
converge if the following conditions hold:
\begin{align}
    |C_1 r| > |C_2 r^2| > |C_3 r^3| > \hdots , \nonumber \\
\lim\limits_{j \to \infty} \left[ C_j r^j \right] = 0.
    \label{eq_alternating_series_theorem_02}
\end{align}
The conditions in Eq.~\eqref{eq_alternating_series_theorem_02} hold for the following cases, but are not restricted to them. \\

\noindent \underline{Polylogarithmic exponential density} \\

\noindent If we define $C_j = \frac{1}{j^{m}} \;\; \forall j\;\;$ then the PDF in Eq.~\eqref{eq_alternating_prior_r_convergence_limit_02} is as in Eq.~\eqref{eq_alternating_shrinking_r_pdf_polylogarithm}, where $\text{Li}_m[ \cdot ]$ is the polylogarithm function of order $m=1,2,3,\hdots$, defined as 
  \begin{equation}
      \label{eq_polylogarithm_series}
      \text{Li}_m\left[ x \right] = \sum_{j=1}^{\infty} \frac{x^{j}}{ j^{m} }
  \end{equation}
  When $m=1$, the series converges to the natural logarithm, i.e., Eq.~\eqref{eq_polylogarithm_to_log}. \\
  
  \noindent \underline{Shifted-geometric exponential density} \\
  
\noindent If we instead define $C_j = \left( \tau \right)^{j}$, with $\;0<\tau < 1, \; \forall j\;\;$ so that $\tau r < 1$, then the distribution in Eq.~\eqref{eq_alternating_prior_r_convergence_limit_02} is obtained as Eq.~\eqref{eq_alternating_shrinking_r_pdf_shifted_geometric} because the exponent is computed as follows:
\begin{align}
   \lim\limits_{n \to \infty} \left[ -\mathpzc{f} \sum_{j=1}^{n} \left(-1\right)^{j+1} \left(\tau r \right)^{j}\right] 
   = & \mathpzc{f} \lim\limits_{n \to \infty} \left[ \sum_{j=1}^{n} \left( -\tau r \right)^{j} \right] \nonumber \\
   = & \mathpzc{f} \lim\limits_{n \to \infty} \left[  \frac{ - \left( \tau r + \left(-1\right)^{n+1} \left( \tau r \right)^{n+1} \right)  }{ 1 + \tau r }  \right] \nonumber \\
   = & -\mathpzc{f} \frac{1}{1 + \frac{1}{\tau r }} \nonumber \\
   = & \mathpzc{f} \left( \frac{1}{1 + \tau r}  - 1 \right).
    \label{eq_shifted_geometric_finite_01}
\end{align}

\newpage
\subsection{Non-positive property for the arguments of the exponential function} \label{APPENDIX_NONPOSITIVE_PROPERTY}

\noindent\underline{Polylogarithmic exponential density} \\

\noindent For $m=1$ we have the function
\begin{equation}
    \label{eq_f_polylogarithm_m_1}
    \mathpzc{f} \text{Li}_{1}\left[ - r \right] = - \mathpzc{f} \log\left( 1 + r \right) \leq 0 \phantom{==} \forall r \in [0,1]
\end{equation}

\noindent because $\log\left(x\right)$ is a strictly increasing function for $x > 0$ and correspondingly $-\mathpzc{f}\log\left(x\right)$ ($\mathpzc{f}>0$) is a strictly decreasing function for $x > 0$, where $-\mathpzc{f} \log\left( 1 + r \right) = 0$ for $r=0$ and $-\mathpzc{f}\log\left( 1 + r \right) = - \mathpzc{f} \log  2 $ for $r=1$. Then Eq.~\eqref{eq_f_polylogarithm_m_1} holds for $r \in [0,1]$ and $-\mathpzc{f} \log\left( 1 + r\right)$ has a maximum at $r=0$.

For $m \in \left\{ 2, 3, 4, \hdots \right\}$, the following alternating series (shown on the right-hand side in the parentheses)
\begin{equation}
    \label{eq_f_polylogarithm_m}
    \mathpzc{f} \text{Li}_{m}\left[ - r \right] = \mathpzc{f} \left(\sum\limits_{j=1}^{\infty} \frac{\left(-r\right)^{j} }{ j^{m}} \right) 
\end{equation}
converge to a finite limit $L$  by the Leibniz criterion. In this case, $L = \text{Li}_{m}\left[-r\right]$, and the convergence proof of the Leibniz criterion for alternating series guarantees the following bounds for any $n>0$
\begin{equation}
    \label{eq_convergent_alternating_series_finite_bounds}
    S_{2n} \leq L \leq S_{2n + 1},
\end{equation}
\noindent where $S_{2n}$ denotes the partial sum of an even number of terms and $S_{2n+1}$ denotes the partial sum of an odd number of terms in the alternating series. Then we have the partial sum with an even number of terms ($2n=2$)
\begin{align}
 S_2 &= \sum\limits_{j=1}^{2} \frac{\left(-r\right)^{j}}{j^{m}} \nonumber \\
 &= -r + \frac{r^{2}}{2^{m}} \leq  0 \phantom{==} \forall r \in [0,1] ,
    \label{eq_partial_sum_polylogarithm_s2}
\end{align}

\noindent where the last inequality holds because the coefficient of the first negative term is such that $r \geq r^2 \geq \frac{r^2}{2^{m}} \phantom{==}\forall r \in [0,1]$. Similarly, we have the following partial sum with an odd number of terms ($2n+1=3$)
\begin{align}
    S_3 &= \sum\limits_{j=1}^{3} \frac{ \left(-r\right)^{j} }{j^{m}} \nonumber \\
    &= S_2 - \frac{r^3}{3^{m}} \leq 0 \phantom{==} \forall r \in [0,1],
    \label{eq_partial_sum_polylogarithm_s3}
\end{align}
\noindent where the last inequality holds because $S_2 \leq 0$ by Eq.~\eqref{eq_partial_sum_polylogarithm_s2} and since $-\frac{r^3}{3^{m}} \leq 0 $ for $r \in [0,1]$. Using inequalities Eqs.~\eqref{eq_convergent_alternating_series_finite_bounds} and \eqref{eq_partial_sum_polylogarithm_s3} we have for $m \in \left\{ 2, 3, 4, \hdots \right\}$ that
\begin{equation}
    \label{eq_polylogarithm_m_geq_2_nonpositive}
    \text{Li}_m\left[-r\right] \leq 0 \phantom{==} \forall r \in [0,1]
\end{equation}
\noindent and also for $\mathpzc{f}>0$ we have $ \mathpzc{f} \text{Li}_m\left[-r\right] \leq 0$ . Consequently by Eqs.~\eqref{eq_polylogarithm_m_geq_2_nonpositive} and \eqref{eq_f_polylogarithm_m_1} the function in Eq.~\eqref{eq_f_polylogarithm_m} is non-positive for $r \in [0,1]$ with $m \in \left\{1, 2, 3, \hdots \right\}$. \\

\noindent \underline{Shifted-geometric exponential density} \\

\noindent For the shifted-geometric function, we have the following inequality
\begin{equation}
    \label{eq_shifted_geometric_nonpositive}
    \mathpzc{f} \left( \frac{1}{1 + \tau r} - 1 \right) \leq 0 \phantom{==} \forall r \in [0,1]
\end{equation}

\noindent because for $\tau > 0$ it holds that $\frac{1}{1 + \tau r} \leq 1$ for $r \in [0,1]$. Equality in Eq.~\eqref{eq_shifted_geometric_nonpositive} is reached only when $r=0$ and hence $\mathpzc{f}\left( \frac{1}{1+\tau r} -1 \right)$ is non-positive with a maximum at $r=0$.

\newpage
\subsection{Decreasing property for the arguments of the exponential function} \label{APPENDIX_DECREASING_PROPERTY}

\noindent\underline{Polylogarithmic exponential density} \\
We now prove the decreasing property for the polylogarithmic function in the polylogarithmic exponential density in Eq.~\eqref{eq_alternating_shrinking_r_pdf_polylogarithm}. For $m=1$, the derivative of the function is
\begin{align}
\frac{d}{dr}
\left[ \mathpzc{f} \text{Li}_1 \left[ -r \right] \right]  &= \frac{d}{dr}\left[ - \mathpzc{f} \log\left(1 + r\right)  \right] \nonumber \\
&= -\frac{\mathpzc{f}}{ 1 + r } < 0 \phantom{==}\forall r \in [0,1],
    \label{eq_polylogarithmic_form_m_1_derivative}
\end{align}
\noindent since $\mathpzc{f} > 0$. On the other hand, for $m \in \left\{ 2, 3, 4, \hdots  \right\}$ the derivative is
\begin{align}
\frac{d}{dr} \left[ \mathpzc{f} \text{Li}_m \left[ -r \right] \right]  &= \mathpzc{f} \frac{d}{dr} \left[  \text{Li}_m \left[ -r \right] \right]  \nonumber \\
&=\mathpzc{f} \left[ \frac{1}{r} \text{Li}_{m-1} \left[-r\right] \right] \nonumber \\
&= \mathpzc{f} \left[ \sum\limits_{j=1}^{\infty}\left(-1\right)^{j} \left( \frac{ r^{j-1}}{j^{m-1}}\right)\right],
    \label{eq_polylogarithmic_form_m_geq_2_derivative}
\end{align}
\noindent where $\frac{ r^{j-1}}{j^{m-1}} > \frac{r^{j}}{\left(j+1\right)^{m-1}}$ and $\lim_{j \to \infty}\left[ \frac{r^{j-1}}{j^{m-1}} \right] = 0$. Then, by the Leibniz criterion, the alternating series of $\frac{d}{dr}\left[  \text{Li}_m \left[ -r \right] \right] $ converge to a finite limit $L $. Using the bounds in Eq.~\eqref{eq_convergent_alternating_series_finite_bounds} for $n=1$ it holds that
\begin{equation}
    \label{eq_convergent_alternating_series_finite_bounds_n_1}
    S_{2} \leq L \leq S_{3}.
\end{equation}
The partial sum for the even number of terms is
\begin{align}
S_2 &= -\frac{r^{0}}{1^{m-1}} + \frac{r^{1}}{2^{m-1}} \nonumber \\
 &= -1 + \frac{r}{2^{m-1}} < 0 \phantom{==}\forall r\in [0,1] 
    \label{eq_polylogarithm_m_S2}
\end{align}
\noindent because $S_2$ increases linearly in $r$ but is always negative since we have at the extremes $S_2 = -1$ when $r=0$ and $S_2 = -1 + \frac{1}{2^{m-1}} < 0$  when $r=1$ for $m \in \left\{ 2,3,4, \hdots \right\}$. Similarly, we have for the partial sum with an odd number of terms
\begin{align}
S_3 &= -1 + \frac{r}{2^{m-1}} - \frac{r^{2}}{3^{m-1}} \nonumber \\
&= S_2 - \frac{r^2}{3^{m-1}} < 0
    \label{eq_polylogarithm_m_S3}
\end{align}
\noindent because $S_2 < 0$ and $-\frac{r^2}{3^{m-1}} < 0 \;\;\forall r \in [0,1], m \in \left\{ 2, 3, 4, \hdots \right\}$. Consequently, we have that
\begin{align}
    \label{eq_polylogarithm_m_ge_2_derivative}
    \frac{d}{dr}\left[ \mathpzc{f} \text{Li}_m \left[-r\right] \right] &=  \mathpzc{f} \frac{d}{dr}\left[\text{Li}_m \left[-r\right] \right]  \nonumber \\
    &= \mathpzc{f} L < 0 \phantom{==} \forall r  \in [0,1],
\end{align}
\noindent with $m \in \left\{ 2, 3, 4, \hdots \right\}$. By the inequalities, Eqs.~\eqref{eq_polylogarithmic_form_m_1_derivative} and \eqref{eq_polylogarithm_m_ge_2_derivative}, it holds that
\begin{align}
    \label{eq_polylogarithmic_form_m_derivative}
    \frac{d}{dr}\left[ \mathpzc{f} \text{Li}_m \left[-r\right] \right] < 0  & \phantom{==}\forall r \in [0,1], \\
    & \phantom{==}  m \in \left \{ 1, 2, 3, \hdots \right\}, \nonumber 
\end{align}
\noindent and hence the function $\mathpzc{f} \text{Li}_m \left[-r\right]$ is strictly decreasing for $r\in [0,1]$ and $m \in \left\{1, 2, 3, \hdots \right\}$. \\

\noindent \underline{Shifted-geometric exponential density} \\

\noindent Next, we prove the decreasing property for the shifted-geometric function in the shifted-geometric exponential density in Eq.~\eqref{eq_alternating_shrinking_r_pdf_shifted_geometric}. The derivative of the function is

\begin{align}
    \frac{d}{dr}\left[ \mathpzc{f} \left( \frac{1}{1 + \tau r} - 1 \right) \right] &=
    \mathpzc{f} \frac{d}{dr}\left[ \left(\frac{1}{1 + \tau r} - 1 \right) \right] \nonumber \\
    &= \mathpzc{f} \frac{d}{dr}\left[ \left( 1 + \tau r \right)^{-1} \right] \nonumber \\
    &= -\frac{\mathpzc{f} \tau }{\left( 1 + \tau r \right)^{2}} < 0 \phantom{==} \forall r \in[0,1] 
\end{align}

\noindent because $\mathpzc{f} > 0$, $\tau > 0$ and $\left( 1 + \tau r \right)^2 > 0$. It follows that the function $\mathpzc{f} \left( \frac{1}{1 + \tau r} - 1 \right)$ is strictly decreasing for $r\in [0,1]$ and $0 < \tau <  1$.

\newpage
\subsection{Mean and variance} \label{APPENDIX_MEAN_AND_VARIANCE}
  
The mean of the polylogarithmic exponential distribution for $m=1$ is given by
\begin{align}
    \label{eq_mean_polylogarithmic_m1}
    \mu_{R} &= \frac{1}{Z} \int_{0}^{1} \left(1 + r \right)^{-\mathpzc{f}} r dr \nonumber \\
        &= \left\{ \begin{array}{ll}
        \frac{1}{ 1 - 2^{-\mathpzc{f}+1} } \left[ \frac{ 1 - 2^{-\mathpzc{f}+2} }{\mathpzc{f}-2} - 2^{-\mathpzc{f} + 1} \right] &  \text{for} \; \mathpzc{f} \neq 1, 2\\
        \; & \; \\
        \frac{1 - \log2}{\log2} & \text{for} \; \mathpzc{f} = 1, \\
        \; & \; \\
        2 \log 2 - 1 & \text{for} \; \mathpzc{f} = 2,
    \end{array} \right .
\end{align}
and the variance is 
\begin{align}
    \sigma^2 _{R} &= \frac{1}{Z} \int_{0}^{1} \left(1 + r \right)^{-\mathpzc{f}} r^2 dr - \mu^2 _{R} \nonumber \\
    &= \left\{ \begin{array}{ll}
         \frac{1}{ 1 - 2^{-\mathpzc{f}+1} } \left[ \frac{2 \left( 1 - 2^{-\mathpzc{f} + 3}\right) }{\left( \mathpzc{f}-3\right)\left(\mathpzc{f}-2\right)} - 2^{-\mathpzc{f} + 1} \left( 1 + \frac{2^2}{\mathpzc{f} - 2 } \right) \right] - \mu_{R} ^2 &  \text{for} \; \mathpzc{f} \neq 1, 2, 3\\
        \; & \; \\
        \frac{\log2 - \frac{1}{2}}{\log2} - \left[\frac{1 - \log2}{\log 2 }\right]^2 & \text{for} \; \mathpzc{f} = 1, \\
        \; & \; \\
        3-(2 \log 2 -1)^2 - 4 \log 2 & \text{for} \; \mathpzc{f} = 2, \\
        \; & \; \\
        \frac{8}{3}\log 2 - \frac{16}{9} & \text{for} \; \mathpzc{f} = 3,
\end{array} \right .
    \label{eq_variance_polylogarithmic_m1}
\end{align}

The mean and variance of the shifted-geometric exponential distribution are given by
\begin{align}
    \mu_{R} &= \frac{1}{Z}\int_{0}^{1} \exp\left[ \mathpzc{f} \left( \frac{1}{1 + \tau r} - 1 \right) \right] r dr ,
    \label{eq_mean_shifted_geometric}
\end{align}
and
\begin{align}
    \sigma^2 _{R} &= \frac{1}{Z} \int_{0}^{1} \exp\left[ \mathpzc{f}\left( \frac{1}{1 + \tau r} - 1\right) \right] r^2 dr - \mu^2 _{R},
    \label{eq_variance_shifted_geometric}
\end{align}
respectively. We provide the explicit expressions for the integrals in these equations as  Eqs.~\eqref{eq_mean_shifted_geometric_02} and \eqref{eq_variance_shifted_geometric_02} in the 
\ref{APPENDIX_SHIFTED_GEOMETRIC_DISTRIBUTION_FUNCTION}. The normalization constant $Z$ is given in  Eq.~\eqref{eq_F_r_shifted_geometric_form_proof_02}.

\newpage
\subsection{Integrals for the shifted-geometric exponential distribution} \label{APPENDIX_SHIFTED_GEOMETRIC_DISTRIBUTION_FUNCTION}
  
We now verify the definite integral result from Eq.~\eqref{eq_F_r_shifted_geometric_form}. Since $F\left( u | \mathpzc{f}, \tau  \right)$ denotes the distribution function evaluated at $u$ we have that
\begin{equation}
  \label{eq_F_r_shifted_geometric_form_proof_01}
  F\left(1 | \mathpzc{f}, \tau \right) = \frac{1}{Z}\int_{0}^{1} \exp\left[ \mathpzc{f} \left( \frac{1}{1 + \tau r} - 1 \right)\right] dr = 1.
\end{equation}
From Eq.~\eqref{eq_F_r_shifted_geometric_form_proof_01}, we obtain the normalization constant
\begin{align}
  Z &= \int_{0}^{1} \exp\left[ \mathpzc{f}\left( \frac{1}{1 + \tau r} - 1 \right) \right] dr  \nonumber \\
  &= \left( \frac{1 + \tau r }{\tau} \exp\left[ \mathpzc{f}\left( \frac{1}{1+\tau r} - 1 \right) \right] - \frac{\mathpzc{f} e^{-\mathpzc{f}} }{\tau} \text{Ei}\left( \frac{\mathpzc{f}}{1 + \tau r} \right) \right) \Bigg |_0 ^{1}.
  \label{eq_F_r_shifted_geometric_form_proof_02}
\end{align}

The last right hand side of Eq.~\eqref{eq_F_r_shifted_geometric_form_proof_02} contains the indefinite integral of $\exp\left[ \mathpzc{f}\left( \frac{1}{1 + \tau r} - 1\right) \right] dr $ to be evaluated from $0$ to $1$. We now obtain the derivative of such an indefinite integral. For this we make use of the series representation of the special exponential integral function $\text{Ei}\left( x \right)$ when its argument $x$ is real, defined in Eq.~\eqref{eq_exponential_integral_function_real_series_representation}. The derivative is then
\begin{align}
   & \frac{d}{dr}\left[ \frac{ 1 + \tau r }{\tau} \exp\left[ \mathpzc{f}\left( \frac{1}{1+\tau r} - 1 \right) \right] - \frac{\mathpzc{f} e^{-\mathpzc{f}} }{\tau} \text{Ei}\left( \frac{\mathpzc{f}}{1 + \tau r} \right) \right]  \nonumber \\
   &= \frac{d}{dr}\left[ \frac{ 1 + \tau r }{\tau} \exp\left[ \mathpzc{f}\left( \frac{1}{1+\tau r} - 1 \right) \right] - \frac{\mathpzc{f} e^{-\mathpzc{f}} }{\tau} \left( \gamma + \log\left(\frac{\mathpzc{f}}{1 + \tau r}\right) + \sum\limits_{k=1}^{\infty} \frac{\left(\frac{\mathpzc{f}}{1 + \tau r}\right)^{k}}{k \; k!} \right) \right]   \nonumber \\
   &= \left( -\frac{\mathpzc{f}}{\left( 1 + \tau r \right)^2} + 1 - \tau r \frac{\mathpzc{f}}{ \left( 1 + \tau r \right)^2 } \right)\exp\left[ \mathpzc{f}\left( \frac{1}{1+\tau r} - 1 \right) \right]  + \phantom{================}  \nonumber \\
   & \phantom{==} \left( -\frac{\mathpzc{f} e^{-\mathpzc{f}} }{\tau} \left[ -\frac{\tau \mathpzc{f}^{0} \left( 1+ \tau r \right)^{-1}}{0!} - \frac{\tau \mathpzc{f}^{1} \left(1 + \tau r\right)^{-2}}{ 1! } - \frac{\tau \mathpzc{f}^{2} \left(1 + \tau r\right)^{-3}}{ 2! } -  \hdots   \right] \right)  \nonumber \\
   &= \left( 1 - \left( \frac{\mathpzc{f} }{\left(1 + \tau r\right)^{2}} \left(1 + \tau r\right) \right) \right)\exp\left[ \mathpzc{f}\left( \frac{1}{1+\tau r} - 1 \right) \right]  +   \nonumber \\
   & \phantom{================} \left( \frac{\mathpzc{f}}{1 + \tau r} \right) e^{-\mathpzc{f}}  \sum\limits_{k=1}^{\infty}\frac{ \left( \frac{\mathpzc{f}}{1 + \tau r} \right)^{k-1} }{\left(k-1\right)!}  \nonumber \\
   &= \left( 1 - \frac{\mathpzc{f}}{1 + \tau r} \right)\exp\left[ \mathpzc{f}\left( \frac{1}{1+\tau r} - 1 \right) \right]  + \left( \frac{\mathpzc{f}}{1 + \tau r} \right) e^{-\mathpzc{f}}  \sum\limits_{k=0}^{\infty}\frac{ \left( \frac{\mathpzc{f}}{1 + \tau r} \right)^{k} }{\left(k\right)!}  \nonumber \\
   &= \left( 1 - \frac{\mathpzc{f}}{1 + \tau r} \right)\exp\left[ \mathpzc{f}\left( \frac{1}{1+\tau r} - 1 \right) \right]  + \left( \frac{\mathpzc{f}}{1 + \tau r} \right) e^{-\mathpzc{f}}  \exp\left[  \frac{\mathpzc{f}}{1 + \tau r }  \right]  \nonumber \\
   &= \exp\left[ \mathpzc{f}\left( \frac{1}{1+\tau r} - 1 \right) \right].
    \label{eq_F_r_shifted_geometric_form_proof_03}
\end{align}

Hence by Eq.~\eqref{eq_F_r_shifted_geometric_form_proof_03} we have that
\begin{align}
      & \int_{0}^{u} \exp\left[ \mathpzc{f}\left( \frac{1}{1 + \tau r} - 1 \right) \right] dr  \nonumber \\
      &= \left( \frac{ 1 + \tau r }{\tau} \exp\left[ \mathpzc{f}\left( \frac{1}{1+\tau r} - 1 \right) \right] - \frac{\mathpzc{f} e^{-\mathpzc{f}} }{\tau} \text{Ei}\left( \frac{\mathpzc{f}}{1 + \tau r} \right) \right) \Bigg |_0 ^{u},
      \label{eq_F_r_shifted_geometric_form_proof_04}
\end{align}
which is the integral used to obtain $Z$ in Eq.~\eqref{eq_F_r_shifted_geometric_form_proof_02} and which also defines the distribution function in Eq.~\eqref{eq_F_r_shifted_geometric_form}. \\

\noindent \underline{Integrals to compute the mean and the variance } \\

\noindent The integral for the mean value in Eq.~\eqref{eq_mean_shifted_geometric} is
\begin{align}
    \mu_{R} &= \frac{1}{2 Z } \left[\frac{ \tau r - 1 }{\tau } \int \exp\left[ \mathpzc{f}\left( \frac{1}{1 + \tau r} - 1 \right) \right] dr  + \frac{\mathpzc{f} e^{-\mathpzc{f}} }{\tau }  \int \text{Ei}\left( \frac{\mathpzc{f}}{1 + \tau r} \right) dr  \right] \Bigg | _0 ^{1},
    \label{eq_mean_shifted_geometric_02}
\end{align}
\noindent where the normalization constant $Z$ is given in Eq.~\eqref{eq_F_r_shifted_geometric_form_proof_02} and the improper integral over the exponential shifted-geometric function is given in Eq.~\eqref{eq_F_r_shifted_geometric_form_proof_04} (without evaluating over the limits). The improper integral over the special exponential integral function is 
\begin{align}
    \int \text{Ei}\left( \frac{\mathpzc{f}}{1 + \tau r} \right) dr &=
    \gamma r  + \frac{1 +  \tau r }{\tau} \log\left( \frac{f}{ 1 + \tau r  } \right) + r \nonumber\\
    &\phantom{=} + \frac{f}{\tau} \log\left(1 + \tau r\right) -\sum\limits_{k=2}^{\infty} \frac{\mathpzc{f}^{k}  \left(1 + \tau r\right)^{-\left(k-1\right)} }{k k! \tau \left(k-1\right)} + C_1,
    \label{eq_special_exp_int_function_integral}
\end{align}
\noindent with $C_1$ an integration constant. \\

The integral for the variance in Eq.~\eqref{eq_variance_shifted_geometric} is
\begin{align}
    \sigma^2 _{R} &= \frac{1}{3Z}\left[  \frac{1 + \tau r^2}{\tau}  \int \exp\left[ \mathpzc{f}\left( \frac{1}{1 + \tau r} - 1\right) \right] dr   \right . \nonumber \\
    &\phantom{=====}+2\frac{\mathpzc{f}e^{-\mathpzc{f} } }{\tau} \int \text{Ei} \left( \frac{\mathpzc{f}}{1 + \tau r }\right) r dr \nonumber \\
    &\phantom{=====} -\frac{1}{\tau} \int \exp\left[ \mathpzc{f}\left( \frac{1}{1 + \tau r } - 1 \right) \right] dr \nonumber \\
    &\phantom{=====} \left . -\frac{2}{\tau} \int \exp\left[  \mathpzc{f} \left( \frac{1}{1 + \tau r } - 1 \right)  \right] r dr  \;\;\right] \Bigg | _0^{1} - \mu^2_{R},
    \label{eq_variance_shifted_geometric_02}
\end{align}
\noindent where the normalization constant $Z$ is given in Eq.~\eqref{eq_F_r_shifted_geometric_form_proof_02}, the improper integrals over the exponential shifted-geometric function are given by Eq.~\eqref{eq_F_r_shifted_geometric_form_proof_04} (without evaluating over the limits). Also, the improper integral over the first moment for the shifted-geometric distribution is given in Eq.~\eqref{eq_mean_shifted_geometric_02} (without evaluating the limits). Further, the improper integral of $r$ multiplied by the special exponential integral function is 
\begin{align}
   \int \text{Ei}\left( \frac{\mathpzc{f}}{1 + \tau r} \right) r  dr &= r \int \text{Ei}\left(\frac{\mathpzc{f}}{1 + \tau r }\right)dr  - \int  \int \text{Ei}\left( \frac{\mathpzc{f}}{1 + \tau r} \right)  dr'  dr.
    \label{eq_r_time_special_exp_int_function_integral}
\end{align}
\noindent In turn, the iterated integral in Eq.~\eqref{eq_r_time_special_exp_int_function_integral} is
\begin{align}
\int \int \text{Ei}\left( \frac{\mathpzc{f}}{1 + \tau r} \right)  dr' dr &= \frac{1}{\tau} \int \log\left( \frac{\mathpzc{f}}{ 1 + \tau r } \right) dr + \int r \log\left( \frac{ \mathpzc{f} }{ 1 + \tau r } \right) dr \nonumber \\
& \phantom{=} +\frac{\mathpzc{f}}{\tau} \int \log\left( 1 + \tau r \right) dr \nonumber \\
& \phantom{=} + \frac{1}{2}\left( r^2 + \gamma r^2 \right) - \frac{\mathpzc{f}^2}{ 4 \tau ^2 } \log\left( 1 + \tau r \right) \nonumber \\
& \phantom{=} + \sum_{k=3}^{\infty} \frac{ \mathpzc{f}^{k} \left( 1 + \tau r \right)^{-\left(k-2\right)} }{ \tau^2 k \left( k-1 \right) \left( k - 2 \right) k!  },
\label{eq_special_exp_int_function_iterated_second_integral}
\end{align}
\noindent with the integrals over the logarithmic functions
\begin{align}
\frac{1}{\tau}\int \log\left( \frac{\mathpzc{f}}{1 + \tau r} \right) dr &= \frac{1}{\tau }\left[  \frac{1 + \tau r}{\tau} \log\left( \frac{\mathpzc{f}}{1 + \tau r} \right)  + r \right] + C_2,
    \label{eq_logarithmic_function_integral}
\end{align}
\begin{align}
 \int \log\left( \frac{\mathpzc{f}}{1 + \tau r} \right) r dr &= \frac{1}{2}\left[ \frac{ \tau r - 1}{\tau} \int \log\left( \frac{\mathpzc{f}}{1 + \tau r }  \right) dr - \frac{1}{2} r^2 \right] + C_3,
    \label{eq_logarithmic_function_times_r_integral}
\end{align}
\noindent and
\begin{align}
 \frac{\mathpzc{f}}{\tau} \int \log\left( 1 + \tau r \right) dr = \frac{\mathpzc{f}}{\tau}\left[ r \log\left( 1 + \tau r \right) -  r + \frac{1}{\tau} \log\left( 1 + \tau r \right)  \right] + C_4,
    \label{eq_logarithmic_function_neg_integral}
\end{align}
\noindent with $C_2, C_3$ and $C_4$ integration constants. \\

\newpage
\section{Experimental verification, neural interpretation, and statistical properties}

\subsection{Neural interpretation}
  \label{APPENDIX_NEURAL_INTERPRETATION}
The conditional probability that the $i$th neuron spikes given the rest of the neurons in the population is derived as follows

\begin{align}
    \mathcal{P}(x_i = 1 | \mathbf{x}_{\backslash i}, \boldsymbol{\omega} ) 
    &= \frac{ \mathcal{P}\left( x_i = 1 , \mathbf{x}_{\backslash  i} , \boldsymbol{\omega} \right) }{ \mathcal{P}\left( \mathbf{x}_{\backslash i} , \boldsymbol{\omega} \right) }  \nonumber \\
    &= \frac{ \mathcal{P}\left( x_i = 1, \mathbf{x}_{\backslash  i} , \boldsymbol{\omega} \right) }{ \mathcal{P}\left( x_i = 1, \mathbf{x}_{\backslash  i} , \boldsymbol{\omega} \right) + \mathcal{P}\left( x_i = 0, \mathbf{x}_{\backslash  i} , \boldsymbol{\omega} \right) } \nonumber \\
    &= \frac{1}{ 1 + \frac{ \mathcal{P}\left( x_i = 0, \mathbf{x}_{\backslash  i} , \boldsymbol{\omega} \right) }{ \mathcal{P}\left( x_i = 1, \mathbf{x}_{\backslash  i} , \boldsymbol{\omega} \right) } } \nonumber \\
    &= \frac{1}{1 + \exp \left( - \log
              \frac{  \mathcal{P}(x_i = 1, \mathbf{x}_{\backslash i},\boldsymbol{\omega}) }
              { \mathcal{P}(x_i = 0, \mathbf{x}_{\backslash i},\boldsymbol{\omega}) }
              \right) }.
    \label{eq_conditional_distrib_single_neuron_02}
\end{align}

The log ratio is obtained as
\begin{align}
   & \log\left( \frac{ \mathcal{P}\left(x_i = 1, \mathbf{x}_{\backslash i}, \boldsymbol{\omega} \right) }{  \mathcal{P}\left( x_i = 0, \mathbf{x}_{\backslash i} , \boldsymbol{\omega} \right) } \right) \nonumber \\
      &=  \log \left( \frac{h\left( 1 + \sum_{j\neq i} x_j \right)}{Z} \exp\left[ -\mathpzc{f} \sum_{j=1}^{N} (-1)^{j+1} C_j \left( \frac{1 + \sum_{k\neq i} x_k}{N} \right)^{j} \right]  \right)\nonumber \\
      & \phantom{=} - \log \left( \frac{h\left( 0 + \sum_{j\neq i} x_j \right)}{Z} \exp\left[ -\mathpzc{f} \sum_{j=1}^{N} (-1)^{j+1} C_j \left( \frac{0 + \sum_{k\neq i} x_k}{N} \right)^{j} \right]  \right)\nonumber \\
      &= \log\left( \frac{ h\left( 1 + \sum_{j\neq i} x_j \right) }{ h\left( \sum_{j\neq i} x_j \right) } \right) \nonumber \\
      & \phantom{=} + \log\left( \exp\left[ -\mathpzc{f} \sum_{j=1}^{N} (-1)^{j+1} C_j \left( \frac{1 + \sum_{k\neq i} x_k}{N} \right)^{j} \right] \right) \nonumber \\
      & \phantom{=} - \log\left( \exp\left[ -\mathpzc{f} \sum_{j=1}^{N} (-1)^{j+1} C_j \left( \frac{\sum_{k\neq i} x_k}{N} \right)^{j} \right] \right) \nonumber \\
      &= \log\left( \frac{ 1 + \sum_{j\neq i} x_j }{ N - \sum_{j\neq i} x_j } \right) \nonumber \\
      & \phantom{=}  -\mathpzc{f} \left( \sum_{j=1}^{N} \left( -1 \right)^{j+1} C_j \left( \frac{1 + \sum_{k \neq i} x_k }{ N } \right)^j   - \sum_{j=1}^{N} \left( -1 \right)^{j+1} C_j \left( \frac{ \sum_{k \neq i} x_k }{ N } \right)^j \right) \nonumber \\
   \; &= \widetilde{h}_{i}\left( \mathbf{x} \right) + \widetilde{Q}_{i} \left( \mathbf{x} ; \boldsymbol{\omega} \right),
       \label{eq_log_ratio_single_neuron_02}
\end{align}

\noindent where the simplified argument for the ratio between the base measure function values comes from the following, 
\begin{align}
\frac{ h\left( 1 + \sum_{j\neq i} x_j \right) }{ h\left( \sum_{j\neq i} x_j \right) } 
& = \frac{ \left( \begin{array}{c}
     N\\
     \sum_{j\neq i} x_j 
\end{array} \right) }{ \left( \begin{array}{c}
     N\\
     1 + \sum_{j\neq i} x_j 
\end{array} \right) }  \nonumber \\
&=  \frac{ \frac{N!}{\left( \sum_{j\neq i} x_j \right)! \left( N - \sum_{j\neq i} x_j\right)!} }{ \frac{N!}{\left( 1 + \sum_{j\neq i} x_j \right)! \left( N - 1 - \sum_{j\neq i} x_j\right)!} } \nonumber \\
&= \frac{ \left( 1+ \sum_{j\neq i} x_j \right) \left( \sum_{j\neq i} x_j \right)! \left( N - \sum_{j\neq i} x_j - 1 \right)! }{ \left( \sum_{j\neq i} x_j \right)! \left( N - \sum_{j\neq i} x_j \right) \left( N - \sum_{j\neq i} x_j - 1 \right)! } \nonumber \\
&= \frac{ 1 + \sum_{j\neq i} x_j }{ N - \sum_{j\neq i} x_j }.
    \label{eq_ration_base_measure_function}
\end{align}

Finally, the derivation of Eq.~\eqref{eq_conditional_prob_x_k_1_given_n_active} is as follows
\begin{align}
    \label{eq_conditional_prob_x_k_1_given_n_active_02}
    & \mathcal{P}\left( x_i = 1 \left | \sum_{j\neq i} x_j = n \right. \right) \nonumber \\
    & = \frac{ \mathcal{P}\left( x_i = 1 , \sum_{j\neq i} x_j = n \right) }{ \mathcal{P}\left( \sum_{j\neq i} x_j = n \right) } \nonumber \\
     & = \frac{ \left( \begin{array}{c}
   N - 1\\
   n 
\end{array} \right) \frac{ h\left( 1 + \sum_{j\neq i} x_j \right) }{Z} \exp\left[ \sum_{k=1}^{n+1} \left( \begin{array}{c}
   n+1 \\
   k 
\end{array} \right) \theta_k \right] }{ \sum_{s \in \left\{0,1 \right \} }  \left( \begin{array}{c}
   N - 1\\
   n 
\end{array} \right) \frac{ h\left( s + \sum_{j\neq i} x_j \right) }{Z} \exp\left[ \sum_{k=1}^{n+s} \left( \begin{array}{c}
   n+s \\
   k 
\end{array} \right) \theta_k \right]   } \nonumber \\
 & = \frac{ \left( \begin{array}{c}
   N - 1\\
   n 
\end{array} \right) \frac{1}{ \left( \begin{array}{c}
   N \\
   n + 1
\end{array} \right) } \frac{ 1 }{Z} \exp\left[ \sum_{k=1}^{n+1} \left( \begin{array}{c}
   n+1 \\
   k 
\end{array} \right) \theta_k \right] }{ \sum_{s \in \left\{0,1 \right \} }  \left( \begin{array}{c}
   N - 1\\
   n 
\end{array} \right) \frac{1}{ \left( \begin{array}{c}
   N\\
   n + s 
\end{array} \right) } \frac{ 1 }{Z} \exp\left[ \sum_{k=1}^{n+s} \left( \begin{array}{c}
   n+s \\
   k 
\end{array} \right) \theta_k \right]   } \nonumber \\
& = \frac{ \left( n + 1 \right) \exp\left[ \sum_{k=1}^{n+1} \left( \begin{array}{c}
   n+1 \\
   k 
\end{array} \right) \theta_k \right]  }{ \left(n+1\right) \exp\left[ \sum_{k=1}^{n+1} \left( \begin{array}{c}
   n+1 \\
   k 
\end{array} \right) \theta_k \right]  + \left(N-n\right) \exp\left[ \sum_{k=1}^{n} \left( \begin{array}{c}
   n \\
   k 
\end{array} \right) \theta_k \right]  } \nonumber \\
& = \frac{ 1 }{ 1  + \frac{N-n}{n+1} \exp\left[ \sum_{k=1}^{n} \left( \begin{array}{c}
   n \\
   k 
\end{array} \right) \theta_k 
- 
\sum_{k=1}^{n+1} \left( \begin{array}{c}
   n+1 \\
   k 
\end{array} \right) \theta_k
\right]  } \nonumber\\
& = \frac{ 1 }{ 1  + \exp\left[ \log \frac{N-n}{n+1} + \theta_{n+1}  + \sum_{k=1}^{n} \theta_k  \left\{ \left( \begin{array}{c}
   n \\
   k 
\end{array} \right) 
- 
   \left( \begin{array}{c}
   n+1 \\
   k 
\end{array} \right) \right\}
\right]  }. \nonumber\\
& = \frac{ 1 }{ 1  +  \exp\left[ \log \frac{N-n}{n+1} + \theta_{n+1}  -  \sum_{k=1}^{n} \theta_k  
   \left( \begin{array}{c}
   n \\
   k-1 
\end{array} \right) 
\right]  }, 
\end{align}
where we used the Pascal's equality to obtain the last equality:
\begin{align}
    \left( \begin{array}{c}
   n \\
   k 
\end{array} \right) + 
\left( \begin{array}{c}
   n \\
   k-1 
\end{array} \right)
=
\left( \begin{array}{c}
   n+1 \\
   k 
\end{array} \right).
\end{align}

\newpage

\newpage
\subsection{Sparsity} \label{APPENDIX_LIMIT_DISTRIBUTIONS}

We consider the shifted-geometric distribution, 
$$
p(r|\mathpzc{f}, \tau) = \frac{1}{Z} \exp\left( -\mathpzc{f} \left( 1 - \frac{1}{1 + \tau r} \right) \right),
$$
where $Z$ is a normalization constant, and show that it becomes a Dirac delta distribution at zero as $\mathpzc{f}$ becomes large. 

First, we rewrite the density as follows:
\begin{align}
    \label{eq_pdf_shifted_geometric_limit_f_sparsity_01}
    p(r|\mathpzc{f}, \tau) & = \frac{ \exp\left( -\mathpzc{f} \left( 1 - \frac{1}{1 + \tau r} \right) \right)  }{ \int_0 ^{1} \exp\left( -\mathpzc{f} \left( 1 - \frac{1}{1 + \tau r'} \right) \right) dr' } \nonumber \\
    &= \frac{1}{  \exp\left( \mathpzc{f}\left( 1 - \frac{1}{1 + \tau r} \right) \right) \int_0 ^{1} \exp\left( -\mathpzc{f} \left( 1 - \frac{1}{1 + \tau r'} \right) \right) dr'  }.
\end{align}
Since the normalization constant is independent of $r$, we can absorb the exponential factor from the numerator into the integrand of the denominator, yielding:
\begin{align}
    \label{eq_pdf_shifted_geometric_limit_f_sparsity_02}
    p(r|\mathpzc{f}, \tau) &= 
    \left(
    \int_0 ^{1} \exp\left( \mathpzc{f} \left(  \frac{1}{1 + \tau r'} - \frac{1}{1 + \tau r} \right) \right) dr'  
    \right)^{-1}
    \nonumber \\
    &= 
    \left( \int_0 ^{r} \exp\left( \mathpzc{f} z_{r, r'} \right) dr' + \int_r ^{1} \exp\left( \mathpzc{f} z_{r, r'} \right) d r' \right)^{-1}
    .
\end{align}
where 
\begin{align}
    \label{eq_z_r}
    z_{r, r'} &= \frac{1}{1 + \tau r'} - \frac{1}{1 + \tau r}. 
\end{align}
Since $z_{r, r'}>0$ for $r' < r$,  the exponential term in the first integral approaches infinity as $\mathpzc{f} \to \infty$: $\exp\left( \mathpzc{f} z_{r, r'} \right) \to \infty$. On the other hand, since we have $z_{r, r'}<0$ for $r' > r$, the exponential term in the second integral approaches zero as $\mathpzc{f} \to \infty$: $\exp\left( \mathpzc{f} z_{r, r'} \right) \to 0$. 

This leads to the following limiting behavior of the density. For $r>0$, we have
\begin{align}
    \label{eq_pdf_shifted_geometric_limit_f_sparsity_03}
    \lim_{\mathpzc{f} \to +\infty} 
    p(r|\mathpzc{f}, \tau) & = \lim_{\mathpzc{f} \to +\infty}  
    \left(
    \int_0 ^{r} \exp\left( \mathpzc{f} z_{r, r'} \right) dr' + \int_r ^{1} \exp\left( \mathpzc{f} z_{r, r'} \right) d r' 
    \right)^{-1}
    \nonumber \\
    & =   
    \left(
    \lim_{\mathpzc{f} \to +\infty}  \int_0 ^{r} \exp\left( \mathpzc{f} z_{r, r'} \right) dr' + \lim_{\mathpzc{f} \to +\infty} \int_r ^{1} \exp\left( \mathpzc{f} z_{r, r'} \right) d r' 
    \right)^{-1}
    \nonumber \\
    &= \left( \infty + 0 \right)^{-1} 
    \nonumber \\
    &= 0.
\end{align}
We note that, since the integrand is non-negative and increases monotonically in $\mathpzc{f}$, the monotone convergence theorem allows us to interchange limit and integral. 

When $r = 0$, however, we have
\begin{align}
    \label{eq_pdf_shifted_geometric_limit_f_sparsity_04}
    \lim_{\mathpzc{f} \to +\infty} 
    p(r|\mathpzc{f}, \tau) & =   
    \left(
    \lim_{\mathpzc{f} \to +\infty}  \int_0 ^{0} \exp\left( \mathpzc{f} z_{r, r'} \right) dr' + \lim_{\mathpzc{f} \to +\infty} \int_0 ^{1} \exp\left( \mathpzc{f} z_{r, r'} \right) d r' 
    \right)^{-1}
    \nonumber \\
    &=  
    \left( 0 + 0 \right)^{-1} 
    \nonumber \\
    &= + \infty.
\end{align}

Considering both cases in Eq.~\eqref{eq_pdf_shifted_geometric_limit_f_sparsity_03} and \eqref{eq_pdf_shifted_geometric_limit_f_sparsity_04},  we conclude that the density converges in the distributional sense to
\begin{align}
    \label{eq_pdf_shifted_geometric_limit_f_sparsity_05}
    \lim_{\mathpzc{f} \to +\infty} 
    p(r|\mathpzc{f}, \tau) & =  \delta\left( r \right),
\end{align}
i.e., a Dirac delta distribution centered at $r=0$. 

For the polylogarithmic exponential distribution, we also obtain the limiting behavior given by $\lim_{\mathpzc{f} \to \infty} p\left( r | \mathpzc{f} , m \right) =\delta\left( r \right) $. For the latter, we omit the proof, but it follows analogously by considering the non-positive strictly decreasing function $\mathpzc{f} \text{Li}_m[-r]$ instead of $-\mathpzc{f} \left( 1 - \frac{1}{1 + \tau r} \right)$.

\newpage
\subsection{Heavy-tailedness}
\label{APPENDIX_HEAVY_TAILEDNESS_PROOF}

In this subsection, we use the definition in Subsection \ref{SUBSECTION_HEAVY_TAILEDNESS} and Appendix \ref{APPENDIX_HEAVY_TAILEDNESS} to formally prove that the distributions in our alternating shrinking models are heavy-tailed.

For a given finite non-decreasing non-negative $u\left(r \right)$, the upper and lower bounds as $\mathpzc{f}$ varies remain unchanged regardless of the distribution (Eq.~\eqref{eq_alternating_shrinking_expectation_bounds}). Since such expectations are strictly decreasing functions of $\mathpzc{f}$ (Lemma \ref{lemma_expectation_decreasing_property}), when comparing expectations between two different distributions, they either intersect at all values of $\mathpzc{f}$ or they can only intersect at the upper and lower bounds for $\mathpzc{f}$. This is the case when we compare the expectations between the (truncated) exponential distribution and our proposed distributions. More explicitly, for any finite non-decreasing non-negative $u\left(r\right)$, the comparison between both expectations can only fall in one of the three following mutually exclusive cases:
\begin{align}
    \label{eq_expectations_comparison_case_01}
    \mathbb{E}_{R | \boldsymbol{\lambda} } \left[ u\left(r\right) \right] > \mathbb{E}^{exp} _{R | \mathpzc{f}}\left[ u\left( r \right) \right] \;\;\forall \mathpzc{f} \in \left(0,+\infty\right),
\end{align}
\begin{align}
    \label{eq_expectations_comparison_case_02}
    \mathbb{E}_{R | \boldsymbol{\lambda} } \left[ u\left(r\right) \right] < \mathbb{E}^{exp} _{R | \mathpzc{f}}\left[ u\left(r\right)\right] \;\;\forall \mathpzc{f} \in \left(0,+\infty\right),
\end{align}
or
\begin{align}
    \label{eq_expectations_comparison_case_03}
    \mathbb{E}_{R | \boldsymbol{\lambda} } \left[ u\left(r\right) \right] = \mathbb{E}^{exp} _{R | \mathpzc{f}}\left[ u\left(r\right) \right] \;\;\forall \mathpzc{f} \in \left(0,+\infty\right).
\end{align}

We note that the case of Eq.~\eqref{eq_expectations_comparison_case_01} (for all $r \in (0,1)$) is equivalent to the first-order stochastic dominance of $F_{\boldsymbol{\lambda}}\left(r\right)$ over $F_{\boldsymbol{\mathpzc{f}}} ^{exp} \left( r \right)$ (see Eq.~\eqref{eq_expectations_comparison_case_01_main}) for our definition of heavy-tailedness for $F_{\boldsymbol{\lambda}}\left(r\right)$.

\begin{itemize}
    \item{\textbf{Polylogarithmic exponential density} \\
    We analyze the following three cases based on the parameter values of this distribution. \\
    \underline{$\mathpzc{f}=1$, $m=1$} \\
    For this case, we find
    \begin{align}
        \label{eq_expectation_poly_f1_m1_vs_exponential_f1}
        \mathbb{E}_{R | \boldsymbol{\lambda} } \left[ r \right] & = \frac{1 - \log\left(2 \right)}{\log\left(2 \right)} \nonumber \\
        & = 0.4427 + \mathcal{O}\left( 1 \times 10^{-5} \right) \nonumber \\
        & > 
        0.4180 + \mathcal{O}\left( 1 \times 10^{-5} \right) \nonumber \\
        & = \frac{1}{1 - e^{-1}} - \frac{e^{-1}}{1 - e^{-1}} \left(2\right) \nonumber \\
        & = \mathbb{E}^{exp} _{R | \mathpzc{f}}\left[ r\right],
    \end{align}
    which falls directly under the case Eq.~\eqref{eq_expectations_comparison_case_01}. \\

    \underline{$\mathpzc{f}\neq 1$, $m=1$} \\

    We consider two exemplary values of $\mathpzc{f} \neq 1$, one below $1$ and the other above $1$. For $\mathpzc{f} = \frac{1}{2}$, we have
    \begin{align}
        \label{eq_expectation_poly_fneq1_m1_vs_exponential_f_01}
        \mathbb{E}_{R | \boldsymbol{\lambda} } \left[ r \right] & = \frac{ 1 }{ 1 - 2^{-\mathpzc{f} +1} } \left( \frac{1 - 2^{-\mathpzc{f} + 2}}{ \mathpzc{f} - 2 } - 2^{-\mathpzc{f} + 1} \right) \nonumber \\        
        & = 0.4714 + \mathcal{O}\left( 1 \times 10^{-5} \right) \nonumber \\
        & > 
        0.4585 + \mathcal{O}\left( 1 \times 10^{-5} \right) \nonumber \\
        & = \frac{1}{ \mathpzc{f} \left( 1 - e^{-\mathpzc{f}} \right) } - \frac{e^{-\mathpzc{f}}}{1 - e^{-\mathpzc{f}}} \left( 1 + \frac{1}{\mathpzc{f}}\right) \nonumber \\
        &=  \mathbb{E}^{exp} _{R | \mathpzc{f}}\left[ r\right],
    \end{align}
    and for $\mathpzc{f} = 5$
    \begin{align}
        \label{eq_expectation_poly_fneq1_m1_vs_exponential_f_02}
        \mathbb{E}_{R | \boldsymbol{\lambda} } \left[ r \right] & = 0.2444 + \mathcal{O}\left( 1 \times 10^{-5} \right) \nonumber \\
        & > 
        0.1932 + \mathcal{O}\left( 1 \times 10^{-5} \right) \nonumber \\
        &=  \mathbb{E}^{exp} _{R | \mathpzc{f}}\left[ r\right],
    \end{align}
    where it is evident that both inequalities, Eqs.~\eqref{eq_expectation_poly_fneq1_m1_vs_exponential_f_01} and \eqref{eq_expectation_poly_fneq1_m1_vs_exponential_f_02}, also belong to the case Eq.~\eqref{eq_expectations_comparison_case_01}. 
    Considering all three inequalities, Eqs.~\eqref{eq_expectation_poly_f1_m1_vs_exponential_f1}, \eqref{eq_expectation_poly_fneq1_m1_vs_exponential_f_01} and \eqref{eq_expectation_poly_fneq1_m1_vs_exponential_f_02}, fall into the only possible case and continuity of the distribution with respect to $\mathpzc{f}$, Eq.~\eqref{eq_expectations_comparison_case_01} holds strictly for all $\mathpzc{f} \in \left(0,\infty\right)$, 
    which proves that the polylogarithmic exponential distribution is heavy-tailed for $m=1$.
    
    \underline{$m>1$} \\
    For this case, we first consider the function $q\left( r ; \boldsymbol{\lambda}_m \right) = \mathpzc{f}\text{Li}_{m}\left[-r \right] $,  with $\boldsymbol{\lambda}_m =\left\{ m, \mathpzc{f} \right\}$, as the parameter $m \in \left\{1, 2, 3, \cdots \right\}$ increases. We then note the following (setting $\mathpzc{f}=1$ to simplify the notation)
    \begin{align}
        \label{eq_polylogarithmic_function_comparison_m_01}
        \text{Li}_{m}\left[-r \right] &= \sum_{j=1}^{\infty} \frac{ \left(-1\right)^{j} r^{j} }{ j^{m} } \nonumber \\
        &= -r + \sum_{j=2}^{\infty} \frac{ \left(-1\right)^{j} r^{j} }{ j^{m} } \nonumber \\
        & > 
        -r + \sum_{j=2}^{\infty} \frac{ \left(-1\right)^{j} r^{j} }{ j^{m+1} } \nonumber \\
        & = \text{Li}_{m+1}\left[-r \right],
    \end{align}
    which holds because, starting at $j=2$, the summation of all remaining terms is positive due to the alternating shrinking property and the positivity of the term at $j=2$, and also because the denominators are bigger for $m+1$ compared to those at $m$. Based on inequality Eq.~\eqref{eq_polylogarithmic_function_comparison_m_01}, the following can be noted
    \begin{align}
        \label{eq_polylogarithmic_function_comparison_m_02}
        \text{Li}_{1}\left[-r \right] 
        = -\log\left(1 + r \right) & >  \cdots >  \text{Li}_{m}\left[-r \right] \nonumber \\
        \; & >  \cdots >  \lim_{k \to \infty} \text{Li}_{k}\left[-r \right] = -r,
    \end{align}
    which means that as $m$ increases, the sparsity of the underlying distribution also increases by slowly changing the nonlinearity in $\text{Li}_{m}\left[-r \right]$ from the sublinear function $-\log\left(1 + r\right)$ up to the linear function $-r$. The modulation by $m$ is different from that of the $\mathpzc{f}$ parameter. Note that the modulation by $m$ starts from the second-order term, while that of $\mathpzc{f}$ affects the terms at all orders. 
    
    We can now obtain a similar result to Eq.~\eqref{eq_alternating_shrinking_expectation_bounds}. Note the unimodality of the PDF $p\left( r | m, \mathpzc{f}\right)$ with a peak at $r=0$, the strictly decreasing property of the function $\mathpzc{f}\text{Li}_{m}\left[-r \right]$ in $r$ and the fact that $m$ makes $\text{Li}_{m}\left[-r \right]$ decrease faster along $r$. Specifically, $m$ makes this polylogarithmic function go from a sublinear to a linear decreasing function. Based on the previous, and since $\exp\left( \mathpzc{f}\text{Li}_{1}\left[-r\right] \right)$ is a positive integrable function which preserves this decreasing property along $r$, the following holds for any non-negative non-decreasing function $u\left( r \right)$ as $m$ varies
    \begin{eqnarray}
      \label{eq_alternating_shrinking_expectation_poly_bounds_in_m}
    \int_{0} ^{1} \frac{\exp\left( \mathpzc{f}\text{Li}_{1}\left[-r\right] \right)}{Z_1} u\left( r \right) dr & = & \int_{0} ^{1} \frac{\exp\left( -\mathpzc{f}\log\left(1+r\right) \right)}{Z_1} u\left( r \right) dr \nonumber \\
    & > & \int_{0} ^{1} \frac{\exp\left( \mathpzc{f}\text{Li}_{2}\left[-r\right] \right)}{Z_2} u\left( r \right) dr \nonumber \\
    & > & \cdots \nonumber \\
    & > & \int_{0} ^{1} \lim_{k \to \infty}  \frac{\exp\left( \mathpzc{f}\text{Li}_{k}\left[-r\right] \right)}{Z_k} u\left( r \right) dr \nonumber \\
    & = & \mathbb{E}^{exp} _{R | \mathpzc{f}}\left[ u\left( r \right) \right],
    \end{eqnarray}
    where $Z_k$ is the normalization constant for the corresponding distribution with $m=k$. It is evident from inequalities in Eq.~\eqref{eq_alternating_shrinking_expectation_poly_bounds_in_m} that all polylogarithmic exponential distributions fall in the case of inequality Eq.~\eqref{eq_expectations_comparison_case_01} for all $\mathpzc{f}>0$. Therefore, all of our proposed polylogarithmic exponential distributions are heavy-tailed except at the limit of $m \to \infty$, where the higher-order effects disappear and the underlying distribution becomes exponential.
    }
    
    \item{\textbf{Shifted-geometric density} \\

    For this case we have the function $q\left( r ; \boldsymbol{\lambda} \right) = \mathpzc{f}\left( \frac{1}{1 + \tau r} - 1 \right) $. For some $r\in \left(0,1\right)$ and for $\tau_L = 0 < \tau_k < \tau_{k+1} < 1 = \tau_U$ the following relation can be seen (setting $\mathpzc{f}=1$ to simplify)
    \begin{align}
        \label{eq_shifted_geometric_function_comparison_tau_01}
        \frac{1}{1 + \tau_L r} - 1 & =  0  \nonumber \\
                    & > \cdots > \frac{1}{1 + \tau_{k} r} - 1 > \frac{1}{1 + \tau_{k+1} r} - 1 > \cdots \nonumber \\
                    & > \frac{1}{1 + \tau_U r} - 1 = \frac{1}{1 + r} - 1,
    \end{align}
    because larger values for $\tau_k$ decrease the value of the first term in each $q\left( r ; \boldsymbol{\lambda} \right)$ function. Similarly to the previous case in the polylogarithmic exponential distribution, the inequalities in Eq.~\eqref{eq_shifted_geometric_function_comparison_tau_01} also imply that the sparsity of the underlying distribution increases with increasing values of $\tau_k$. This modulation by $\tau_k$ affects terms at all underlying orders of interaction (see Eq.~\eqref{eq_polylogarithm_series}) but in a different way when compared to the modulation by $\mathpzc{f}$ because it modifies the shape of the decreasing function as opposed to $\mathpzc{f}$, which modulates the strength of the function. 
    Similarly to the polylogarithmic exponential distribution, due to the unimodality of the PDF $p\left( r | \tau, \mathpzc{f}\right)$ with a peak at $0$, the strictly decreasing property of the function $\mathpzc{f}\left( \frac{1}{1 + \tau r} - 1 \right)$ in $r$ and because $\tau$ modifies the shape of the decreasing function such that it increases sparsity, the following holds for a given $r \in \left(0,1\right)$ and $\tau_L = 0 < \tau < 1  = \tau_U$
    \begin{align}\label{eq_alternating_shrinking_expectation_SG_bounds_in_tau}
    \int_{0} ^{1} \frac{\exp\left( \mathpzc{f}\left( \frac{1}{1 + \tau_L r} - 1 \right) \right)}{Z_{\tau_L}} r dr 
    & = \frac{1}{2} \nonumber \\
    & > \int_{0} ^{1} \frac{\exp\left( \mathpzc{f}\left( \frac{1}{1 + \tau r} - 1 \right) \right)}{Z_{\tau}} r dr \nonumber \\
    & > \int_{0} ^{1} \frac{\exp\left( \mathpzc{f}\left( \frac{1}{1 + \tau_{U} r} - 1 \right) \right)}{Z_{\tau_U}} r dr \nonumber \\
    & = \int_{0} ^{1} \frac{\exp\left( \mathpzc{f}\left( \frac{1}{1 + r} - 1 \right) \right)}{Z_{\tau_U}} r dr,
    \end{align}
    where $Z_{\tau}$ is the normalization constant for the corresponding distribution with parameter $\tau$. The previous inequalities in Eq.~\eqref{eq_alternating_shrinking_expectation_SG_bounds_in_tau} hold for some $\mathpzc{f}>0$. We will now compare the expectation of the shifted-geometric exponential distribution with that of the exponential distribution, using the lower bound expectation value for $\tau$, which corresponds to the highest value $\tau=1$. We use Eq.~\eqref{eq_mean_shifted_geometric_02} to compute this expectation value. This is (setting $\mathpzc{f} = 5$)
    \begin{align}
        \label{eq_expectation_SG_vs_exponential_f_tau1}
        \mathbb{E}_{R | \boldsymbol{\lambda} } \left[ r \right] & = 0.2993 + \mathcal{O}\left( 1 \times 10^{-5} \right) \nonumber \\
        & > 
        0.1932 + \mathcal{O}\left( 1 \times 10^{-5} \right) \nonumber \\
        &=  \mathbb{E}^{exp} _{R | \mathpzc{f}}\left[ r\right],
    \end{align}
    which means that the shifted-geometric distribution for $\tau=1$ can only belong to the case Eq.~\eqref{eq_expectations_comparison_case_01}. However, since the expectation at $\tau=1$ is the lower bound expectation value for this distribution Eq.~\eqref{eq_alternating_shrinking_expectation_SG_bounds_in_tau}, case Eq.~\eqref{eq_expectations_comparison_case_01} applies also to all other values of $0<\tau<1$ and therefore the shifted-geometric exponential distribution is heavy-tailed for all $\mathpzc{f}>0$ and $0<\tau<1$.
    }
\end{itemize}

\newpage
\subsection{Entropy} \label{APPENDIX_ENTROPY}
\noindent \underline{Entropy for the polylogarithmic exponential distribution} \\

For the polylogarithmic exponential distribution ($m=1$), the entropy is
\begin{align}
& \mathbb{E}_{R}\left[ -\log\left( p\left( r | \mathpzc{f} , m=1 \right) \right) \right] \nonumber \\
   \phantom{==} 
   &= -\int_{0}^{1} \frac{1}{Z} \exp\left[ -\mathpzc{f} \log\left(1 + r\right) \right] \log\left( \frac{1}{Z} \exp\left[ -\mathpzc{f} \log\left(1 + r\right) \right] \right) dr \nonumber \\
   &=  \int_{0}^{1} \frac{1}{Z} \frac{1}{\left( 1 + r \right)^{\mathpzc{f}}} \log\left(\left(1 + r \right)^{\mathpzc{f}}\right) dr + \log\left( Z \right) \nonumber \\
   &=  \frac{1}{Z} \left[ \left( -\frac{ \mathpzc{f} \log\left( 1 + r \right) }{ \left( \mathpzc{f}-1 \right) \left( 1 + r \right)^{\mathpzc{f}-1} } - \frac{ \mathpzc{f} }{ \left( \mathpzc{f} - 1 \right)^2 \left( 1 + r  \right)^{\mathpzc{f} -1 } } \right)  \Bigg |_{0} ^{1} \right] + \log\left( Z \right) \nonumber \\
   &= \frac{\mathpzc{f}}{ 1 - 2^{- \mathpzc{f} + 1 } } \left[ \frac{ 1 }{ \left( \mathpzc{f} -1 \right)  } - \frac{\log 2}{ 2^{\mathpzc{f}-1} } - \frac{ 1 }{ \left( \mathpzc{f} - 1 \right) 2^{\mathpzc{f} - 1} } \right] \nonumber \\
   & \phantom{=============} + \log\left( \frac{ 1 - 2^{-\mathpzc{f} + 1 } }{ \mathpzc{f}-1 } \right)  \phantom{====}  \mathpzc{f} \neq 1,
   \label{eq_entropy_exp_log_modulated_dist_f_neq_1}
\end{align}
\noindent and where
\begin{align}
& \mathbb{E}_{R}\left[ -\log\left( p\left( r | \mathpzc{f}=1 , m=1 \right) \right) \right] \nonumber \\
   \phantom{==} &= \frac{1}{Z}\left[ \frac{1}{2} \left( \log\left( 1 + r \right) \right)^2  \Bigg |_{0}^{1} \right] + \log\left(Z \right) \nonumber \\
   &= \frac{1}{2} \log 2 + \log\left( \log 2  \right)  \phantom{====} \mathpzc{f} = 1.
    \label{eq_entropy_exp_log_modulated_dist_f_eq_1}
\end{align}

\noindent \underline{Entropy for the shifted-geometric exponential distribution}    \\

The entropy of the shifted-geometric exponential distribution is 
\begin{align}
& \mathbb{E}_{R}\left[ -\log\left( p\left( r | \mathpzc{f} , \tau \right) \right) \right] \nonumber \\
   \phantom{==} 
   &= -\int_{0}^{1} \frac{1}{Z} \exp\left[ \mathpzc{f} \left( \frac{1}{1 + \tau r} - 1 \right) \right] \log\left( \frac{1}{Z} \exp\left[ \mathpzc{f} \left( \frac{1}{1 + \tau r} - 1 \right) \right] \right) dr \nonumber \\
   &= \log\left(Z \right) - \frac{\mathpzc{f}}{Z} \int_{0}^{1} \left( \exp\left[ \mathpzc{f}\left( \frac{1}{1 + \tau r} - 1 \right)  \right] \frac{1}{1 + \tau r} \right) dr + \mathpzc{f} \nonumber \\
   &= \log\left(Z \right) - \frac{ \mathpzc{f} e^{-\mathpzc{f}} \left( \text{Ei}\left(\mathpzc{f}\right) - \text{Ei}\left( \frac{\mathpzc{f}}{ 1 + \tau } \right)   \right) }{ \left( 1 + \tau \right) \exp\left[ \mathpzc{f}\left( \frac{1}{1 + \tau} - 1 \right) \right] + \mathpzc{f} e^{-\mathpzc{f}} \left( \text{Ei}\left( \mathpzc{f} \right) - \text{Ei}\left( \frac{\mathpzc{f}}{1 + \tau} \right) \right) - 1  } + \mathpzc{f},
   \label{eq_entropy_exp_shifted_geometric_dist}
\end{align}

To obtain the last equality, we computed the integral in the second term as follows 
\begin{align}
\phantom{=} & \frac{f}{Z} \int_{0}^{1}  \exp\left[ \mathpzc{f} \left( \frac{1}{1 + \tau r} - 1 \right) \right] \frac{1}{ 1 + \tau r } dr \nonumber \\
&= \frac{1}{Z} \left( \frac{ 1 + \tau r }{ \tau } \exp\left[ \mathpzc{f} \left( \frac{1}{ 1 + \tau r } - 1 \right)  \right] - \frac{\mathpzc{f} e^{-\mathpzc{f}}  }{ \tau } \text{Ei}\left( \frac{\mathpzc{f}}{ 1 + \tau r } \right)  \right . \nonumber \\
& \phantom{======================} \left . -\frac{ 1 + \tau r  }{ \tau } \exp\left[ \mathpzc{f} \left( \frac{1}{ 1 + \tau r } - 1 \right) \right] \right) \Bigg |_0 ^{1} \nonumber \\
&= \frac{1}{Z} \left[  \frac{ 1 + \tau }{ \tau } \exp\left[ \mathpzc{f} \left( \frac{1}{1 + \tau} - 1 \right) \right] - \frac{ \mathpzc{f} e^{-\mathpzc{f}}  }{ \tau } \text{Ei}\left( \frac{\mathpzc{f}}{ 1 + \tau } \right)  \right .  \nonumber \\
& \phantom{==================}  - \frac{ 1 + \tau }{ \tau } \exp\left[ \mathpzc{f}\left( \frac{1}{ 1 + \tau } - 1 \right) \right]  \nonumber \\
& \phantom{==================} \left . - \frac{1}{\tau} + \frac{\mathpzc{f} e^{-\mathpzc{f}} }{ \tau } \text{Ei}\left( \mathpzc{f} \right) + \frac{1}{\tau}   \right ] \nonumber \\
&= \frac{  \mathpzc{f} e^{-\mathpzc{f} } \left( \text{Ei}\left( \mathpzc{f} \right) -  \text{Ei}\left( \frac{\mathpzc{f}}{ 1 + \tau } \right) \right)  }{  \left( 1 + \tau \right) \exp\left[ \mathpzc{f}\left( \frac{1}{1 + \tau} - 1 \right) \right] + \mathpzc{f} e^{-\mathpzc{f}}  \left( \text{Ei}\left( \mathpzc{f} \right) - \text{Ei}\left( \frac{\mathpzc{f}}{ 1 + \tau } \right) \right) - 1   } .
\label{eq_entropy_second_term_shifted_geometric_integral}
\end{align}

\newpage
\subsection{Heat capacity} \label{APPENDIX_HEAT_CAPACITY}

\noindent \underline{Heat capacity for the polylogarithmic exponential distribution} \\

\noindent The normalization constant and the derivatives of the heat capacity (Eq.~\eqref{eq_heat_capacity_definition}) corresponding to the polylogarithmic exponential distribution ($m=1$) with $\mathpzc{f} \neq 1$ are as follows
\begin{align}
    Z &= \frac{1 - 2^{-\mathpzc{f} + 1} }{ \mathpzc{f}-1 },
    \label{eq_heat_capacity_exp_log_modulated_dist_01_01}
\end{align}
\begin{align}
    \frac{d Z}{d \mathpzc{f}} &= \frac{2^{-\mathpzc{f}+1 } \log 2  }{ \mathpzc{f}-1 } - \frac{ \left( 1 - 2^{-\mathpzc{f}+1 } \right) }{ \left( \mathpzc{f} - 1 \right)^2 } ,
    \label{eq_heat_capacity_exp_log_modulated_dist_01_02}
\end{align}
\noindent and
\begin{align}
    \frac{ d^2 Z  }{\left( d \mathpzc{f} \right)^2  } &= -\frac{ 2^{-\mathpzc{f}+1} \left( \log 2  \right)^2 }{ \left( \mathpzc{f}-1 \right) } - \frac{ 2^{-\mathpzc{f}+2 } \log 2  }{ \left( \mathpzc{f} - 1 \right)^2 } + \frac{ 2\left( 1 - 2^{- \mathpzc{f}+1 } \right) }{ \left( \mathpzc{f}-1 \right)^3 }.
    \label{eq_heat_capacity_exp_log_modulated_dist_01_03}
\end{align}

We now show that the limit is $1$ for the heat capacity of the polylogarithmic exponential distribution ($m=1$) as $\mathpzc{f} \to \infty$.
\begin{align}
   \lim_{\mathpzc{f} \to \infty} C\left(\mathpzc{f}\right) &= \lim_{\mathpzc{f} \to \infty} \left[ \frac{ \mathpzc{f}^2 \frac{d^2 }{\left( d \mathpzc{f}\right)^2} Z - \mathpzc{f}^2 \left( \frac{d Z}{d \mathpzc{f}} \right)^2 }{ Z^2 } \right],
   \label{eq_limit_f_to_infty_heat_capacity}
\end{align}
\noindent where
\begin{align}
\mathpzc{f}^2 \frac{d^2 Z}{\left( d\mathpzc{f} \right)^2} Z 
&=  \frac{2 \mathpzc{f}^2 Z^2}{ \left( \mathpzc{f} -1 \right)^2 } - \mathpzc{f}^2 Z \left(  \frac{2^{-\mathpzc{f}+1} \log^2 2 }{ \mathpzc{f}-1 } + \frac{ 2^{-\mathpzc{f}+1} \log 4 }{ \left( \mathpzc{f}-1 \right)^2 } \right),
    \label{eq_limit_f_to_infty_heat_capacity_polylogarithmic_m1_nume_01}
\end{align}
\noindent and
\begin{align}
   \mathpzc{f}^2 \left( \frac{d Z}{d \mathpzc{f}} \right)^2 
   &= \mathpzc{f}^2 \left( \frac{2^{-2 \mathpzc{f}+2} \log^2 2}{ \left( \mathpzc{f} - 1 \right)^2 } - \frac{ 2^{-\mathpzc{f}+1} \log 4 }{ \left( \mathpzc{f}-1 \right)^2 } Z + \frac{Z^{2}}{ \left( \mathpzc{f} - 1 \right)^2 } \right).
   \label{eq_limit_f_to_infty_heat_capacity_polylogarithmic_m1_nume_02}
\end{align}
\noindent Then, continuing with Eq.~\eqref{eq_limit_f_to_infty_heat_capacity} we obtain the following limit
\begin{align}
\lim_{\mathpzc{f} \to \infty} C\left(\mathpzc{f}\right) &= \lim_{\mathpzc{f} \to \infty}\left[ \frac{ \frac{\mathpzc{f}^2 }{\left(\mathpzc{f}-1\right)^2} Z^2 -\frac{ \mathpzc{f}^2 2^{-\mathpzc{f}+2} \log^2  2 }{ \left( \mathpzc{f}-1 \right) } Z - \frac{ \mathpzc{f}^2 2^{-2 \mathpzc{f} + 2} \log^2 2 }{ \left( \mathpzc{f} -1\right)^2 } }{ Z^2 } \right] \nonumber \\
&= \frac{ \lim_{\mathpzc{f} \to \infty} \frac{ \mathpzc{f}^2 Z^2 }{ \left( \mathpzc{f}-1 \right)^2 } - \lim_{\mathpzc{f} \to \infty} \left( \frac{\mathpzc{f}^2 2^{-\mathpzc{f}+1} \log^2 2 }{ \mathpzc{f}-1 } Z + \frac{ \mathpzc{f}^2 2^{-2 \mathpzc{f}+2} \log^2 2 }{ \left( \mathpzc{f} - 1 \right)^2 } \right)  }{ \lim_{\mathpzc{f} \to \infty} Z^2 } \nonumber \\
&= \lim_{\mathpzc{f} \to \infty} \frac{ \mathpzc{f}^2 Z^2 }{\left( \mathpzc{f}-1 \right)^2 Z^2} - 0 = 1.
    \label{eq_limit_f_to_infty_heat_capacity_polylogarithmic_m1_02}
\end{align}

\noindent \underline{Heat capacity for the shifted-geometric exponential distribution}    \\

\noindent Similarly to the polylogarithmic case, for the heat capacity of the shifted-geometric exponential distribution, we obtain
\begin{align}
Z &= \frac{ 1 + \tau  }{\tau} \exp\left[ \mathpzc{f}\left( \frac{1}{1+\tau} - 1 \right) \right] + \frac{ \mathpzc{f} e^{-\mathpzc{f}}  }{ \tau } \left( \text{Ei}\left( \mathpzc{f} \right) - \text{Ei}\left( \frac{\mathpzc{f}}{1 + \tau} \right) \right) -\frac{1}{\tau},
    \label{eq_heat_capacity_exp_shifted_geometric_dist_1_1}
\end{align}
\begin{align}
\frac{d Z}{d \mathpzc{f}} &= -\frac{ 1 + \tau }{\tau} \exp\left[ \mathpzc{f}\left( \frac{1}{1 + \tau} - 1 \right) \right]  + \frac{1}{\tau}  + \frac{ 1 - \mathpzc{f} }{\tau} e^{-\mathpzc{f}}  \left( \text{Ei}\left( \mathpzc{f} \right) - \text{Ei}\left( \frac{\mathpzc{f} }{ 1 + \tau } \right) \right) 
    \label{eq_heat_capacity_exp_shifted_geometric_dist_1_2}
\end{align}
\noindent and 
\begin{align}
\frac{ d^2 Z }{ \left( d \mathpzc{f} \right)^2 } &=  \frac{1 + \tau - \mathpzc{f}^{-1}}{\tau}  \exp\left[ \mathpzc{f} \left( \frac{1}{1 + \tau} - 1 \right) \right] + \frac{1-\mathpzc{f}}{\mathpzc{f} \tau} \nonumber \\
& \phantom{=} +\frac{ e^{-\mathpzc{f}}  }{\tau} \left( \mathpzc{f} - 2   \right) \left( \text{Ei}\left( \mathpzc{f} \right) - \text{Ei}\left( \frac{\mathpzc{f}}{1 + \tau} \right)  \right) .
    \label{eq_heat_capacity_exp_shifted_geometric_dist_1_3}
\end{align}

\end{document}